\documentclass[a4paper,11pt]{article}

\usepackage{latexsym}
\usepackage{a4wide}
\usepackage{graphicx, subfigure}
\usepackage{cite}
\usepackage{tabularx}
\usepackage{array}
\usepackage{graphics}     
\usepackage{amsmath}
\usepackage{amssymb}
\usepackage{longtable} 
\usepackage{verbatim} 
\usepackage[table]{xcolor}
\usepackage{booktabs}
\usepackage{hyperref} 
\usepackage[utf8]{inputenc}

\definecolor{light-gray}{gray}{0.90}

\renewcommand{\arraystretch}{1.2} 
 
\newcommand{\cen}[1]{\multicolumn{1}{c}{#1}}
\newcommand{\ra}[1]{\renewcommand{\arraystretch}{#1}}
\newcommand{\rb}[1]{\renewcommand{\tabcolsep}{#1}}
\newcommand{\av}[1]{\langle #1 \rangle}

\newcommand{\err}[2]{^{+#1}_{-#2}}

\begin{document}

\hfill {\tt CERN-TH-2016-046, IPM/P.A-421, MITP/16-003}  

\def\thefootnote{\fnsymbol{footnote}}
 
\begin{center}

\vspace{3.cm}

{\huge\bf {On the anomalies in the latest LHCb data} }

\setlength{\textwidth}{11cm}
                    
\vspace{2.cm}
{\Large\bf  
T.~Hurth\footnote{Email: tobias.hurth@cern.ch}$^{,a}$,
F.~Mahmoudi\footnote{Also Institut Universitaire de France, 103 boulevard Saint-Michel, 75005 Paris, France\\ \hspace*{0.49cm} Email: nazila@cern.ch}$^{,b,c}$,
S.~Neshatpour\footnote{Email: neshatpour@ipm.ir }$^{,d}$
}
 
\vspace{1.cm}
{\em $^a$PRISMA Cluster of Excellence and  Institute for Physics (THEP)\\
Johannes Gutenberg University, D-55099 Mainz, Germany}\\[0.2cm]
{\em $^b$Univ Lyon, Univ Lyon 1, ENS de Lyon, CNRS, Centre de Recherche Astrophysique de Lyon UMR5574, F-69230 Saint-Genis-Laval, France}\\[0.2cm]
{\em $^c$Theoretical Physics Department, CERN, CH-1211 Geneva 23, Switzerland}\\[0.2cm]
{\em $^d$School of Particles and Accelerators,
Institute for Research in Fundamental Sciences (IPM)
P.O. Box 19395-5531, Tehran, Iran}

\end{center}

\renewcommand{\thefootnote}{\arabic{footnote}}
\setcounter{footnote}{0}

\vspace{1.cm}
\thispagestyle{empty}
\centerline{\bf ABSTRACT}
\vspace{0.5cm}
{Depending on the assumptions on the power corrections to the exclusive $b \to s \ell^+\ell^-$ decays, the latest data of the LHCb collaboration - based on the 3 fb$^{-1}$ data set and on two different experimental analysis methods - still shows some tensions with the SM predictions. 
We present a detailed analysis of the theoretical inputs and various global fits to all the available $b\to s \ell^+\ell^-$ data.
This constitutes the first global analysis of the new data of the LHCb collaboration based on the hypothesis that these tensions  can be at least partially explained by new physics contributions. 
In our model-independent analysis we present one-, two-, four-, and also five-dimensional global fits in the space of Wilson coefficients to all available $b \to s \ell^+\ell^-$ data. We also compare the two different experimental LHCb analyses of the angular observables in  $B \to K^{*} \mu^+\mu^-$. 
We explicitly analyse the dependence of our results on the assumptions about power corrections, but also on the errors present in  the form factor calculations.  
Moreover, based on our new global fits we present predictions for ratios of observables which may show a sign of lepton non-universality. 
Their measurements would crosscheck the LHCb result on the ratio $R_K = {\rm BR}(B^+ \to K^+ \mu^+ \mu^-) / {\rm BR}(B^+ \to K^+ e^+ e^-)$ in the low-$q^2$ region 
which deviates from the SM prediction by $2.6\sigma$.

\newpage

%%%%%%%%%%%%%%%%%%%%%%%%%%%%%%%%%%%%%%%%%%%%%%%%
\section{Introduction}
The LHCb collaboration has recently presented the angular analysis of the $B^0 \to K^{*0} \mu^+\mu^-$ decay with the 3 fb$^{-1}$ data set. They use two analysis methods. The observables are  determined using an unbinned maximum likelihood fit and by  the principal angular moments~\cite{Aaij:2015oid}. In addition, a new analysis on the angular observables in $B_s \to \phi \mu^+\mu^-$ has been presented~\cite{Aaij:2015esa}.

These new analyses of the LHCb collaboration have been eagerly awaited in view of the previous  LHCb analysis of the $B^0 \to K^{*0} \mu^+\mu^-$ based on the 1 fb$^{-1}$ data set~\cite{Aaij:2013qta}. The LHCb collaboration had announced a local discrepancy of $3.7 \sigma$ from the Standard Model (SM) predictions in one bin for one of the angular observables~\cite{Aaij:2013qta}. There had been also more, yet smaller tensions with the  SM  predictions in other observables. This announcement was followed by  a large number of theoretical analyses showing that, 
due to the large hadronic uncertainties in exclusive modes,  it is not clear at all whether this anomaly is a first sign for new physics beyond the SM or a consequence of underestimated hadronic 
power corrections or just a statistical fluctuation~\cite{Jager:2012uw,Jager:2014rwa,Descotes-Genon:2013wba,Altmannshofer:2013foa,Hambrock:2013zya,Gauld:2013qba,Buras:2013qja,Gauld:2013qja,Datta:2013kja,Beaujean:2013soa,Horgan:2013pva,Buras:2013dea,Hurth:2013ssa,Mahmoudi:2014mja,Descotes-Genon:2014uoa,Khodjamirian:2010vf,Khodjamirian:2012rm,Lyon:2014hpa,Descotes-Genon:2015uva}.
 
In the recent analysis based on the 3 fb$^{-1}$ data set the LHCb collaboration now announced a $3.4\sigma$ tension with predictions based on the SM within  a global fit to the complete  set of $CP$-averaged observables~\cite{Aaij:2015oid}. They point out that this tension could be explained by contributions from physics beyond the SM or by unexpectedly large hadronic effects that are underestimated in the SM predictions.  

Regarding the latter option it is important to note that there is a significant difference in the theoretical accuracy of the  inclusive and exclusive $b \to s \ell^+\ell^-$ decays in the low-$q^2$ region.  
In the inclusive case, there is a theoretical description of power corrections; they can be calculated or at least {\it estimated} within the theoretical approach (for reviews see~\cite{Hurth:2010tk,Hurth:2007xa,Hurth:2003vb}).~\footnote{In the inclusive case one can show that  if only the leading operator of the electroweak hamiltonian is considered, one is led to a local operator product expansion (OPE). In this case, the leading hadronic power corrections in the decay $\bar B \to X_s \ell^+ \ell^-$ scale  with $1/m_b^2$ and $1/m_b^3$ only and have already been analysed~\cite{Falk:1993dh}. A systematic and careful analysis of hadronic power corrections including all relevant operators has been  performed in the case of the decay $\bar B \rightarrow X_s \gamma$~\cite{Benzke:2010js}.  Such linear power corrections can be analysed within soft-collinear effective  theory (SCET). This  analysis goes beyond the local OPE. 
An additional uncertainty of $\pm 5\%$ has been identified. The analysis in the case of $\bar B \rightarrow X_s \ell^+\ell^-$ is work in progress. There is no reason to expect any large deviation from the  $\bar B \to X_s \gamma$ result.
Nonfactorisable power corrections that scale with $1/m_c^2$ have first  been considered in Ref.~\cite{Buchalla:1997ky}, but can be now included in the systematic analysis of hadronic power corrections and be calculated quite analogously to those in the decay $\bar B \rightarrow X_s \gamma$~\cite{Benzke:2010js}. Moreover, in the KS approach~\cite{Kruger:1996cv,Kruger:1996dt}
one absorbs factorisable long-distance charm rescattering effects (in which the $\bar B \to X_s c\bar c$ transition can be factorised into the product of $\bar s  b$ and $c\bar c$ colour-singlet currents) into the matrix element
of the leading semileptonic operator $O_9$. Following the inclusion of nonperturbative corrections scaling with $1/m_c^2$, the KS approach avoids double-counting.} In contrast, in the exclusive  case  there is no theoretical description of power corrections existing within the theoretical framework of QCD factorisation and SCET which is the standard theoretical framework for these exclusive decay modes in the low-$q^2$ region. Thus, power corrections can only be {\it guesstimated}. This issue makes it rather difficult or even impossible to separate new physics effects from such potentially large hadronic power corrections within these exclusive angular observables. So these tensions might stay unexplained until Belle II will clarify the situation by measuring the corresponding inclusive $b\to s \ell \ell$  observables as was demonstrated 
in Ref.~\cite{Hurth:2013ssa,Hurth:2014zja,Hurth:2014vma}.

Thus, it is also obvious that the significance of the tension depends in principle on the precise guesstimate of the unknown power corrections within the SM prediction. Because  the two sets of SM predictions - LHCb compares with in their first and in their latest analysis - use rather different guesstimates~\cite{Descotes-Genon:2013vna,Beaujean:2013soa} the quoted standard deviations in both  analyses cannot be compared directly.~\footnote{In this sense the significance of the tension with the SM has not been really  reduced within the new 3 fb$^{-1}$ measurement compared to the one based on the 
1 fb$^{-1}$ data set.}

This situation motivated a recent theory analysis in which the unknown power corrections were just fitted to the data~\cite{Ciuchini:2015qxb}
using an ansatz with 18 additional real parameters in the fit. However, this fit  to the data needs very large power corrections in the critical bins.
If one compares the fitted theory predictions in this analysis with the leading contributions based on QCD factorisation then one finds $20\% $ but also $50\% $ or even larger
power corrections relative to the leading contribution. The existence of such large hadronic corrections cannot be ruled out in principle.  
The authors of Ref.\cite{Ciuchini:2015qxb} come to the well-known conclusion that it is difficult to deduce the existence of new physics effects unambiguously from the measurements of exclusive $b \to s \ell^+\ell^-$ observables.

In this respect, another tension in the LHCb data of $b \to s \ell^+\ell^-$  gains importance. The ratio $R_K = {\rm BR}(B^+ \to K^+ \mu^+ \mu^-) / $ ${\rm BR}(B^+ \to K^+ e^+ e^-)$ in the low-$q^2$ region had been measured by LHCb using the full 3 fb$^{-1}$ of data, showing a $2.6\sigma$ deviation from the SM prediction~\cite{Aaij:2014ora}.
This discrepancy has been addressed in many  studies~\cite{Alonso:2014csa,Hiller:2014yaa,Ghosh:2014awa,Biswas:2014gga,Davidtalk:2014,Hurth:2014vma,Glashow:2014iga,Altmannshofer:2015sma,Becirevic:2016zri,Chiang:2016qov,Celis:2015eqs,Crivellin:2016vjc,Sahoo:2015pzk,Bauer:2015knc,Falkowski:2015zwa,Altmannshofer:2015mqa,Guadagnoli:2015nra,Alonso:2015sja,Celis:2015ara,Crivellin:2015era,Hiller:2014ula,Bhattacharya:2014wla,Crivellin:2015mga,Sierra:2015fma,Calibbi:2015kma,Belanger:2015nma}. It is often claimed that the electromagnetic corrections might not  have been fully taken into account in this measurement. Thus one might wonder whether this sign of lepton non-universality could be traced back to logarithmically enhanced QED corrections. These corrections were calculated in the inclusive case in Ref.~\cite{Huber:2015sra}. However, LHCb uses the PHOTOS Monte Carlo to eliminate the impact of collinear photon emissions from the final state electrons. Therefore, such corrections do not seem to apply to the ratio $R_K$, especially if one considers the agreement between the PHOTOS results and the  analytical calculations in the inclusive case in Ref.~\cite{Huber:2015sra}. Nevertheless, it would be an interesting check  to correct for photon radiation using data-driven methods that do not rely on PHOTOS.

In contrast to the anomaly in the rare decay $B \to K^{*} \mu^+\mu^-$ which is affected by power corrections, the ratio  $R_K$ is theoretically rather clean and its tension with the SM cannot be explained by power corrections. But independent of this difference, both tensions might be healed by new physics in the semi-leptonic operator contribution $C_9$. Therefore the measurement of other ratios which could show lepton non-universality would be a very important crosscheck of the present $R_K$ result but also of the anomalies  in the angular observables in $B \to K^{*} \mu^+\mu^-$. Thus, we present predictions for such ratios based on the global fits in this paper. 

Finally, we mention that there is an alternative approach to calculate the nonlocal charm-loop effects. In Ref.~\cite{Khodjamirian:2010vf} a local OPE near the light cone for the gluon emission from the $c$-quark loop at $q^2 \ll m_c^2$  is used to derive a nonlocal effective quark-antiquark gluon operator. Then the LCSR 
approach leads to an effective resummation of the soft-gluon part valid for $q^2 \ll m_c^2$ only. 
But a prediction of the charm loops effects up to the open charm threshold is achieved by a phenomenological model of the charmonium resonances via a dispersion relation. Because not all contributions were included in the dispersion relation this prediction for larger $q^2$ should not be regarded as the final result. 
An analogous complete calculation of the soft-gluon non-factorisable contributions in the case of the decay of $B^0 \to K^0  \ell^+\ell^-$~\cite{Khodjamirian:2012rm} leads to a moderate effect. The corresponding predictions of the branching ratios for the various bins  are given in Appendix A next to our predictions based on QCD factorisation. The two sets of predictions are compatible with each other at the 1$\sigma$ level. 

Based on QCD factorisation there are two strategies to calculate the decay amplitudes:
the so-called soft form factor (soft FF) approach (see~\cite{Kruger:2005ep,Egede:2008uy}) and the full form factor (full FF) approach (see~\cite{Altmannshofer:2008dz}). Both methods have been implemented in {\tt SuperIso v3.5}~\cite{Mahmoudi:2007vz,Mahmoudi:2008tp} which is used for all the calculations presented in this work. 
We discuss in detail the advantages and the disadvantages of these two approaches and  critically analyse the guesstimate of power corrections which are used in the literature (Section 2).  
Among the input parameters and experimental data used in our study, we mainly discuss the theoretical error estimation and the form factor calculations via the light cone sum rule method and also define the statistical method used
for the fitting analysis of the $b\to s$ data in Section 3.
In contrast to many  analyses in the literature, we consider one- and two- but also four- and five-dimensional global fits (in the space of Wilson coefficients) 
within our model-independent analysis of the present data. We analyse the dependence of our results on the theoretical approach used, as well as the assumptions about power corrections, and also on the errors present in the form factor calculations. We also investigate the effect of the $S_5$ and the $R_K$ measurements on the 
global fit results. Moreover, we compare the Wilson coefficient fits when using the 
two different experimental LHCb analysis for $B \to K^{*} \mu^+\mu^-$ angular observables which this collaboration presented recently~\cite{Aaij:2015oid} (Section 4).
Our results represent the first global fit analysis with the latest LHCb results for $B \to K^{*} \mu^+\mu^-$ angular observables assuming that the tensions can be at least partially explained by new physics contributions.
Finally, we use our global fit results to present predictions for other ratios of observables  which may indicate lepton non-universality. As mentioned above, these measurements will be important crosschecks of the anomalies discussed in this paper (Section 5). We give our conclusions in Section 6 and additional analyses and more details on our theoretical predictions in the appendices.

%%%%%%%%%%%%%%%%%%%%%%%%%%%%%%%%%%%%%%%%%%%%%%%%
\section{Soft vs. full form factor approach}\label{app:FullSoft}
The theoretical description of the $B\to K^{(*)} \mu^+ \mu^-$ or    $B_s \to \phi  \mu^+ \mu^-$         decays in the low-$q^2$ region is based on the QCD-improved  factorisation (QCDf) approach and its field-theoretical formulation of Soft-Collinear Effective Theory (SCET)~\cite{Beneke:2001at,Beneke:2004dp}. The combined limit of a heavy $b$ quark and an energetic meson $M$ like $K^*$, $K$, or $\phi$  leads to the schematic form of the decay amplitude
\begin{equation} \label{FACT1}
\mathcal T   =  C  \xi   + \Phi_B \otimes T  \otimes \Phi_{K^{(*)},\phi} + {\cal O}(\Lambda_{\rm QCD} / m_b),
\end{equation}
which is valid to leading order in $\Lambda_{\rm QCD} / m_b$ and to all orders in $\alpha_s$. Thus 
the decay amplitude factorises into process-independent non-perturbative quantities like $B\to M$ soft form factors ($\xi$)  and light-cone distribution amplitudes (LCDAs) of the heavy and light mesons ($\Phi$) and perturbatively calculable quantities ($C$, $T$). The key issue of this factorisation formula is that there are non-factorisable contributions (second term) beyond  the ones described by form factors (first term). 

However, at order $\Lambda_{\rm QCD}/m_b$ the factorisation formula breaks down. This means that  the power corrections cannot be calculated within QCDf in general. As already emphasised in the introduction, this is the main drawback of this approach which makes it difficult to identify new physics effects. 
There are  power corrections to both terms in the factorisation formulae, which are called factorisable and non-factorisable power corrections respectively. 

The large energy limit allows for simplifications~\cite{Charles:1998dr}. The seven independent full QCD  $B \to K^*$ form factors $V,A_{0,1,2}$ and $T_{1,2,3}$, 
for example, are reduced  to two universal soft form factors $\xi_\perp$ and $\xi_\parallel$ in this limit. These relations can also be written via a factorisation formula in a schematic way:
\begin{equation} \label{FACT2}
F_{\rm full} (q^2)  = D \xi_{\rm soft} + \Phi_B \otimes T_F \otimes \Phi_M  + {\cal O}(\Lambda_{\rm QCD}/m_b)\,,
\end{equation}
%%%
where $D$ and $T_F$ are perturbatively calculable functions. There are non-factorisable contributions (second term), but also power corrections (third term).

Based on QCDf there are two strategies to calculate the decay amplitudes:
the so-called soft form factor (soft FF) approach (see~\cite{Kruger:2005ep,Egede:2008uy}) and the full form factor (full FF) approach (see~\cite{Altmannshofer:2008dz}). 

Using the former strategy, one takes advantage of the factorisation formula~(\ref{FACT1}) and the universal soft form factors. The various meson spin amplitudes at leading order in $\Lambda_{\rm QCD}/m_b$ and $\alpha_s$ turn out to be linear in the soft form factors $\xi_{\bot,\|}$ and also in the short-distance Wilson coefficients. As was explicitly shown in Refs.~\cite{Egede:2008uy, Egede:2010zc}, these simplifications allow to design a set of optimised observables, in which any soft form factor dependence (and its corresponding uncertainty) cancels out for all low dilepton mass squared $q^2$ at leading order in $\alpha_s$  and $\Lambda_{\rm QCD}/m_b$. An optimised set of independent observables  was constructed in Refs.~\cite{Matias:2012xw,Descotes-Genon:2013vna}, in which almost all observables are free from hadronic uncertainties which are related to the form factors.

However, there are additional hadronic uncertainties in the factorisation formula~(\ref{FACT1}), namely the  unknown factorisable and non-factorisable power corrections.
They are not calculable and can only be guesstimated through dimensional arguments {within the  QCDf/SCET approach}~\cite{Egede:2008uy,Egede:2010zc}. 

As Eq.~(\ref{FACT2}) indicates, the factorisable power corrections can be avoided by using the full QCD form factors. Thus, within this full FF approach one faces only unknown power corrections to the non-factorisable contributions; this means  one has to guesstimate only power corrections to leading contributions from 4-quark operators $O_{1-6}$ and the chromomagnetic operator $O_8$.~\footnote{We note that there are also additional leading (calculable) non-factorisable corrections due to the rewriting of the soft form factors into full form factors according to Eq.~(\ref{FACT2}).} It is important to note that the operators $O_7,O_9$, and $O_{10}$ do not induce non-factorisable contributions.~\footnote{In Ref.~\cite{Descotes-Genon:2014uoa}, the central value of factorisable power corrections are determined by fitting the soft form factors to the relevant full form factors using Eq.~(\ref{FACT2}), but  the corresponding uncertainties should still be guesstimated. From the principle point of view this procedure does not allow for any advantages in respect to the full FF form factor approach where one uses the full QCD form factors directly with the uncertainties of the light cone sum rule (LCSR) calculation. However, 
in view of the  fact that there is only one independent LCSR calculation including the correlations~\cite{Straub:2015ica}, the authors of Ref.~\cite{Descotes-Genon:2014uoa} 
prefer to use the full form factor calculation of Ref.~\cite{Khodjamirian:2010vf} with much larger uncertainties. But no correlations are given in this calculation, so only this  hybrid approach is possible.}

As it was argued in Ref.~\cite{Straub:2015ica}, the correlations between QCD form factors in the 
large energy limit of the meson can be established also in this full FF approach using the equation of motions. We will discuss this issue further in Section~\ref{sec:form factor}. 

In a global analysis both strategies should lead to similar results if all correlations are taken into account. If one focuses on a single observable, the advantage of the optimised observables on the theory side is unfortunately diminished by the large errors on the experimental side~\cite{Aaij:2015oid}.

In this work we have used the full FF approach for calculating the $B\to K^{(*)} \bar \ell \ell$ and $B_s \to \phi \bar \ell \ell$ observables, as the main approach. For the uncertainty regarding the SM predictions only the non-factorisable power corrections are relevant in the full FF approach. We have guesstimated the effect due to these corrections by considering either 5, 10 or 20\% and also 60\% errors as described in Section~\ref{sec:error}. 

As a crosscheck we use also the soft FF approach for an independent global fit with the same input data. In this case  we make a guesstimate of the factorisable and the non-factorisable power corrections. 

For the sake of completeness, we note that in the high-$q^2$ region a local operator product expansion can be used. One finds small power corrections  below $5\%$ ~\cite{Grinstein:2004vb,Beylich:2011aq}.
Duality violation effects have been estimated within a model of resonances and have been determined  to be again below $5\%$. But new resonances are found in this region (see e.g. Ref.~\cite{Lyon:2014hpa}),
so an update of this calculation would be desirable.  
The different theory framework in the high-$q^2$ region allows for a non-trivial crosscheck, especially of any new physics hypothesis in the low-$q^2$ region.

%%%%%%%%%%%%%%%%%%%%%%%%%%%%%%%%%%%%%%%%%%%%%%%%
\section{Input of model-independent analysis}

We follow the methodology used in Ref.\cite{Hurth:2014vma}, but consider the following important improvement and updates within the experimental and theoretical inputs:

\begin{itemize}
\item Unless otherwise specified, we have used the full form factor approach  for the SM predictions of the  $B\to K^{(*)} \bar \ell \ell$ and $B_s \to \phi \bar \ell \ell$ 
observables,  assuming 10\% error for the 
power corrections (see Sections~\ref{app:FullSoft} and~\ref{sec:error}). Crosschecks are also performed considering the soft form factor approach and different assumptions for the power corrections.

\item We consider the LHCb measurements on the branching ratio as well as the angular observables
of the $B_s \to \phi \bar \ell \ell$ decay.
The theoretical treatment of the  $B_s \to \phi \bar \ell \ell$ decay is similar to the
$B \to K^* \bar \ell \ell$ decay, with the requisite replacements of the masses and hadronic parameters
as well as the necessary changes resulting from the spectator effects.
The required modifications can be found in~\cite{Beneke:2004dp}.
Since the $B_s \to \phi \bar{\ell} \ell$ decay is not self-tagging, unlike the $B \to K^* \bar{\ell} \ell$ decay,
the time integrated untagged average over the $\bar{B_s^0}$ and $B_s^0$ decay distributions, including the effects
of $\bar{B_s^0} - B_s^0$ mixing, should be calculated~\cite{Bobeth:2008ij}.
Details about how to include this effect can be found in
Refs.~\cite{Bobeth:2008ij,Bobeth:2011gi,Descotes-Genon:2015hea}. We have
implemented this effect following Ref.~\cite{Descotes-Genon:2015hea}.

\item For the $B \to K^*$ and $B_s \to \phi$ form factors we have used the combined fit results of
LCSR calculations~\cite{Ball:2004rg,Straub:2015ica} and lattice computations~\cite{Horgan:2013hoa,Horgan:2015vla} given in Ref.~\cite{Straub:2015ica}. The combined fit results are applicable
for the entire kinematic range of $q^2$ (see Section~\ref{sec:form factor}). 

\item For the $B \to K$ form factors we have used  the combined fit results 
of LCSR calculations~\cite{Ball:2004ye,Bartsch:2009qp} and 
lattice computations~\cite{Bouchard:2013pna} given in Ref.~\cite{Altmannshofer:2014rta}.
The combined fit results are applicable for the entire kinematic range of $q^2$.

\item For the branching ratios of $B^{0(+)}\to K^{0(+)} \mu^+ \mu^-$ decays,
the $[1.1, 6.0]$ and $[15.0, 22.0]$ GeV$^2$ bins have been used (see Section~\ref{sec:c9c10}). 

\item We have calculated the theoretical errors and correlations for the $B\to K^{(*)} \mu^+ \mu^-$ and $B_s \to \phi \mu^+ \mu^-$ decays with various assumptions for the power corrections as described in Section~\ref{sec:error}.

\item For the experimental values of $B \to K^* \mu^+ \mu^-$ angular observables,  
we have considered both the LHCb results determined by the maximum likelihood fit method or
the results for the finer bins determined by the method of moments~\cite{Aaij:2015oid} for comparison.
Unless otherwise specified, for our analysis we have considered the experimental results 
obtained by the method of moments and used the recent results of the LHCb collaboration based on 3 fb$^{-1}$ of data~\cite{Aaij:2015oid} for the $S_i$ observables.  
Alternatively, the experimental results of the optimised observables ($P_i$) can be used.
The latter are determined by reparametrising the $S_i$ observables.
The $F_L(1-F_L)$ dependence of the optimised observables results in large and asymmetrical experimental errors in the bins where $F_L$ is close to zero or one, specially for the results which have been obtained using the method of moments (see Table 9 of Ref.~\cite{Aaij:2015oid}  or Table~\ref{tab:BtoKstarMoment} in Appendix~\ref{app:SMpredict} of the present paper).

\item Some of the relevant updated input parameters of this work have been given in Table~\ref{tab:input}.

\item The full list of observables that we have used as well as their 
SM predictions and experimental measurements can be found in Appendix~\ref{app:SMpredict}.

\end{itemize}

\begin{table}[!t]
\begin{center}
\footnotesize{\begin{tabular}{|lr|lr|}\hline
$M_{B^0}=5.27958 (17)$ GeV                  & \cite{Agashe:2014kda} & $M_{K^*}=0.89166 (26)$ GeV                & \cite{Agashe:2014kda} \\
$M_{B^+}=5.27926 (17)$ GeV                  & \cite{Agashe:2014kda} & $M_{K^0}=0.497614 (24)$ GeV               & \cite{Agashe:2014kda} \\ \hline
$ f_B=190.5 \pm 4.2 $ MeV        & \cite{Aoki:2013ldr}            & $ \tau_{B_s} = 1.512 \pm 0.007\ {\rm ps}$ & \cite{Agashe:2014kda} \\
$ f_{B_s} = 227.7 \pm  4.5$ MeV  & \cite{Aoki:2013ldr}            &                                           &                       \\ \hline
$a_{1}^{K}$(1 GeV)$=0.06\pm0.03$            & \cite{Ball:2006fz}    & $f_K=156 \pm 5$ MeV                   & \cite{Bazavov:2009bb} \\
$a_{2}^{K}$(1 GeV)$=0.25 \pm0.15$           & \cite{Ball:2006wn}    &                                           &                       \\ \hline
\end{tabular}}
\caption{Input parameters. \label{tab:input}}
\end{center}
\end{table}

%%%%%%%%%%%%%%%%%%%%%%%%%%%%%%%%%%%%%%%%%%%%%%
\subsection{Form factors}\label{sec:form factor}
As mentioned above, for the $B \to K^*$ and $B_s \to \phi$ form factors
we have used the combined fit results of LCSR and lattice calculations~\cite{Straub:2015ica} which are applicable
for the entire kinematic range of $q^2$. One should consider the following aspects: 
\begin{itemize}
\item 
The form factor uncertainties are drastically reduced
in the BSZ parametrisation compared to the ones in Refs.~\cite{Khodjamirian:2006st,Khodjamirian:2010vf}. 
The main reason is very simple. In the latter reference the authors used another 
LCSR approach in which  the role of the $B$ meson and the light meson is 
exchanged~\cite{DeFazio:2005dx,Khodjamirian:2005ea,DeFazio:2007hw}.
Our knowledge about the $B$ meson distribution amplitude is rather restricted 
compared to the one about the light-meson amplitude. This simple fact leads to 
much larger errors in this alternative LCSR approach somehow by construction. 
 The BSZ parametrisation goes back to the original calculation given in Ref.\cite{Ball:2004rg}.
An independent check of the calculation is missing.
In view of this, we vary the errors given in LCSR calculation of Ref.~\cite{Straub:2015ica} in our global fits in order
to test the impact of the error estimation in this analysis.  
\item 
In Ref.~\cite{Straub:2015ica} strong correlations between the QCD form factors are derived.
In addition to the standard ones due to physical input parameters, the authors  argue that the equation of motions 
imply the  additional correlations between sum rule specific input parameters like the threshold parameters.
This might be problematic because the latter parameters are intrinsic parameters of each separate LCSR
calculations. However, these additional correlations are introduced in a two-step procedure:
First, separate LCSR calculations are done without these additional correlations 
between the sum rule specific parameters in order to verify that the intrinsic  threshold parameters 
of the LCSR of specific tensor and vector form factors have to be equal up to some variations and that also ratios
of the corresponding form factors are equal to one. In a second step these correlations are implemented 
 in the direct LCSR calculation of all QCD form factors. 
These additional correlations have analogous implications as  the form factor relations in the 
large energy limit~\cite{Charles:1998dr}.
We will also test the impact  of these additional correlations on our final theoretical predictions. 

\end{itemize}

An explicit numerical comparison between the form factors calculated 
in Ref.\cite{Khodjamirian:2010vf} (KMPW) and in Ref.\cite{Straub:2015ica} (BSZ) is given in Appendix~\ref{app:FF}.
Moreover, a comparison for some of the angular observables when using 
these two different sets of form factors, as well as when employing the two different theoretical approaches (soft FF and full FF) have been given in Appendix~\ref{app:SoftFullFF}.

%%%%%%%%%%%%%%%%%%%%%%%%%%%%%%%%%%%%%%%%%%%%%%%%
\subsection{Error estimation}
\label{sec:error}
We have used a Monte Carlo program taking 200,000 random points for the error calculation:
\begin{itemize}
\item All the input parameters are varied with Gaussian distributions in their 1$\sigma$ range.

\item The scales $\mu_0$ and $\mu_b$ are varied with flat distributions in the $(\frac{M_W}{2},\, 2 M_W)$ 
and $(\frac{m_b^{\rm pole}}{2}, 2 m_b^{\rm pole})$, respectively.

\item The form factor parameters are varied using distributions reproducing the correlation matrices. 

\item In the soft FF approach for $B \to K^* \bar \ell \ell$ in the low-$q^2$ region, the factorisable and non-factorisable power corrections have been considered collectively at the transversity spin amplitude level. The amplitudes are varied according to $ A_i \to A_i \times \left( 1+ b_i e^{\theta_i} +  c_i (q^2/6 \; {\rm  GeV}^2)e^{\phi_i} \right)$, where $i=\perp,\parallel,0$, with $\theta_i,\phi_i\in (-\pi,\pi)$ and
$b_i$ have been varied in the $(-0.1,+0.1)$, $(-0.2,+0.2)$, $(-0.3,+0.3)$ ranges and  
$c_i$ in the $(-0.25,+0.25)$, $(-0.5,+0.5)$, $(-0.75,+0.75)$ ranges which we  
refer to as 10, 20 and 30\% error for the power corrections, respectively. 
For the high-$q^2$ region the multiplicity factor that we use is $\left( 1+ d_i e^{\alpha_i} \right)$
with $\alpha_i \in (-\pi,\pi)$ and $d_i \in (-0.1,+0.1)$, $(-0.2,+0.2)$, $(-0.3,+0.3)$ for the 10, 20 and 30\% errors, respectively. 

\item For the full FF approach only the non-factorisable power corrections are missing, hence considering an overall 
multiplicative factor for the transversity amplitudes would overestimate these corrections.
To guesstimate  the non-factorisable power corrections 
we  multiply the hadronic terms which remain after putting $C_{7,9,10}^{(\prime)}$ to zero following Ref.~\cite{Descotes-Genon:2014uoa}.
{We have multiplied the three hadronic terms by a multiplicative factor similar to the soft FF case  with
$b_i \in (-0.05,+0.05), (-0.1,+0.1), (-0.2,+0.2),(-0.6,+0.6)$ and $c_i \in (-0.125,+0.125)$, $(-0.25,+0.25)$, $(-0.50,+0.50)$,$(-1.5,+1.5)$ for the 5, 10, 20 and 60\% error, respectively. 
For the high-$q^2$ region, there is no difference between the soft FF and full FF approaches and we have considered 5, 10 and 20\% errors but also 
60\% again.}
\item For $B_s \to \phi \bar \ell \ell$ and $B\to K \bar \ell \ell$ error estimation we have used a procedure  
similar to the $B\to K^* \bar \ell \ell$ decay.

\item We compute the theoretical errors and correlations between all the observables which are given in the appendices and we combine them with the experimental correlation matrices to obtain the $\chi^2$.

\end{itemize}

We finally note that fitting a polynomial ansatz for the non-factorisable power corrections as it has been done in Ref.~\cite{Ciuchini:2015qxb} and varying randomly 
the coefficients of the polynomial for an estimation of the power correction error are two different approaches. However, if the non-factorisable power corrections needed in the fit and the power correction errors used in the FF approach are of the same magnitude, one expects a similar compatibility with the SM predictions. 
As mentioned in the introduction, the fit to the data presented in Ref.~\cite{Ciuchini:2015qxb} needs rather large non-factorisable power corrections in the critical bins.  
A comparison of the fitted theory predictions with the leading contributions based on QCD factorisation finds 
that the power corrections needed in the fit within the critical bin from 4 GeV$^2$ to 6 GeV$^2$
are  at the level of observables for example  +23\% in $S_3$, -42\% in $S_5$, and -44\% in $S_4$ relative to the leading contribution. 
On the other hand, a 5\%, 10\%, or 20\% estimation of the non-factorisable power correction error described above (following  Ref.~\cite{Descotes-Genon:2014uoa}) leads to an error of maximal 1.5\%, 3\% or 6\% at the observable level in $S_3$, $S_4$ and $S_5$ respectively. Nevertheless the 60\%  error estimation in this framework 
leads to 17\%-20\% error in these three observables. And a 150\% estimation of the power correction error is needed to reproduce errors up to 50\% at the observable level which are comparable in size with the corrections needed in the fit presented in Ref.~\cite{Ciuchini:2015qxb}. 
As a consequence, one should not expect a large  impact of the 5\%, 10\%, or 20\%  power correction error in our global analysis.

%%%%%%%%%%%%%%%%%%%%%%%%%%%%%%%%%%%%%%%%%%%%%%%%
\subsection{Statistical methods} 
\label{sec:statistics}
We use the absolute  $\chi^2$ method to verify the goodness of the fit in the first step. In a second step we consider the difference of the $\chi^2$ with the minimum $\chi^2$ to obtain the allowed regions for the Wilson coefficients ($\Delta \chi^2$). 
We always scan over a specific number of Wilson coefficients $\delta C_i$ at the $\mu_b$ scale and include all available correlations. 

We also directly obtain the allowed regions from the absolute $\chi^2$. This procedure leads to larger allowed
regions  for the Wilson coefficients with respect to the use of the $\Delta \chi^2$. One reason for this is that some of the observables are less sensitive to some Wilson coefficients, while they contribute in a democratic way to the number of degrees of freedom. 
However, the statistical meaning of the two dimensional contours within the absolute $\chi^2$ is that for a point in the 1$\sigma$ interval allowed region, there is at least one solution with the corresponding
values of the Wilson coefficients that has a $\chi^2$ probability corresponding to less than one Gaussian standard deviation with respect to the full set of measurements (see Ref. \cite{Hurth:2013ssa} for more details). 

Within  our analysis below we will arrive at the case that the absolute $\chi^2$ fit leads to no compatibility at $68\%$ C.L. and the $95\%$ C.L. regions are small. In this case, 
the $\Delta \chi^2$ metrology does not make much sense. This indicates the possible role of the absolute $\chi^2$ fit as a check of the goodness of the fit.

%%%%%%%%%%%%%%%%%%%%%%%%%%%%%%%%%%%%%%%%%%%%%%%%
\section{Results}

\subsection{Observable dependency on Wilson coefficients}
The tensions between the experimental measurement for $B\to K^* \mu^+ \mu^-$ observables
and their SM predictions with $10\%$ non-factorisable power corrections can be 
explained by modified Wilson coefficients ($C_i= C_i^{SM} + \delta C_i$), where $\delta C_i$ can 
be due to some new physics effects.
In Figs.~\ref{fig:BKstarmumu:AFB}-\ref{fig:BKstarmumu:S7-S9} -- shown in the Appendix~\ref{app:PiShapes} --
the impact of the modification of a single Wilson coefficient for different  
$B \to K^* \mu^+ \mu^-$ observables has been demonstrated for the benchmark values of 
$\delta C_7^{(\prime)}=\pm 0.1$ and $\delta C_{9,10}^{(\prime)}=\pm 1.0$, where in each case all but one of
the Wilson coefficients have been kept to their SM values. The effect of the modified Wilson coefficients on 
the optimised observables are also given in Appendix~\ref{app:PiShapes}.
The benchmark values for  $\delta C_{7}^{(\prime)}$ and $\delta C_{9,10}^{(\prime)}$ 
correspond to a modification of $\sim 33\%$ and $\sim 25\%$ 
with respect to the SM values of $C_{7}$ and $C_{9,10}$, respectively.

The observables are interdependent through the Wilson coefficients. While modifying some Wilson coefficients can reduce the tension with experimental data for a  specific observable/bin it can increase the tension in some other observables/bins as can be seen in Figs.~\ref{fig:BKstarmumu:AFB}-\ref{fig:BKstarmumu:S7-S9} in Appendix~\ref{app:PiShapes}.
For instance, while $\delta C_{7}=-0.1$ reduces the tension with data for $\langle S_5 \rangle_{[2.5-4.0]}$, it increases the tension for $\langle S_4 \rangle_{[2.5-4.0]}$.
Or for example, $\delta C_{10}=+1.0$ reduces the tension for $\langle d {\rm BR}/dq^2 \rangle_{[2.0-4.3]} $ 
while increasing it for $\langle d {\rm BR}/dq^2 \rangle_{[4.30-8.68]}$
or $\langle F_L \rangle_{[4.0-6.0]}$.
Moreover, there are observables which are not very sensitive to any Wilson coefficient such as $S_{7,8,9}$ and there are
those which are dependent to only certain Wilson coefficients such as $S_3$  which is much more sensitive to 
primed Wilson coefficients, most specifically to $C_7^{\prime}$ in $q^2 \lesssim 3$ GeV$^2$ range~\cite{Becirevic:2011bp} and less so to $C_{10}^\prime$ in the $q^2 \gtrsim 3$ GeV$^2$ region.
In addition to the $B \to K^* \mu^+ \mu^-$ observables, the other $b\to s$ transitions are also dependent on the 
Wilson coefficients.
Of course, in order to get the best agreement with data one should find the best value for the Wilson coefficient(s) 
through a fitting with a  method such as the method of least squares.

\subsection{Global fit results assuming new physics in one operator only}
Considering that new physics effects only appear in one operator,
we make a $\chi^2$ fit by scanning over
a single Wilson coefficient while keeping the other Wilson coefficients to their SM values.
{We here use the full FF approach with a 10\% power correction error.} 
In the case of  lepton flavour universality ($C_i^\mu=C_i^e$),
an 18\% reduction to $C_9$ gives the most probable scenario with a $\chi^2_{\rm min}$ of 123.8.
Assuming this scenario to be the correct description of the $b \to s$ data, the SM value for $C_9$  (corresponding
to $\delta C_9=0$) is in $3.0\sigma$ tension with 
the best fit value (${\rm Pull}_{\rm SM}$).
On the other hand, considering contributions from electrons and muons to be different, 
the most probable scenario is for $C_9^\mu$ to receive a 21\% reduction compared to
the SM value for $C_9$. The $\chi^2_{\rm min}$ of this scenario is 115.5, which is significantly reduced compared to the one in which lepton universality is assumed. 
Here the SM value is in 4.2$\sigma$ tension with the best fit value of $C_9^\mu$.
The best fit values of the different single Wilson coefficient fits as well as their 68 and 95\% confidence level regions are given in Table~\ref{tab:OneOpFit}.

\begin{table}[!b]
\begin{center}
\scalebox{0.80}{
\begin{tabular}{l|ccccc}
 & \cen{b.f. value} & $\chi^2_{\rm min}$ & ${\rm Pull}_{\rm SM}$ & 68\% C.L. &95\% C.L. \\ 
\hline \hline
%
%%%%%%%%%%%%%%%%%
$\delta C_{9}/C_{9}^{\rm SM} $            & $-0.18$ & $123.8$ & $3.0\sigma$  & $[-0.25,-0.09]$  & $[-0.30,-0.03]$   \\ 
$\delta C_{9}^\prime/C_{9}^{{\rm SM}} $   & $+0.03$ & $131.9$ & $1.0\sigma$  & $[-0.05,+0.12]$  & $[-0.11,+0.18]$   \\ 
$\delta C_{10}/C_{10}^{\rm SM} $          & $-0.12$ & $129.2$ & $1.9\sigma$  & $[-0.23,-0.02]$  & $[-0.31,+0.04]$   \\ \hline
$\delta C_{9}^{\mu}/C_{9}^{\rm SM} $      & $-0.21$ & $115.5$ & $4.2\sigma$  & $[-0.27,-0.13]$  & $[-0.32,-0.08]$   \\ 
$\delta C_{9}^e/C_{9}^{{\rm SM}} $        & $+0.25$ & $124.3$ & $2.9\sigma$  & $[+0.11,+0.36]$  & $[+0.03,+0.46]$   
%%%%%%
%%%%%%
\end{tabular}}
\caption{Best fit values and the corresponding 68 and 95\% confidence level regions 
in the one operator global fit to the $b \to s$ data as described in the text. 
In the last two rows the $\chi^2$ fits are done when considering lepton non-universality.
\label{tab:OneOpFit}} 
\end{center} 
\end{table}

\subsection{Global two operator fit}}
In this section we have considered new physics effects to appear in two operators by varying 
\{$C_9,C_{10}$\}, \{$C_9,C_{9}^\prime$\} and \{$C_9^\mu,C_9^e$\} separately.\\

\subsubsection{Fit results for \{$C_9, C_{10}$\}}
\label{sec:c9c10}

\begin{figure}[!t]
\begin{center}
\includegraphics[width=7.cm]{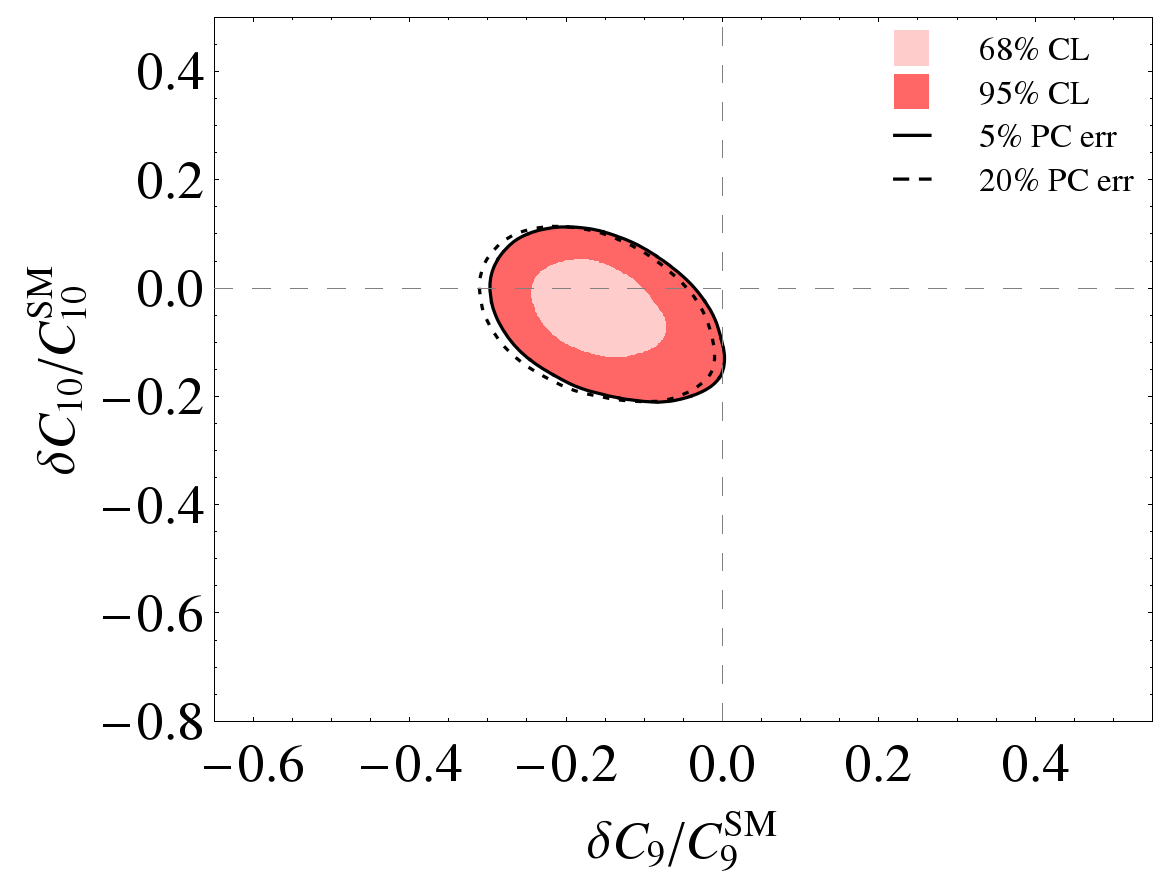}\qquad
\includegraphics[width=7.cm]{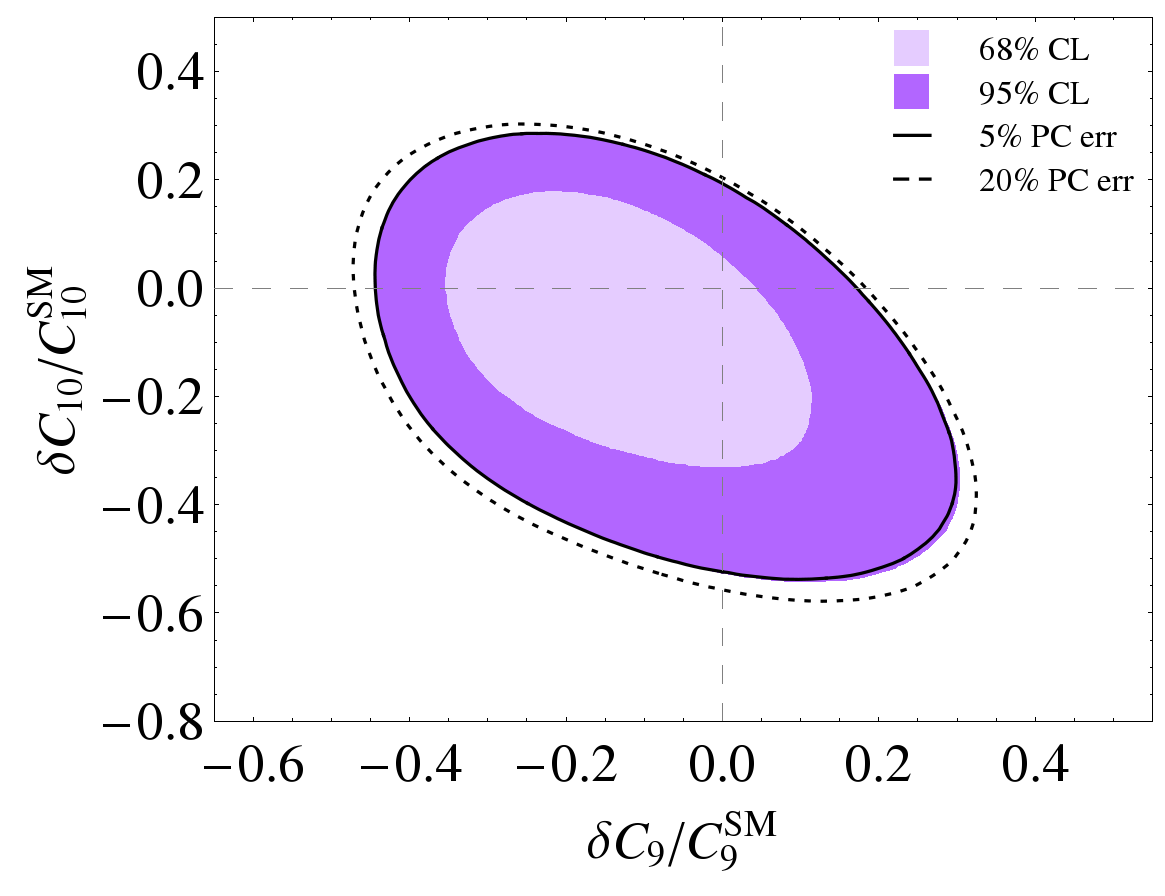}
\caption{Global fit results for $C_9, C_{10}$ by using \emph{full FF} approach and considering 10\% power corrections  when 
employing the $\Delta \chi^2$ (left) and absolute $\chi^2$ method (right). 
The solid (dashed) lines correspond to the  allowed regions at 2$\sigma$ when considering 5\% (20\%) power corrections.
\label{fig:2-full_c9-c10_all}}
\end{center}
\end{figure}

\begin{figure}[!t]
\begin{center}
\includegraphics[width=7.cm]{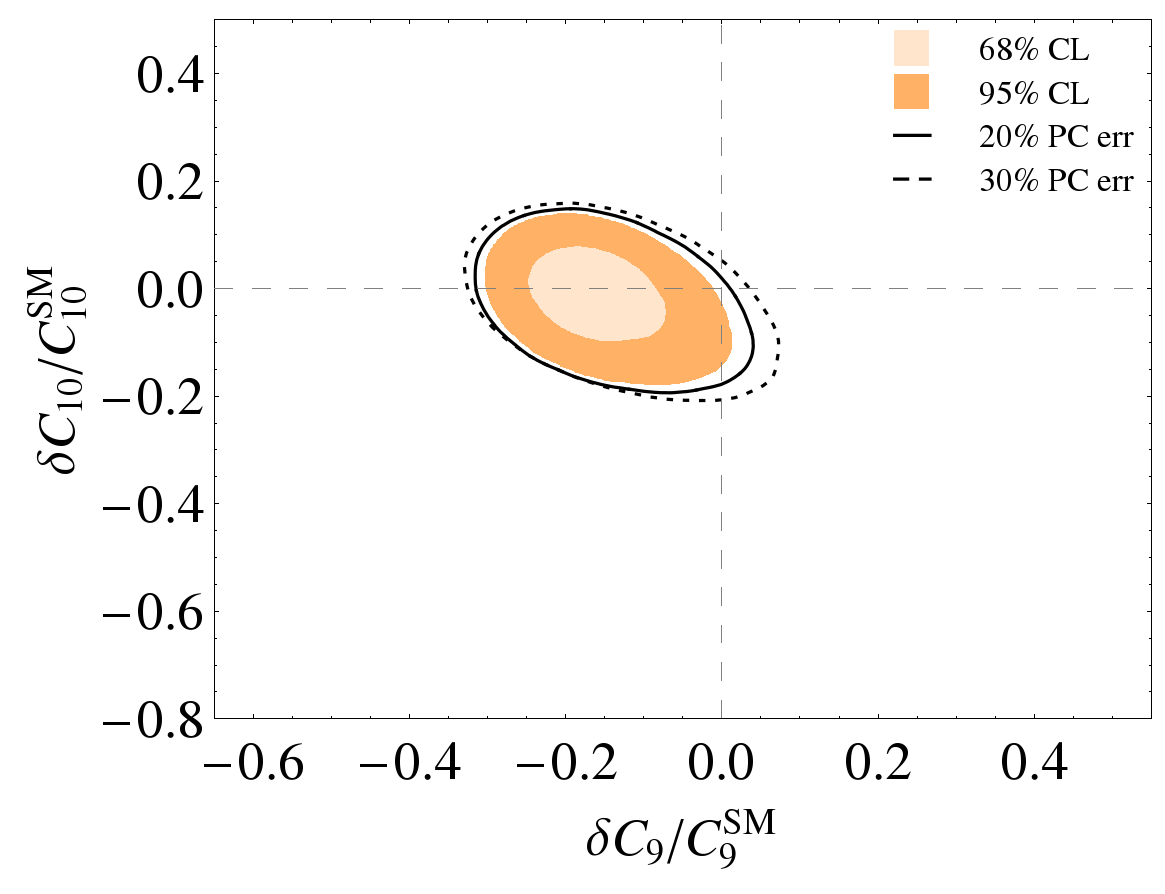}\qquad
\includegraphics[width=7.cm]{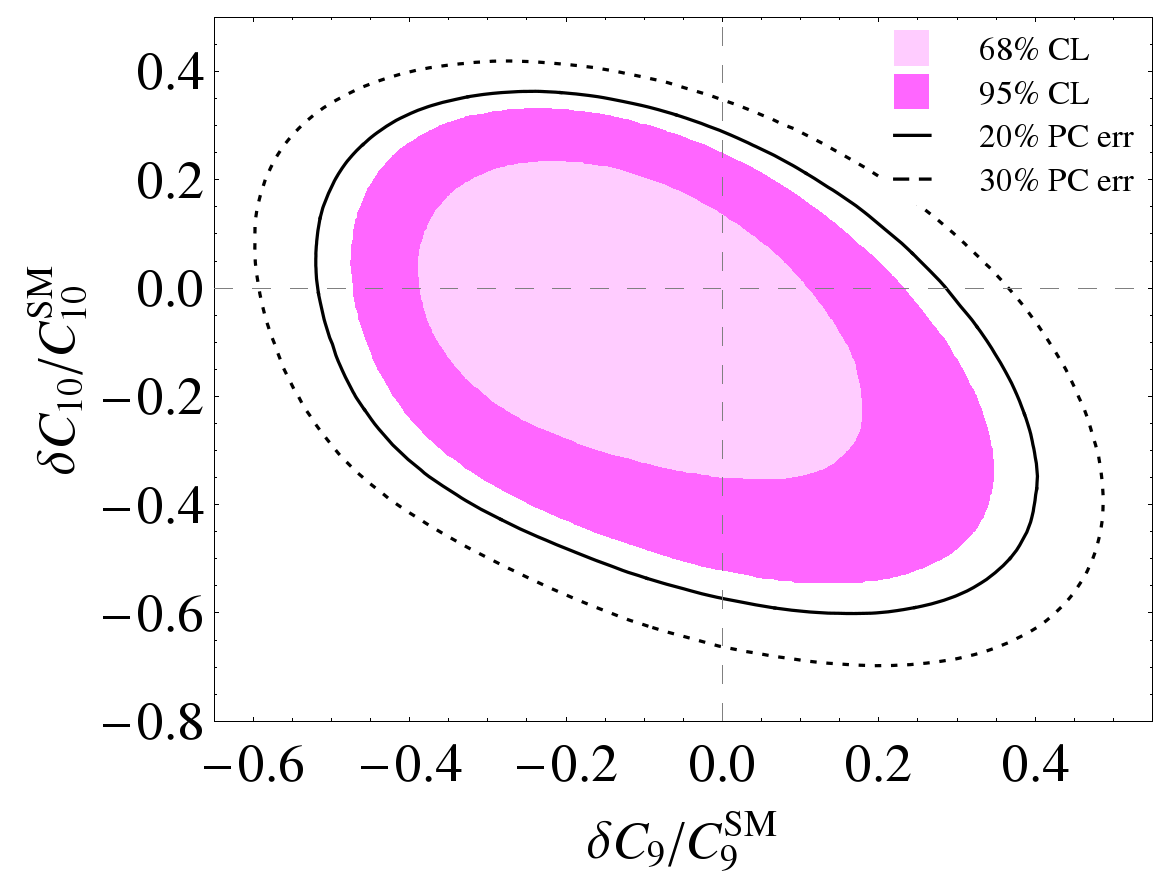}
\caption{Global fit results for $C_9, C_{10}$ by using \emph{soft FF} approach and considering 10\% power corrections when 
employing the $\Delta \chi^2$ (left) and absolute $\chi^2$ (right). 
The solid (dashed) lines correspond to the  allowed regions at 2$\sigma$ when considering 20\%  (30\%) power corrections. 
\label{fig:2-soft_c9-c10_all}}
\end{center}
\end{figure}

We have first obtained the best fit value and the corresponding 1 and 2$\sigma$ allowed regions 
when varying the \{$C_9,C_{10}$\} set of Wilson coefficients,
where for the theoretical predictions of the observables we have used both the full FF 
and the soft FF approaches in Fig.~\ref{fig:2-full_c9-c10_all} and Fig.~\ref{fig:2-soft_c9-c10_all}, respectively.
In the full FF approach, the allowed regions for the \{$C_9,C_{10}$\} set are similar when considering 5, 10 and 20\% error for the power corrections which indicates that  
up to 20\%, the power corrections are sub-dominant compared to the other theoretical uncertainties. As derived in Section~\ref{sec:error}, the 5, 10 and 20\% error for the power corrections at the amplitude level leads to a 6\% error at maximum at the observable level only. 

This is not the case for the soft FF approach since the power corrections affect both the factorisable as well as non-factorisable corrections, while in the full FF approach only the non-factorisable part is affected.
However, in both methods, considering the power corrections to be up to 20\%, the SM is disfavoured at more than $2\sigma$.
For example, assuming a 10\% power correction error within the full FF method leads to a SM pull of $2.6 \sigma$, (meaning that the SM value is in $2.6 \sigma$ tension with the best fit values of $C_9$ and $C_{10}$).  
For comparison we also show the global fits to the Wilson coefficients based on the absolute $\chi^2$ method. 
As expected (see Section~\ref{sec:statistics}) they lead to weaker bounds.

\begin{figure}[!t]
\begin{center}
\includegraphics[width=7.cm]{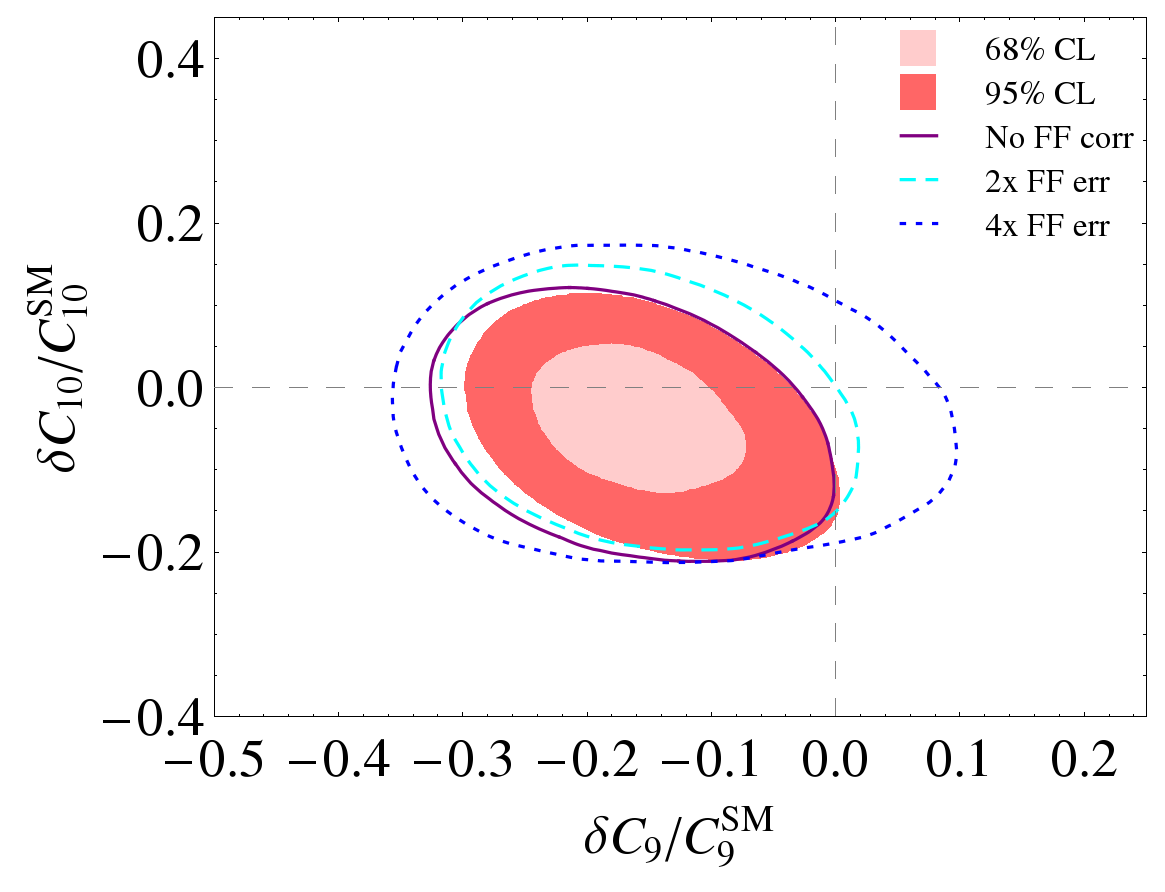}
\includegraphics[width=7.cm]{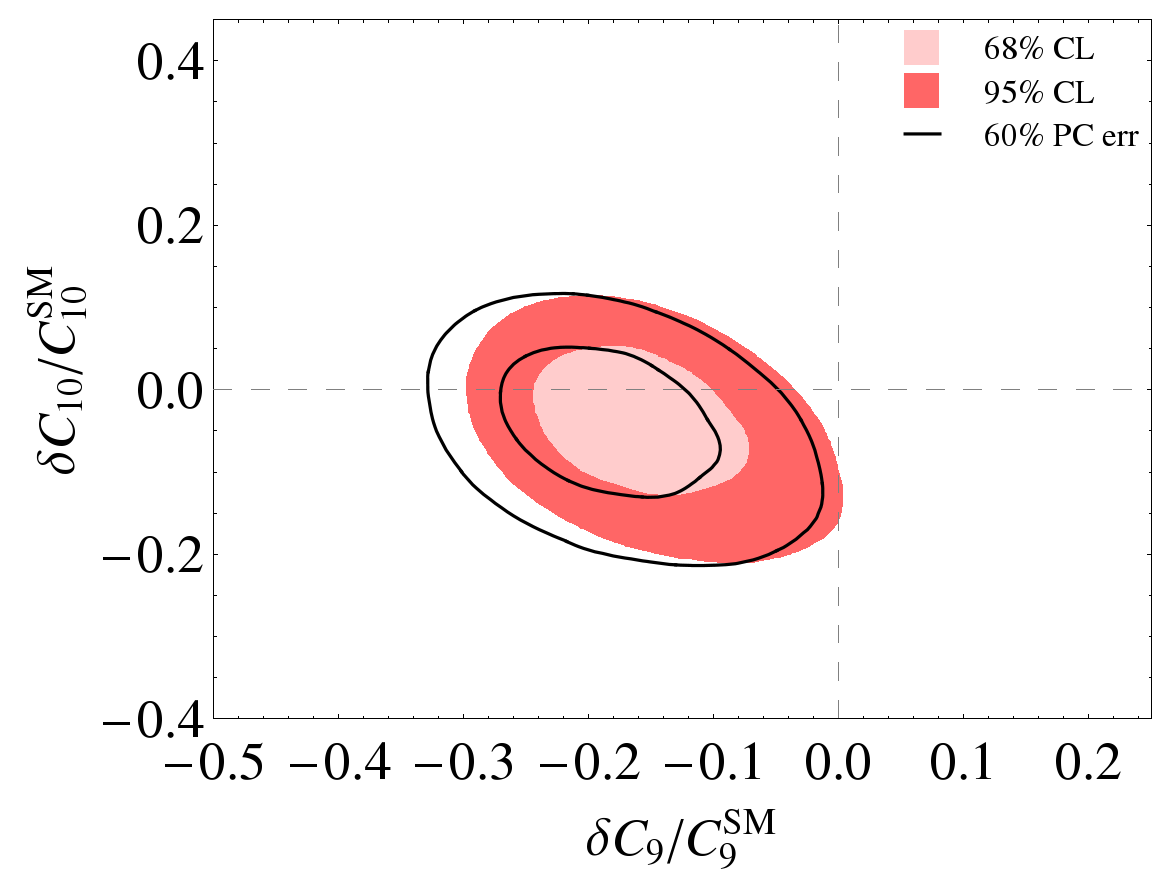}
\caption{Global fit results for $C_9, C_{10}$ by using full form factors, 
with $\Delta \chi^2$ method and 10\% power correction error. 
On the left plot, the 2$\sigma$ contours when removing the form factor correlations, 
as well as when doubling (quadrupling) the form factor errors are shown with solid and dashed (dotted) lines, respectively.
On the right plot, the solid lines correspond to the 1 and 2$\sigma$ contours when considering 60\% power correction error. 
\label{fig:2-full_c9-c10_FFCorrelation}}
\end{center}
\end{figure}

\begin{figure}[!t]
\begin{center}
\includegraphics[width=7.cm]{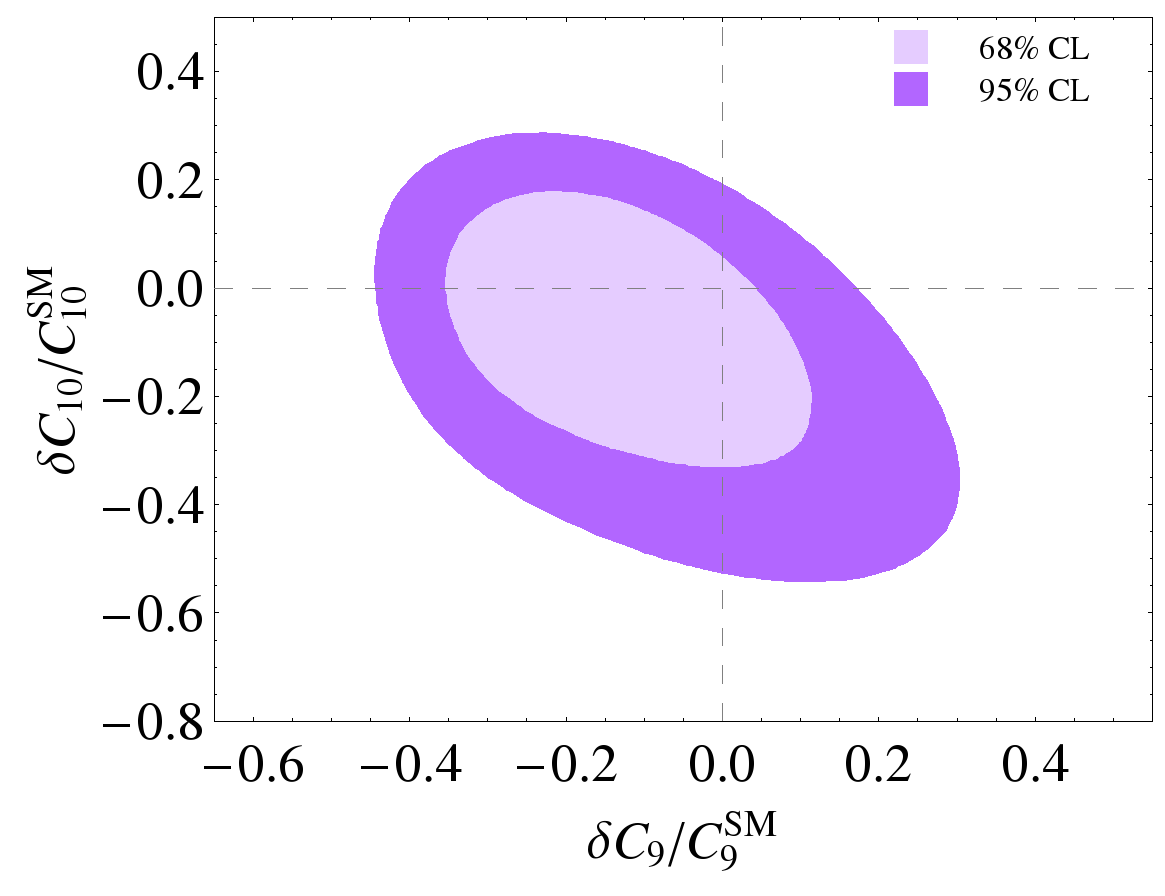}\qquad\includegraphics[width=7.cm]{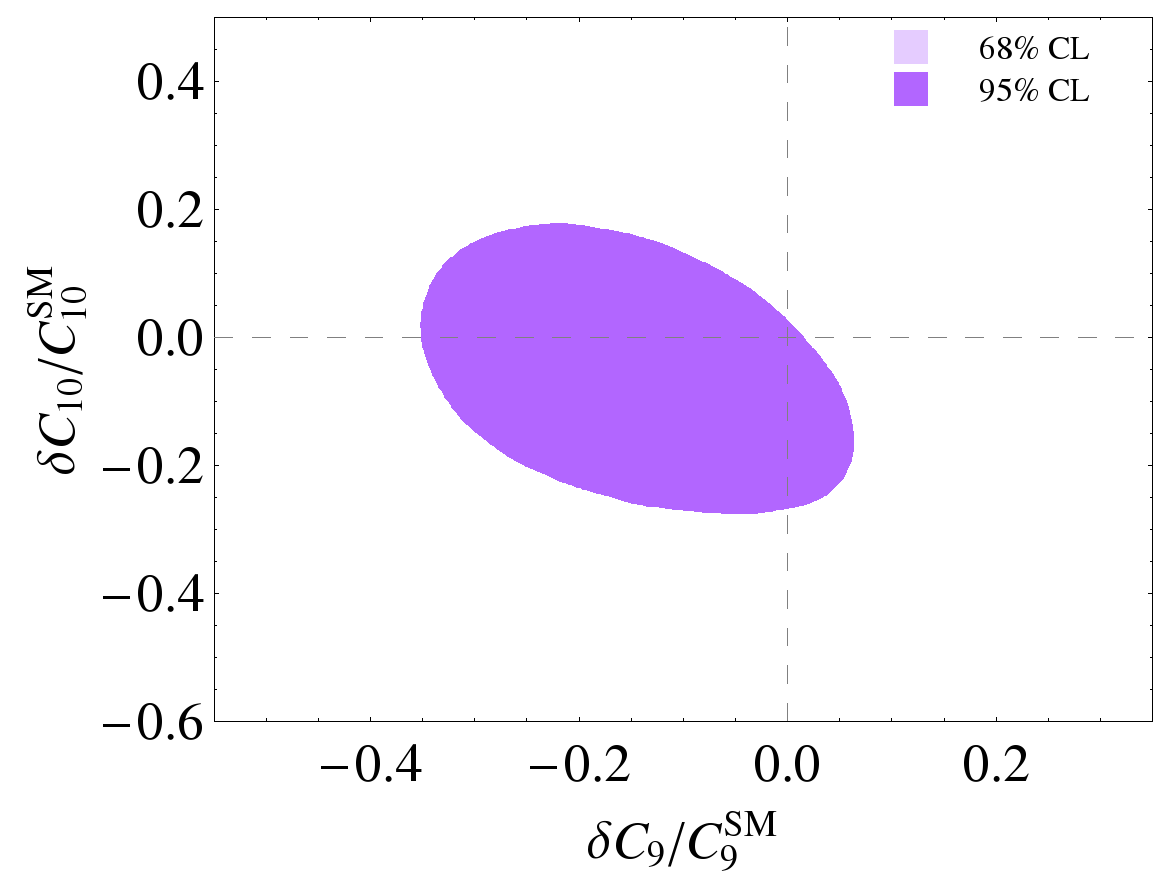}
\caption{Global fit results for $C_9, C_{10}$ by using full form factors and 10\% power correction errors, with absolute $\chi^2$. For $B\to K^+\mu^+\mu^-$ only the low- and high-$q^2$ bins are used in the plot on the left while all the small bins are used in the plot on the right.}
\label{fig:statistics}
\end{center}
\end{figure}

In the left plot of Fig.~\ref{fig:2-full_c9-c10_FFCorrelation} 
the 1 and 2$\sigma$ allowed regions are demonstrated, when removing the form factor correlations or 
doubling as well as quadrupling the form factor errors.
The size of the form factor errors has a crucial role in constraining the allowed region; doubling (quadrupling) the error decreases the tension from $2.6 \sigma$ to $2.1 \sigma$ ($1.4 \sigma$). But  removing the form factor correlations does not have a significant impact. 
This is due to the small uncertainties of the BSZ form factors.
If the quadruple form factor errors were considered, then the correlations would play a more 
important role. Assuming a  60\% power correction error  in the global fit has not a big impact either as the right plot of Fig.~\ref{fig:2-full_c9-c10_FFCorrelation} shows.
As discussed above, such a guesstimate of the 
non-factorisable power correction leads to errors of 20\% at the observable level. That the best fit point gets slightly moved away from the SM point is a consequence of how such a guesstimate is implemented at the amplitude level, namely within terms without dependences on the Wilson coefficients $C_7$, $C_9$, and $C_{10}$ (see Section~\ref{sec:error}).    

Finally, we analyse the global fit to $C_9$ and $C_{10}$ with 10\% power correction using the absolute $\chi^2$ method when we do {\it not} use  only one global low-$q^2$  and one global high-$q^2$, but also smaller bins of the observable $B\to K^+\mu^+\mu^-$.  
 The right plot in Fig.~\ref{fig:statistics} shows the latter option. We see that such a fit does not lead to any  
compatibility at 68\% C.L. which  clearly indicates that the fit is not very good, when the data 
with smaller bins are used. This reflects the well-known observation that 
the large resonance structure in the high-$q^2$ region of  this observable 
 does not allow for a theoretical description of this observable with smaller binning.

\subsubsection{Fit results for \{$C_9, C_{9'}$\}}

\begin{figure}[!t]
\begin{center}
\includegraphics[width=7.cm]{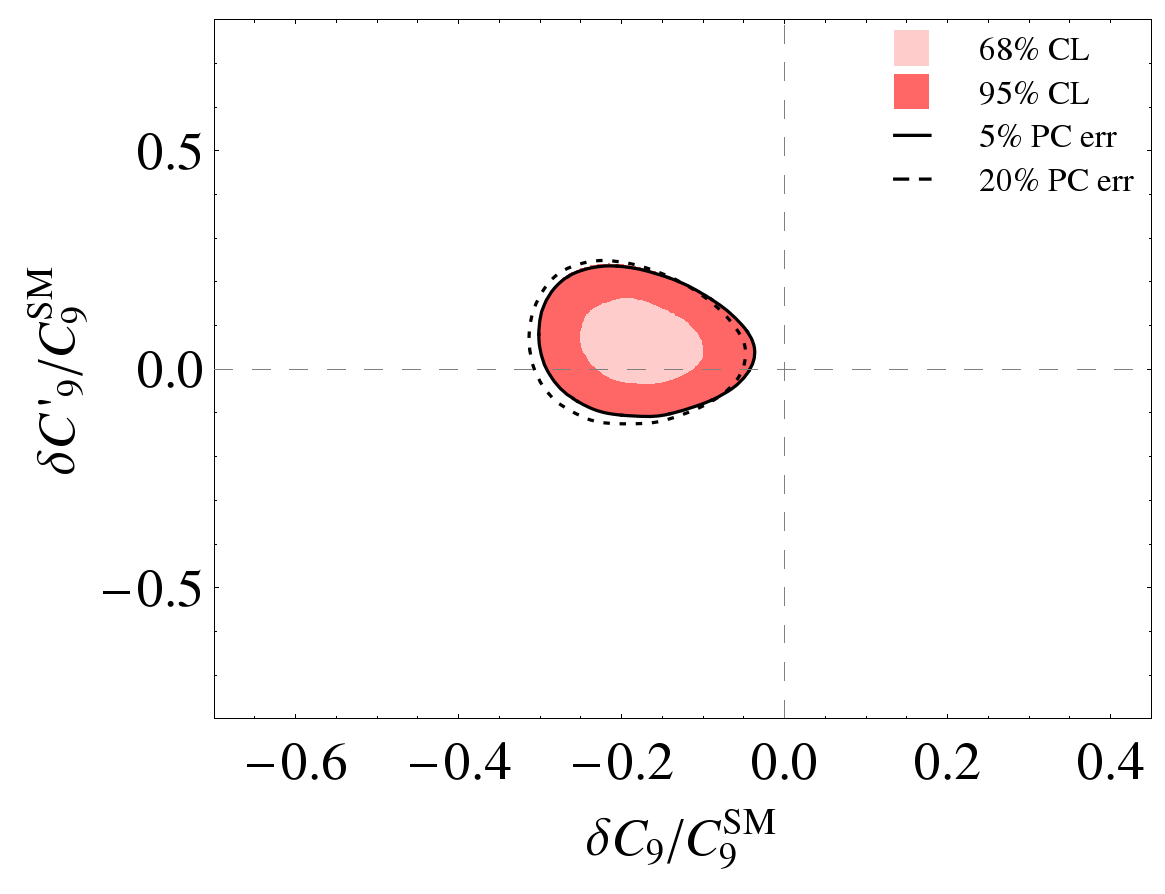}\qquad
\includegraphics[width=7.cm]{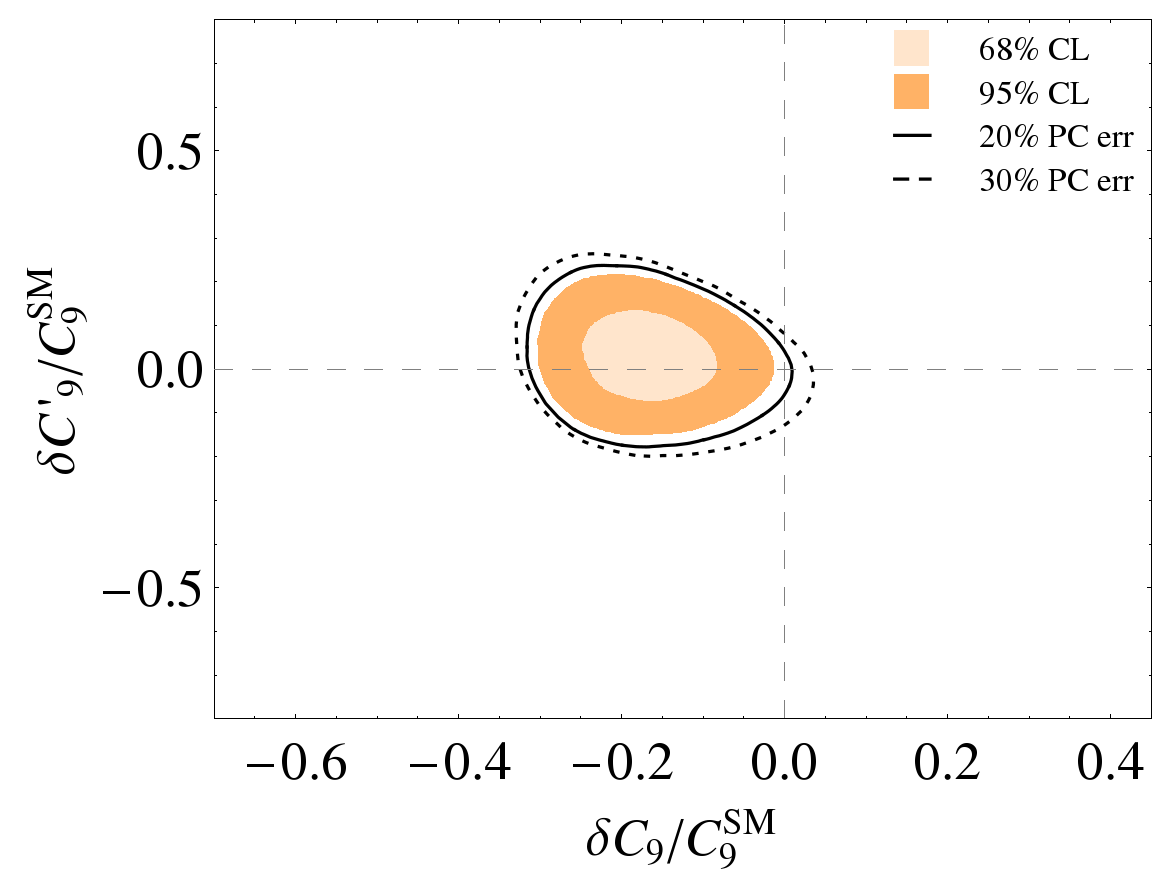}
\caption{Global fit results for $C_9, C_{9'}$, 
with $\Delta \chi^2$ method and by considering 10\% power correction error. On the left the full form factor method has been used, 
where the solid line corresponds to 5\% power correction errors and the dashed line to 20\% power correction errors.
On the right the soft form factor method has been used,
where the solid line corresponds to 20\% power correction errors and the dashed line to 30\% power correction errors.
The solid and dashed lines are at 2$\sigma$.
\label{fig:2_c9-c9p_all}}
\end{center}
\end{figure}

\begin{figure}[!t]
\begin{center}
\includegraphics[width=7.cm]{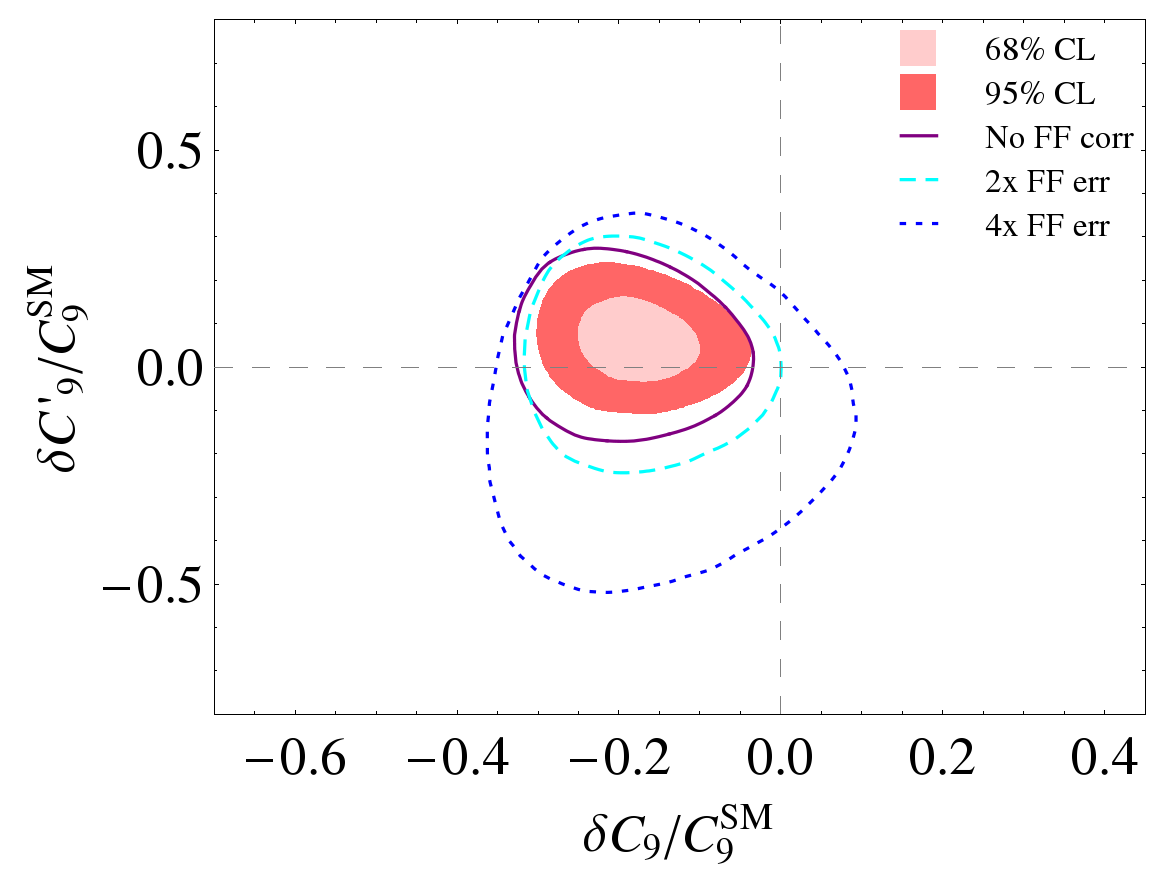}
\includegraphics[width=7.cm]{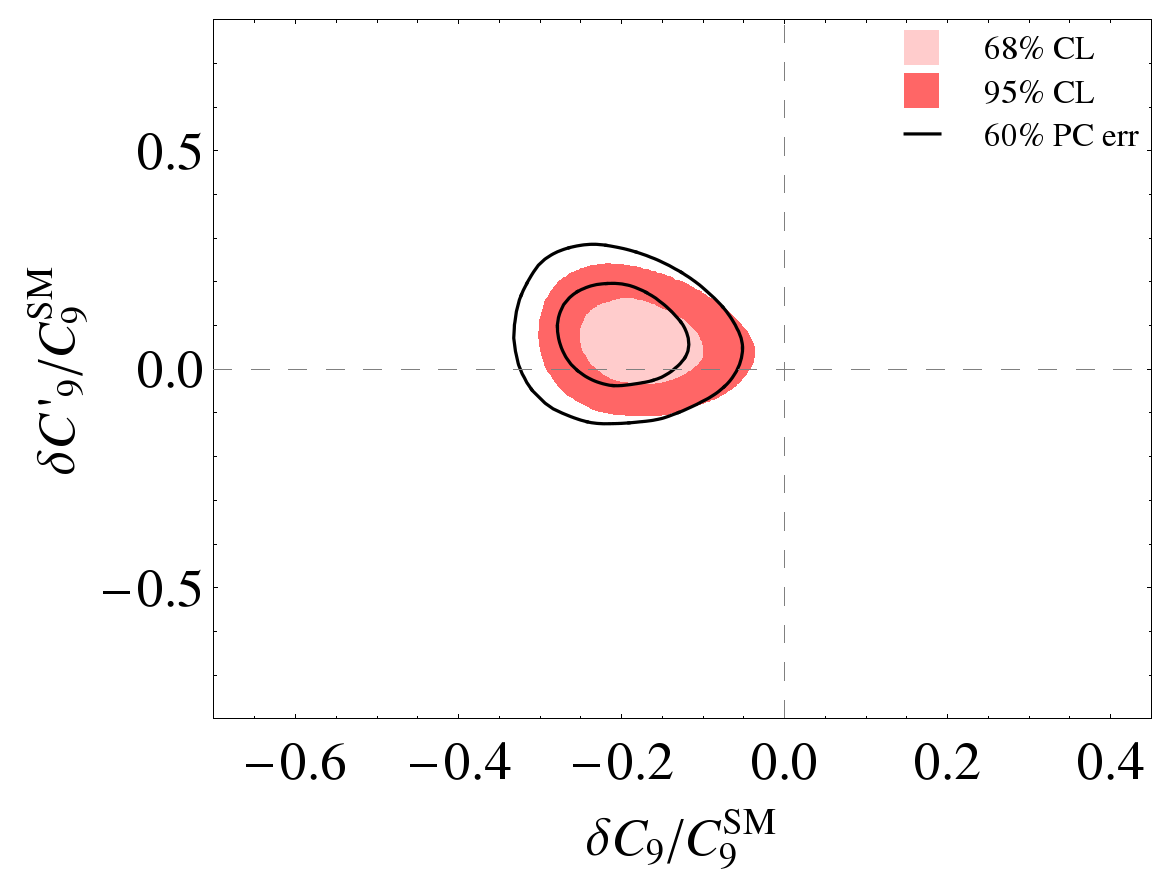}
\caption{Global fit results for $C_9, C_{9'}$ by using full form factors, with $\Delta \chi^2$ method and 10\% power correction errors. 
On the left plot, the 2$\sigma$ contours when removing the form factor correlations, 
as well as when doubling (quadrupling) the form factor errors are shown with solid and dashed (dotted) lines, respectively. 
On the right plot, the solid lines correspond to the 1 and 2$\sigma$ contours when considering 60\% power correction error. 
\label{fig:2_c9-c9p_all_FFCorrelation}}
\end{center}
\end{figure}

In Fig.~\ref{fig:2_c9-c9p_all} we give the 1 and 2$\sigma$ allowed regions for the fit to the \{$C_9, C_{9'}$\} set within both the full FF as well as soft FF approaches.
Assuming 10\% power correction for the SM predictions of $B\to K^{(*)}(\phi) \bar\ell \ell$ decays,
the SM value has a slight tension of more than 2$\sigma$ 
with the best fit point (2.7$\sigma$ (2.1$\sigma$) with $\chi^2=123.1$ (118.9) for the 
full (soft) FF approach), where the tension is mostly in $C_9$.
Compared to  the \{$C_9, C_{10}$\} fit, the very similar $\chi^2$ values indicate that
there is no preference between the fits when considering new physics effects in $C_{10}$ or $C_9^\prime$.
The effects of doubling and quadrupling the form factor errors and of a 60\% power correction error  as shown in Fig.~\ref{fig:2_c9-c9p_all_FFCorrelation} are  also similar to the \{$C_9, C_{10}$\} fit.

\subsubsection{Fit results for \{$C_9^e, C_9^\mu$\}}

\begin{figure}[!t]
\begin{center}
\includegraphics[width=7.cm]{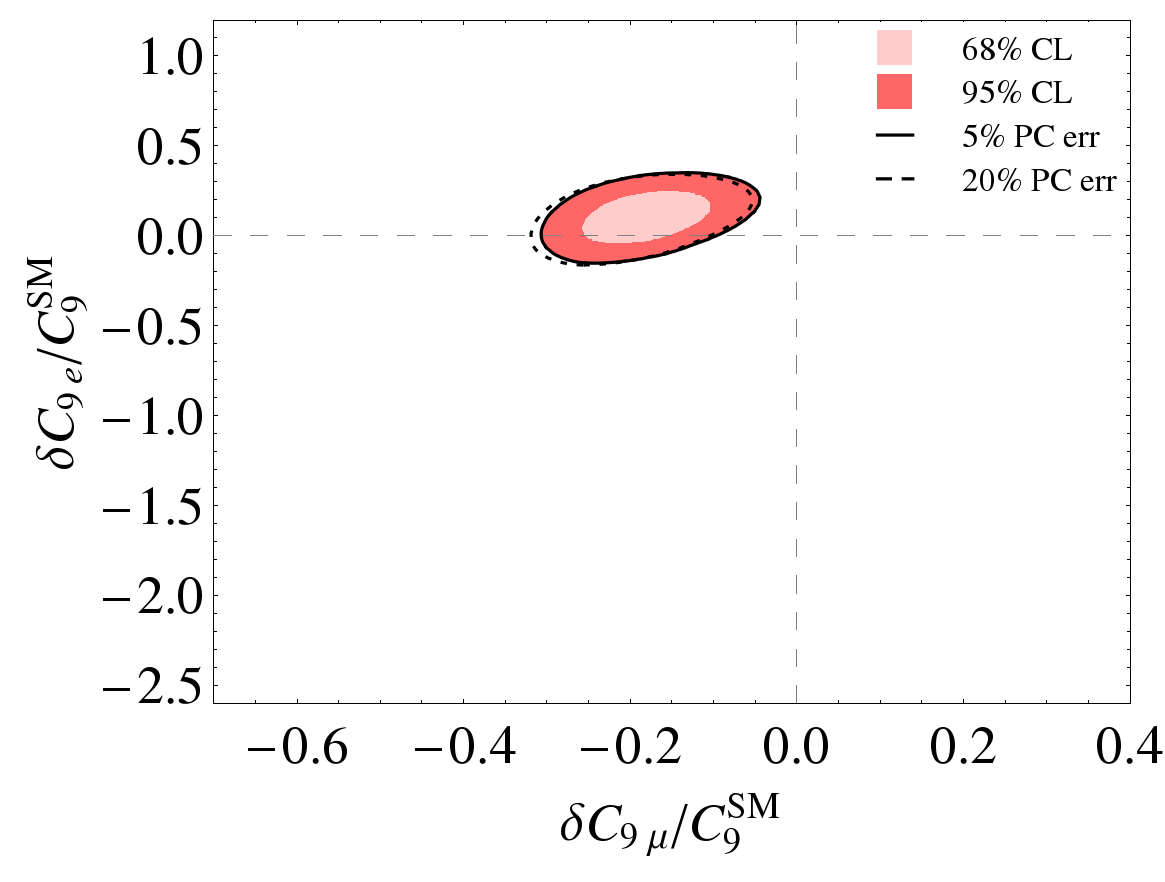}\qquad
\includegraphics[width=7.cm]{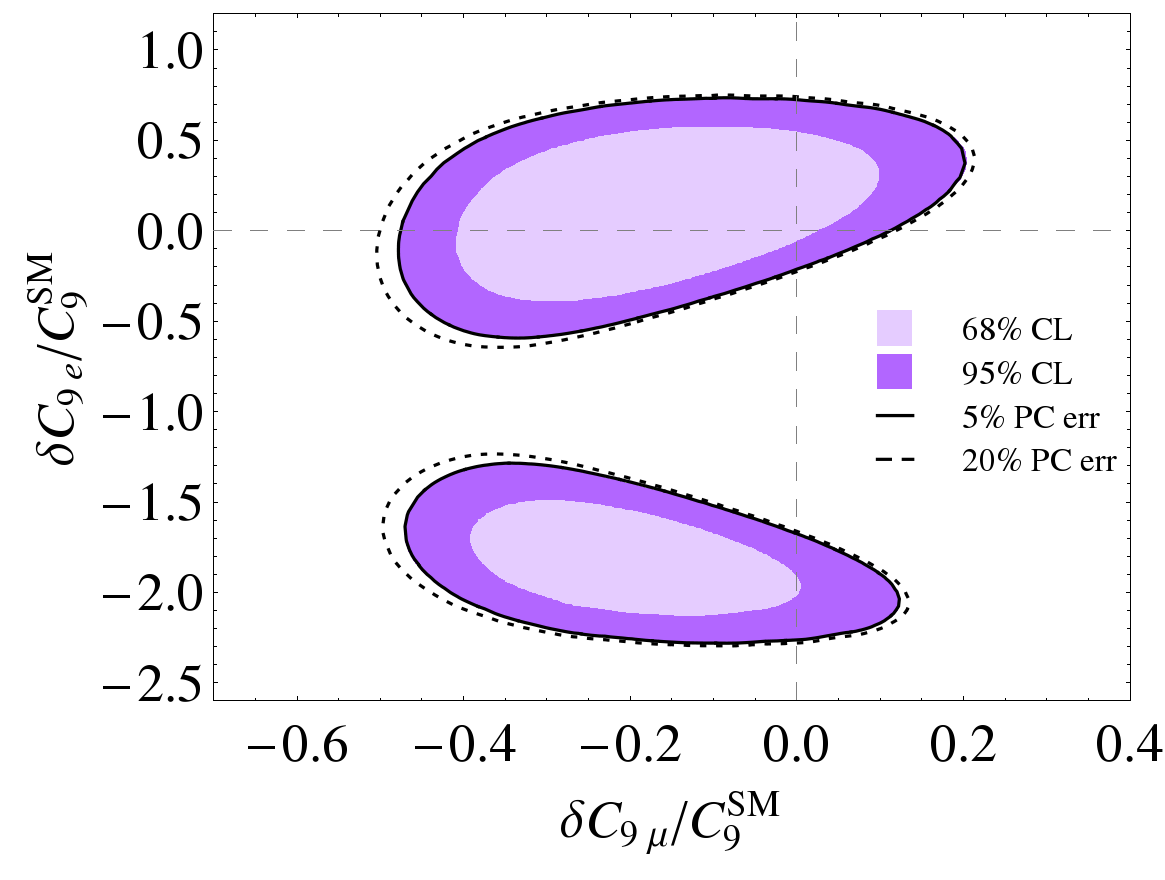}
\caption{Global fit results for $C_9^e, C_9^\mu$ by using \emph{full FF} approach and considering 10\% power correction when 
employing the $\Delta \chi^2$ (left) and absolute $\chi^2$ (right) methods. 
The solid and dashed lines correspond to the allowed regions at 2$\sigma$ when considering 5 and 20\% power corrections, respectively. 
\label{fig:2-full_c9e-c9mu_all}}
\end{center}
\end{figure}

\begin{figure}[!t]
\begin{center}
\includegraphics[width=7.cm]{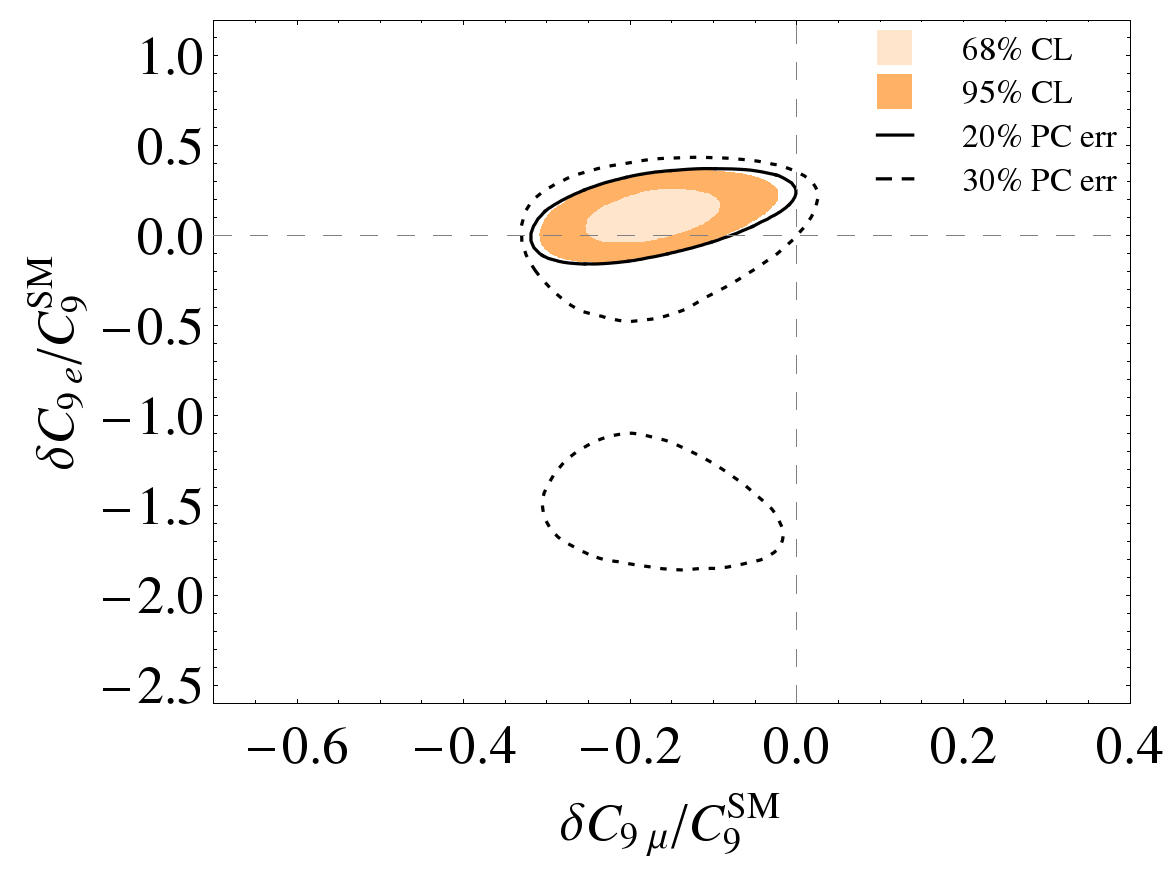}\qquad
\includegraphics[width=7.cm]{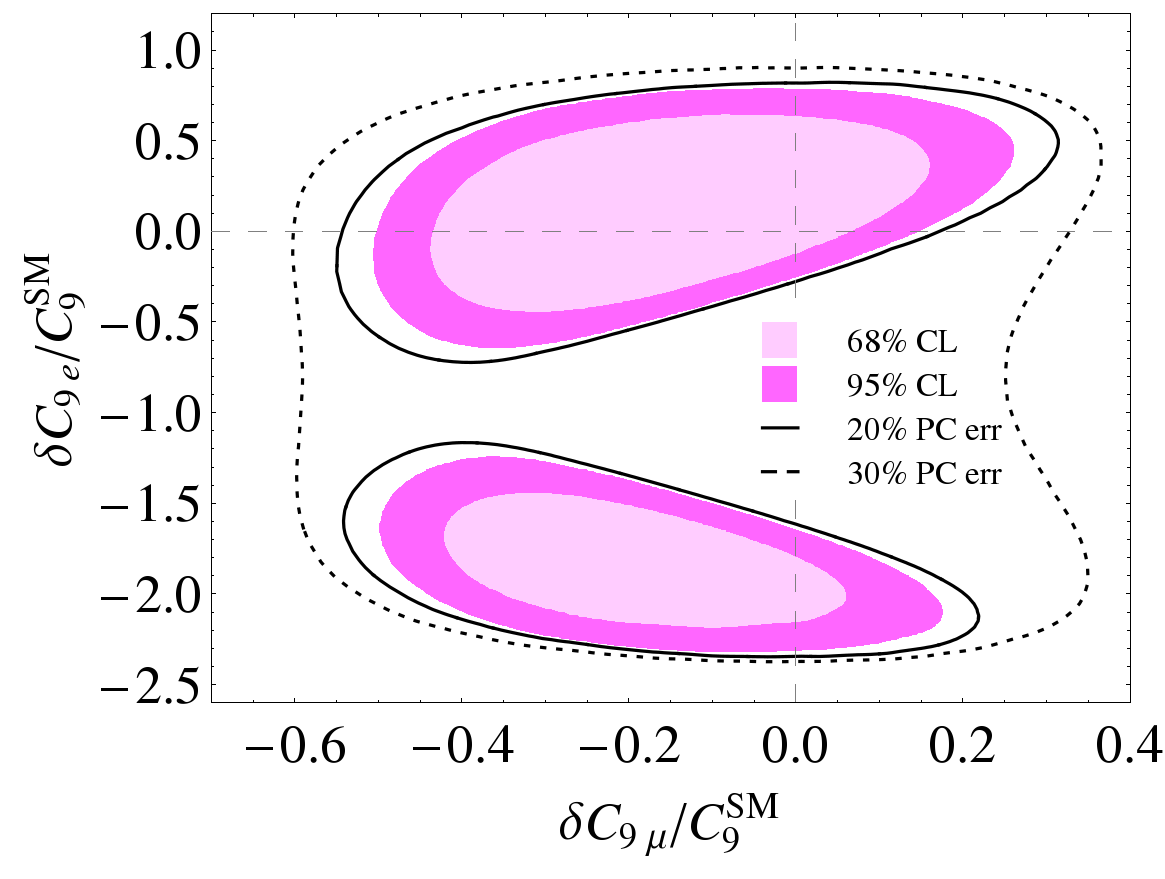}
\caption{Global fit results for $C_9^e, C_9^\mu$ by using \emph{soft FF} approach and considering 10\% power correction when 
employing the $\Delta \chi^2$ (left) and absolute $\chi^2$ (right) methods. 
The solid and dashed lines correspond to the allowed regions at 2$\sigma$ when considering 20 and 30\% power corrections, respectively.
\label{fig:2-soft_c9e-c10me_all}}
\end{center}
\end{figure}

\begin{figure}[!t]
\begin{center}
\includegraphics[width=7.cm]{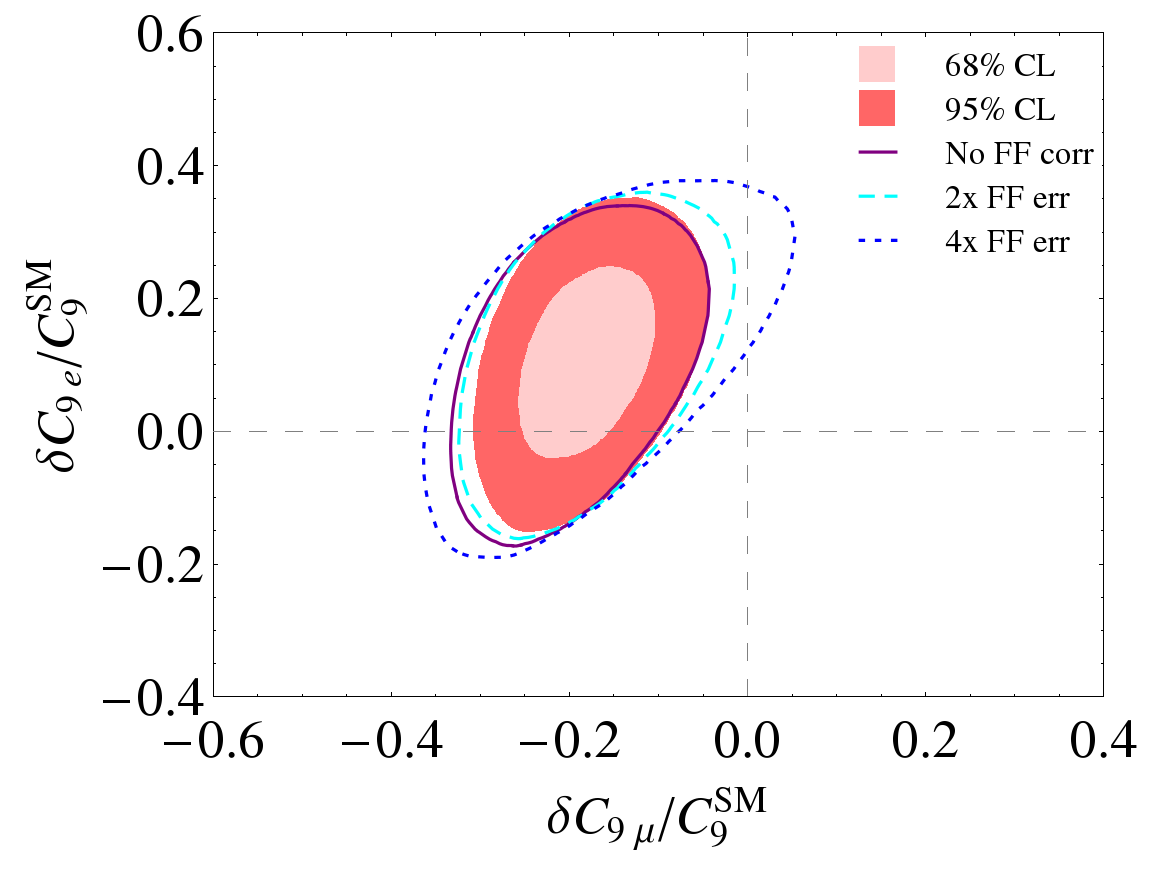}
\includegraphics[width=7.cm]{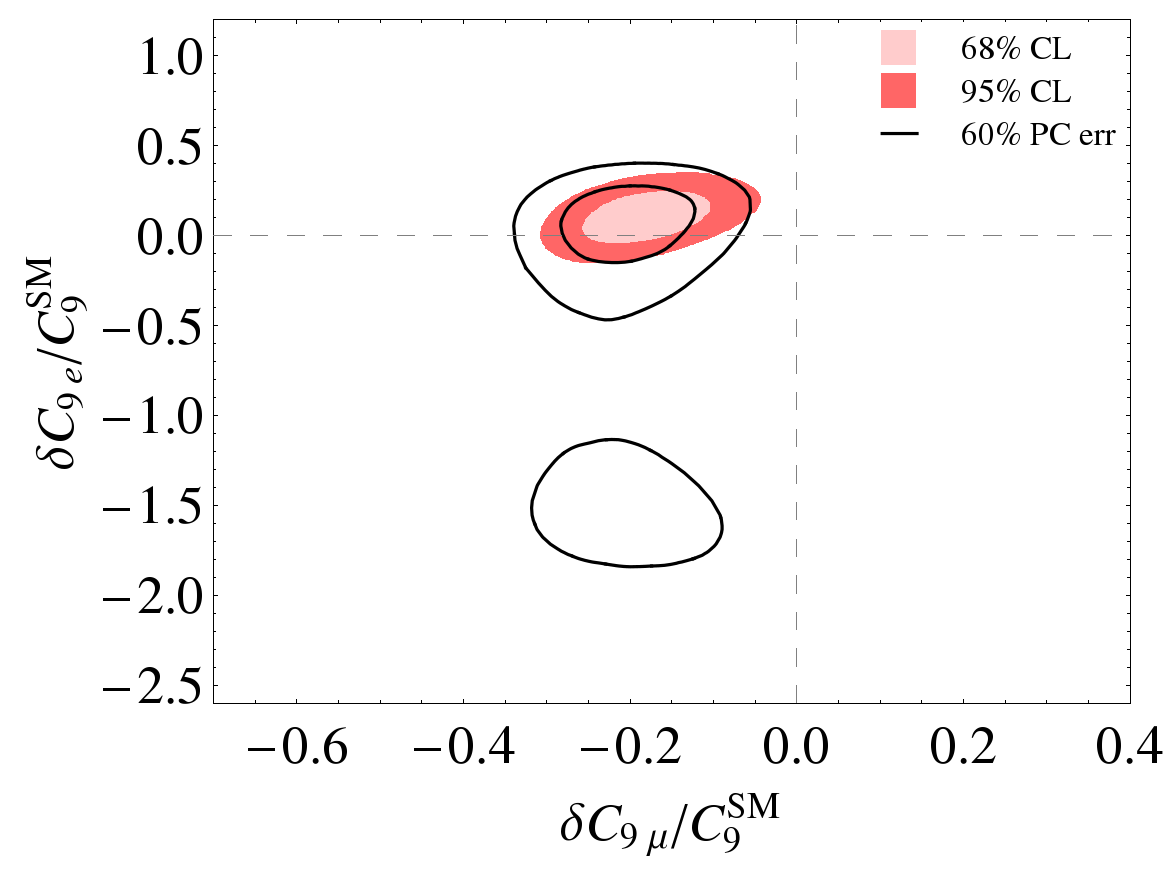}
\caption{Global fit results for $C_9^e, C_9^\mu$ by using full form factors, with $\Delta \chi^2$ method and 10\% power correction errors. 
On the left plot, the 2$\sigma$ contours when removing the form factor correlations, 
as well as when doubling (quadrupling) the form factor errors are shown with solid and dashed (dotted) lines, respectively.
On the right plot, the solid lines correspond to the 1 and 2$\sigma$ contours when considering 60\% power correction error. 
\label{fig:2_c9m-c9e_all_FFCorrelation}}
\end{center}
\end{figure}

In Figs.~\ref{fig:2-full_c9e-c9mu_all},\ref{fig:2-soft_c9e-c10me_all}, the allowed regions of the fit at 1 and 2$\sigma$ are presented when considering different contributions for muons and electrons in $C_9$.
Using the full (soft) FF approach with 10\% power correction, the $\chi^2$ of the best fit point is 114.6 (109.8), which indicates a considerable improvement of the fit compared to the \{$C_9, C_{10}$\} and \{$C_9, C_9^\prime$\} fits.
The SM value which respects lepton universality has a tension of $3.9\sigma$ ($3.6\sigma$) with the best fit point where most of the tension appears in $C_9^\mu$. 
Within the full FF approach, the  Pull$_{\rm SM}$ is reduced from $3.9\sigma$ to $3.1\sigma$ only by quadrupling the form factor error within the full FF approach (see Fig.~\ref{fig:2_c9m-c9e_all_FFCorrelation}).

The left plots in Figs.~\ref{fig:2-full_c9e-c9mu_all},\ref{fig:2-soft_c9e-c10me_all} show the global fit results with absolute $\chi^2$. 
The important new feature is that there is a second minimum in the absolute $\chi^2$ which is not visible at all when using the $\delta \chi^2$ method. 

\begin{figure}[!t]
\begin{center}
\includegraphics[width=0.49\textwidth]{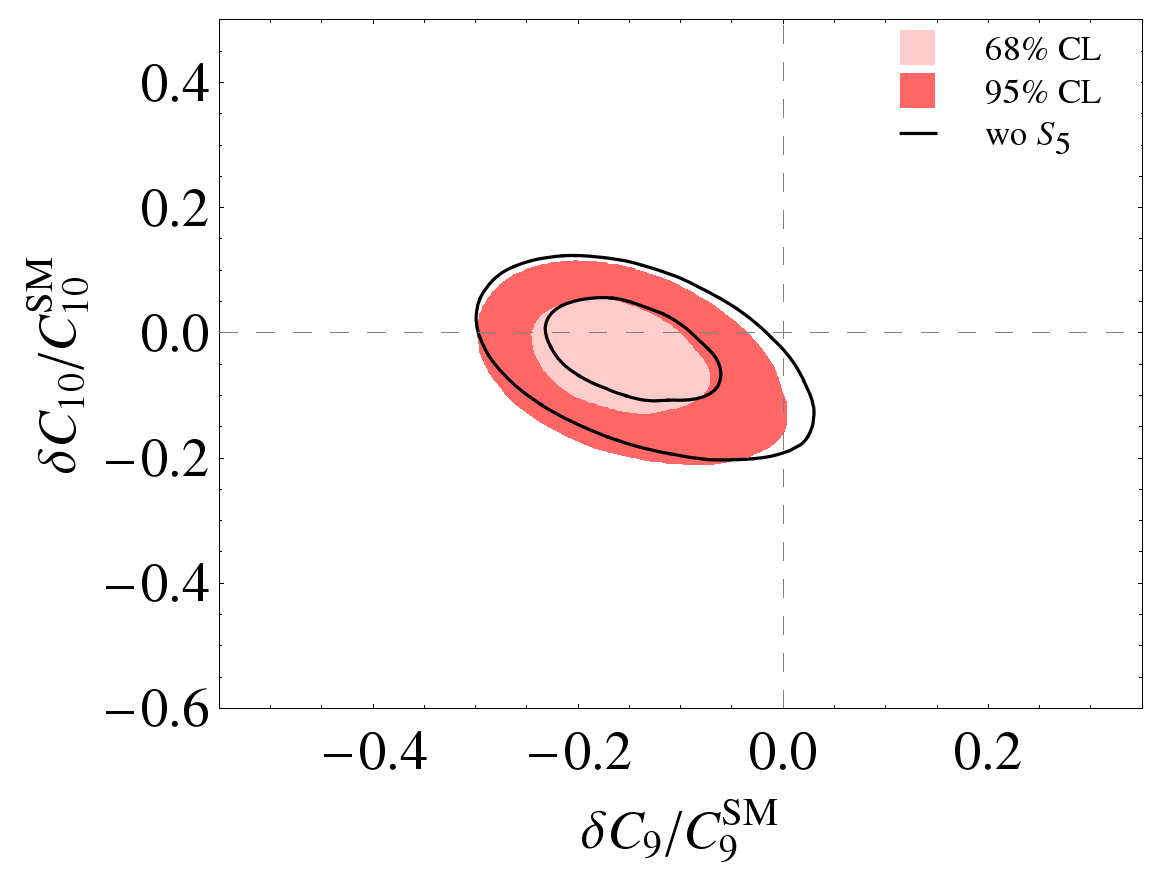}\\[1.mm]
\includegraphics[width=0.49\textwidth]{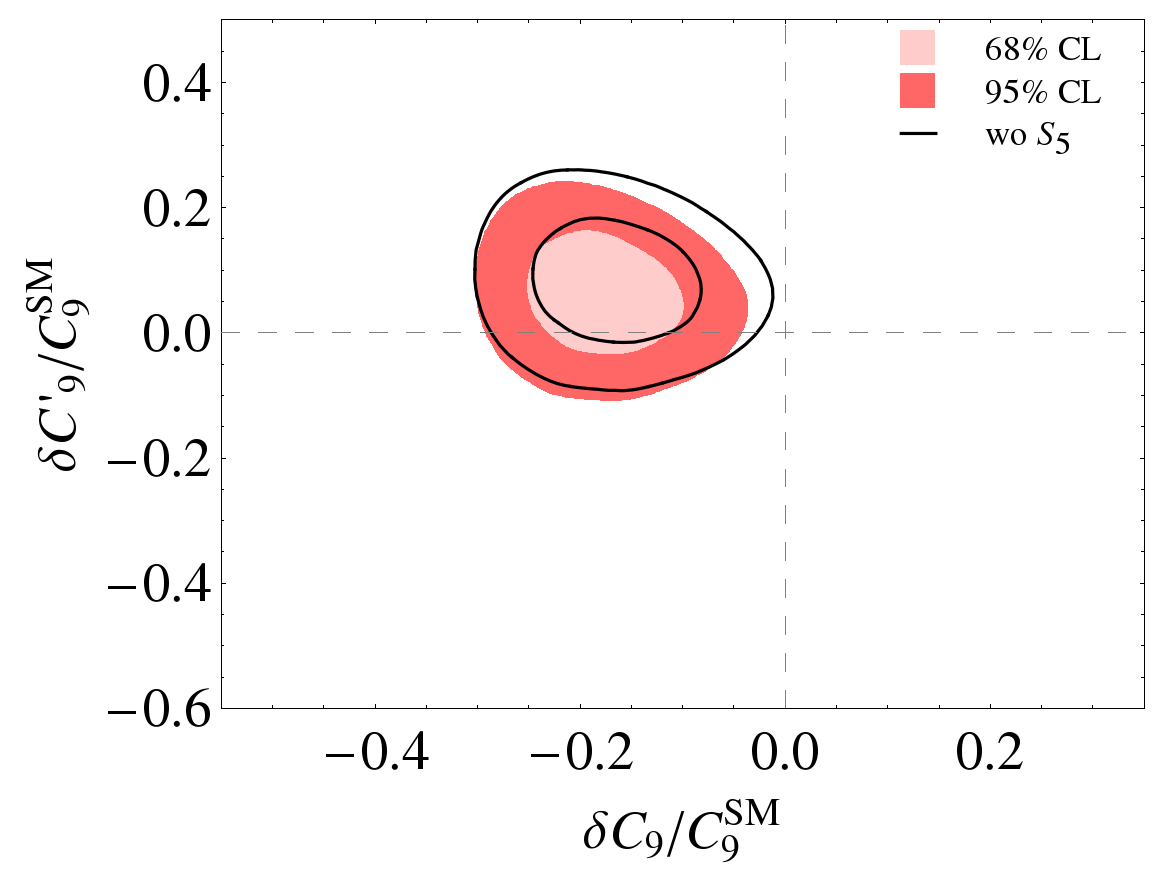}
\includegraphics[width=0.49\textwidth]{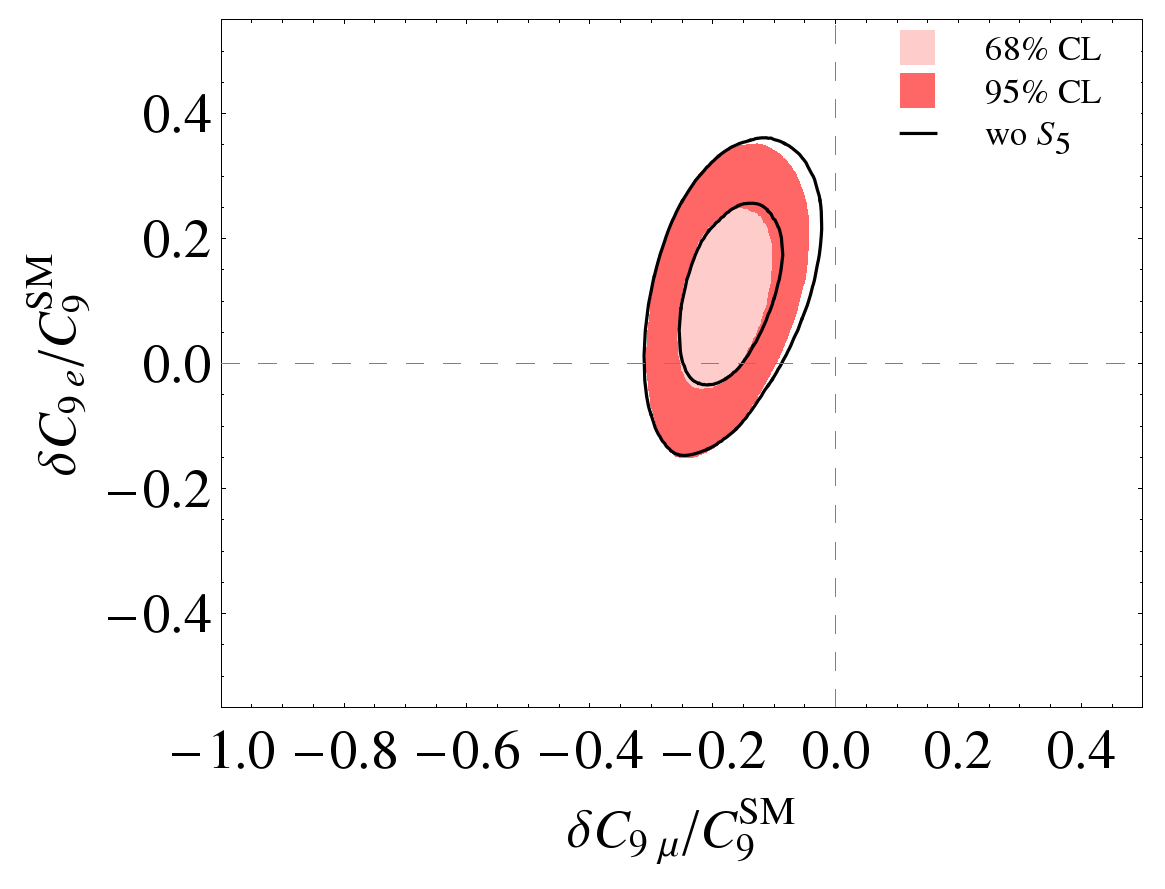}
\caption{Global fit results using full form factors, with $\Delta \chi^2$ method. 
The solid lines correspond to the fit results when omitting $S_5$.\label{fig:omit-S5}}
\end{center}
\end{figure}

\subsection{Role of $R_K$ and $S_5$}
\label{sec:without}

\begin{figure}[!t]
\begin{center}
\includegraphics[width=0.49\textwidth]{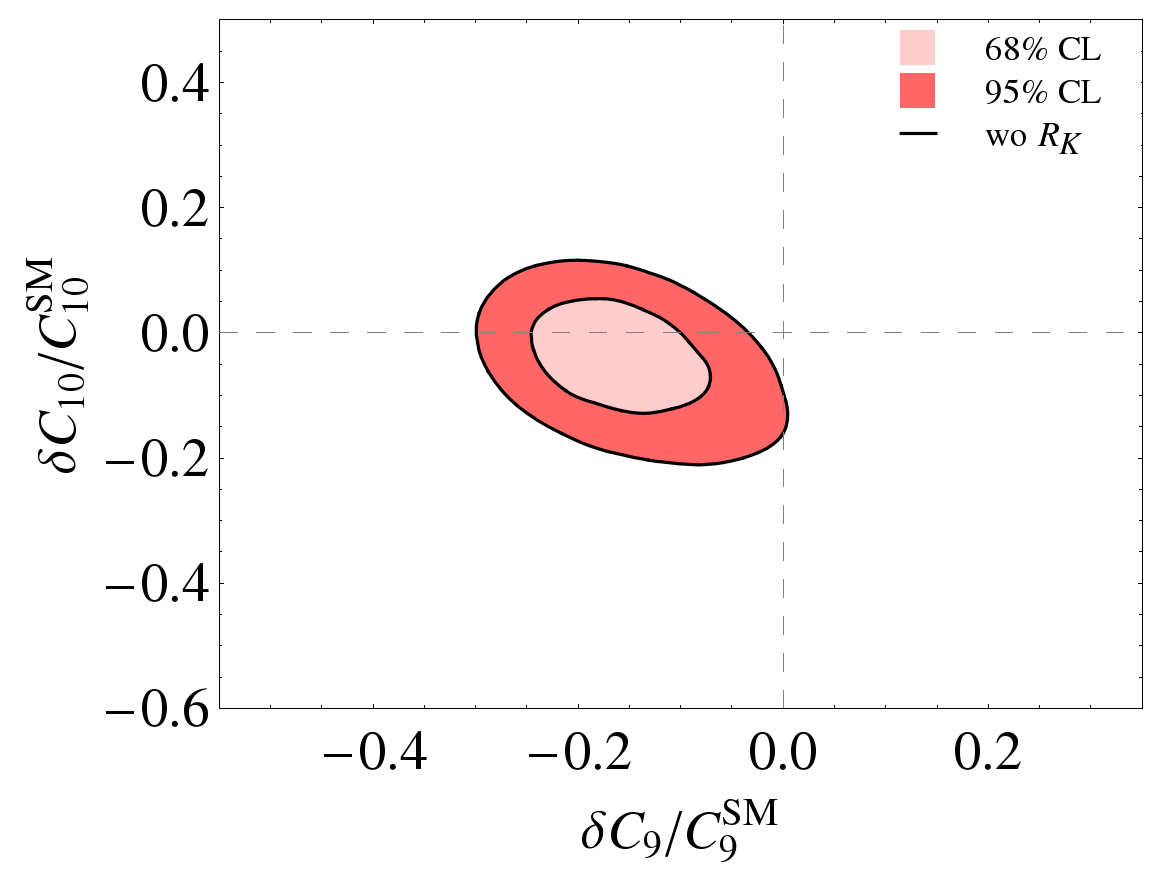}\\[5.mm]
\includegraphics[width=0.49\textwidth]{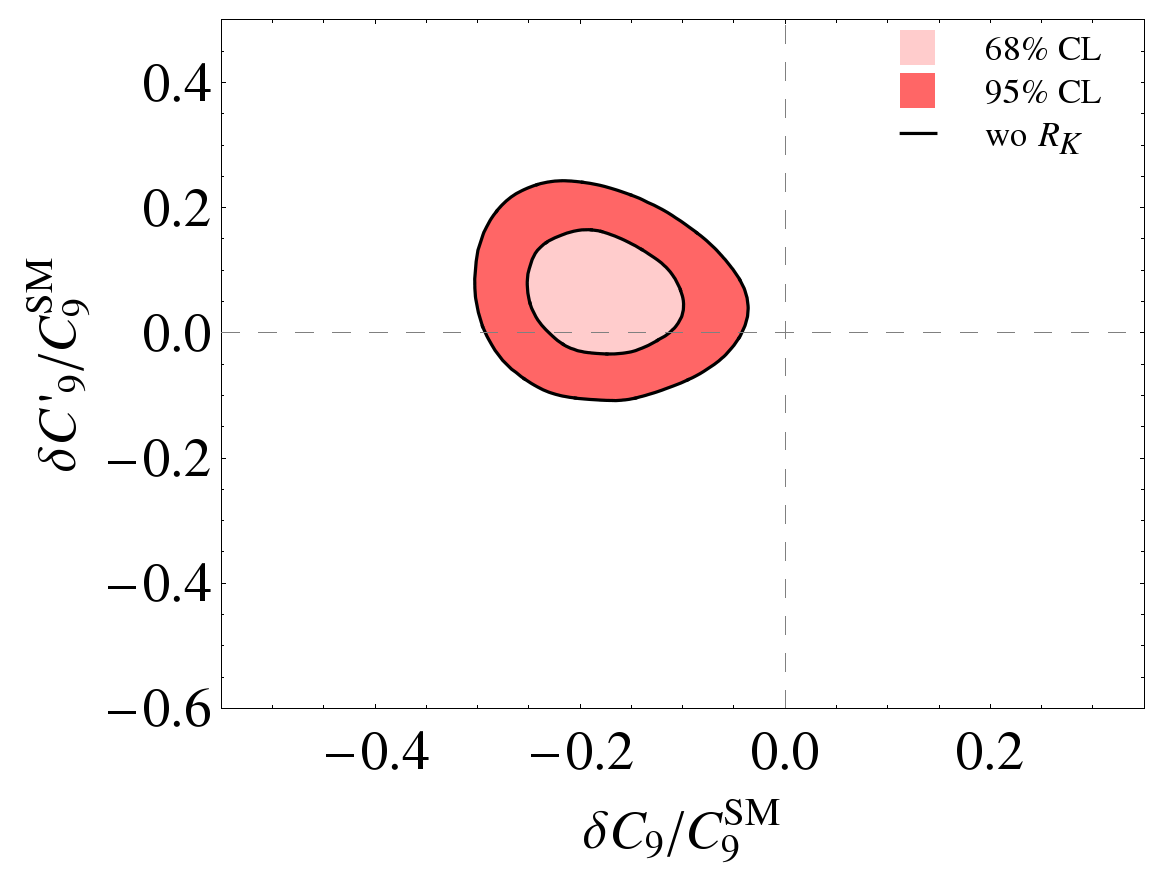}
\includegraphics[width=0.49\textwidth]{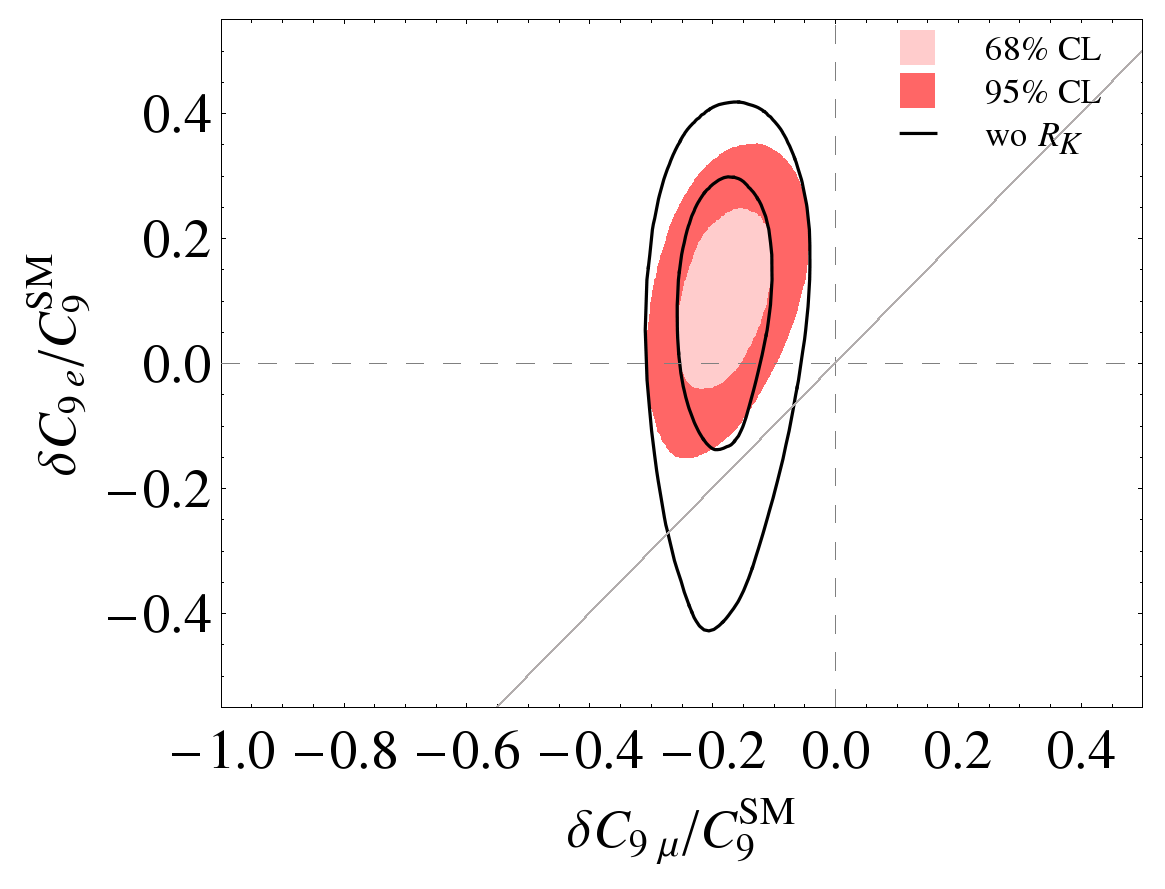}
\caption{Global fit results using full form factors, with $\Delta \chi^2$ method. 
The solid lines correspond to the fit results when omitting $R_K$. The gray line corresponds to the lepton flavour universality condition.} 
\label{fig:omit-RK}
\end{center}
\end{figure}

While the experimental measurement of $S_5$ and its tension with the SM prediction 
seems to be the main reason for a  best fit point of $C_9$ about 20\% less than its SM value, 
the $S_5$ turns out to be not the only observable which drives $\delta C_9/C_9^{\rm SM}$ to negative values.
In Fig.~\ref{fig:omit-S5} we have given the two operator fits when removing the data on $S_5$ while keeping all the other 
$b\to s$ data. It can be seen that while the tension of $C_9^{\rm SM}$ and  best fit point value of $C_9$ is slightly reduced in
the various two operator fits, still the tension exists at more than 2$\sigma$. 
This feature indicates that $S_5$ is not the only observable that drives $\delta C_9$ to negative values and other observables play a similar role.
However, the situation is different in the case of the observable $R_K$. 
In Fig.~\ref{fig:omit-RK} from the lower right figure it can be seen that 
$R_K$ is the main measurement resulting in the best fit values for $C_9^\mu$ and $C_9^e$ which are in more than $2\sigma$
tension with lepton-universality (and with $3.9 \sigma$ tension with the SM). Removing the data on $R_K$ from the fit, lepton-universality can be restored at slightly larger than $1\sigma$. So at present $R_K$ is the only observable within  the $b \to s \ell^+\ell^-$ transitions which shows some sign of lepton non-universality.

\subsection{Fit results considering only $B\to K^* \mu^+ \mu^-$ observables}
\begin{figure}[!t]
\begin{center}
\includegraphics[width=0.47\textwidth]{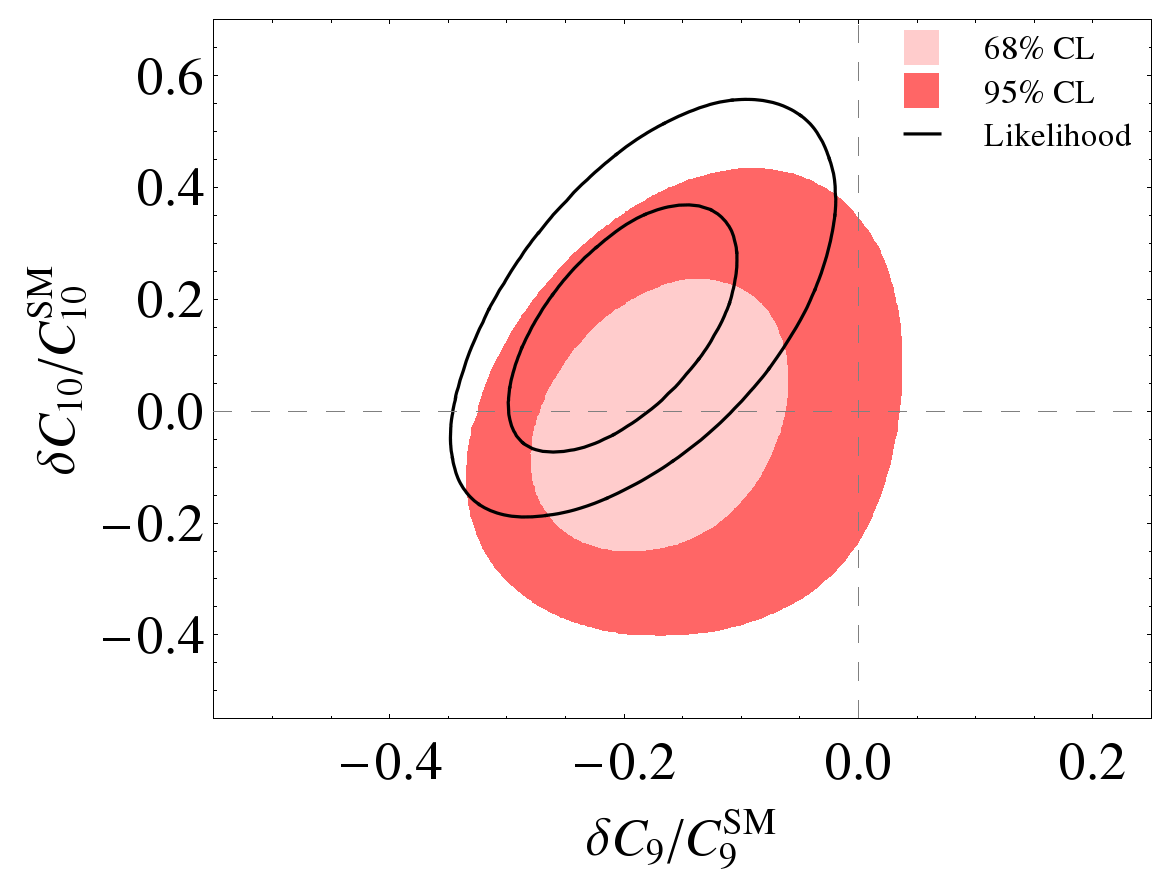}
\includegraphics[width=0.47\textwidth]{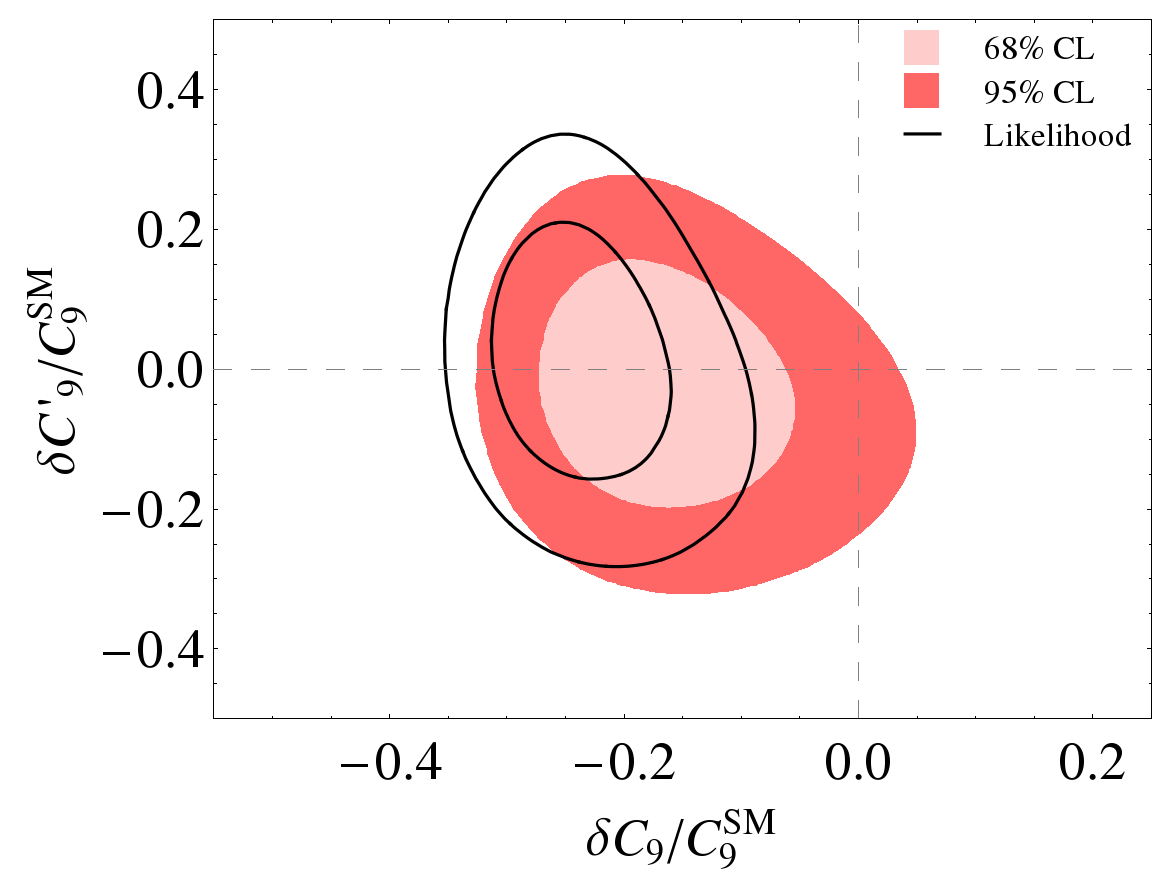}
\caption{Global fit results with the $\Delta \chi^2$ method, using \emph{only} the $B\to K^* \mu^+ \mu^-$ data in the full FF approach with 10\% power correction. 
The coloured regions correspond to the best fit value using the method of moment results  at 1 and 2$\sigma$. The solid lines depict the 1 and 2$\sigma$ allowed regions using the likelihood method results for the $B\to K^* \mu^+\mu^-$ observables.
\label{fig:OnlyBtoKstar}}
\end{center}
\end{figure}

The experimental measurements of the $B \to K^* \mu^+ \mu^-$ angular observables have been obtained 
using the method of moments as well as the most likelihood method.
While the former gives less precise results, it is more robust compared to the latter one, specially for low signal
yields. 
The new physics analysis clearly depends on the 
method which has been used to obtain the $B\to K^* \mu^+ \mu^-$ experimental measurements.
To compare the best fit points when using the two different results we have done two operator fits
using \emph{only} the data on $B\to K^* \mu^+ \mu^-$ observables. 
In Fig.~\ref{fig:OnlyBtoKstar}, the coloured areas show the allowed regions at
1 and 2$\sigma$ when using the method of moment results while the solid and dashed lines correspond to 
the 1 and 2$\sigma$ contours by using the method of most likelihood results. 
We see that when using the method of moment results, the agreement of the fit with the SM is better. 
E.g., in the $C_{9},C_{10}$ plane the SM has a pull of 1.7$\sigma$ with the best fit point when using the 
method of moments results, compared to 3.3$\sigma$ for the most likelihood results.
The smaller tension between the SM value and the best fit point within the method of moments 
is partly due to small shifts in the central values and mostly due to the larger experimental errors
associated with the results of this method.

\subsection{Global four operator fits}
There is in principle no reason to assume that new physics shows up only in one or two Wilson coefficients. Hence, it is of importance to check the global agreement of the experimental data when allowing new physics contributions to several Wilson coefficients at the time.
In this section we consider four operator fits, where in all the cases the
full FF approach with 10\% power correction is used. 

\subsubsection{Global fit results for four operators \{$C_9^e, C_9^\mu, C_9^{\prime e}, C_9^{\prime\mu}$\}}
\begin{figure}[!t]
\begin{center}
\includegraphics[width=0.49\textwidth]{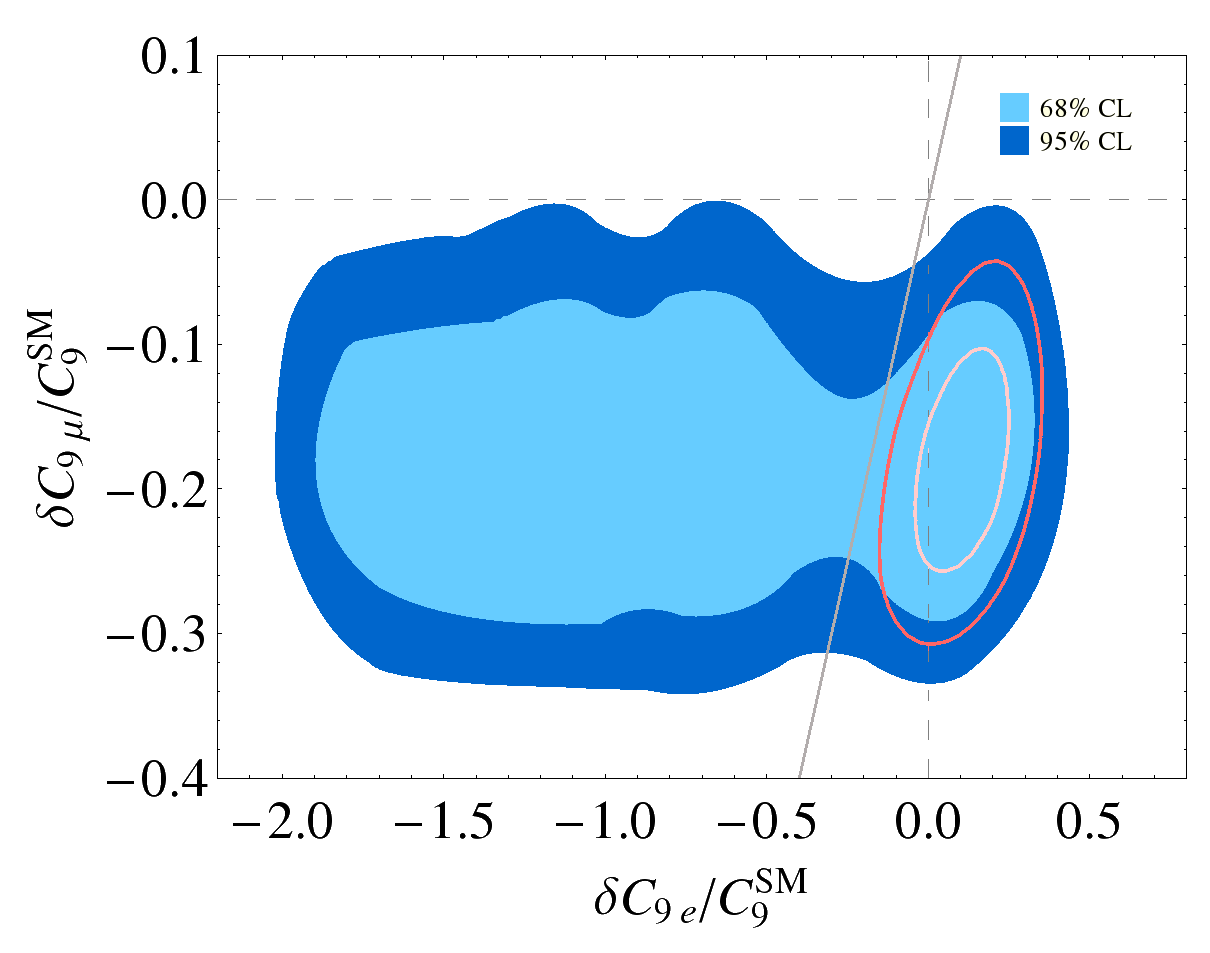}
\includegraphics[width=0.49\textwidth]{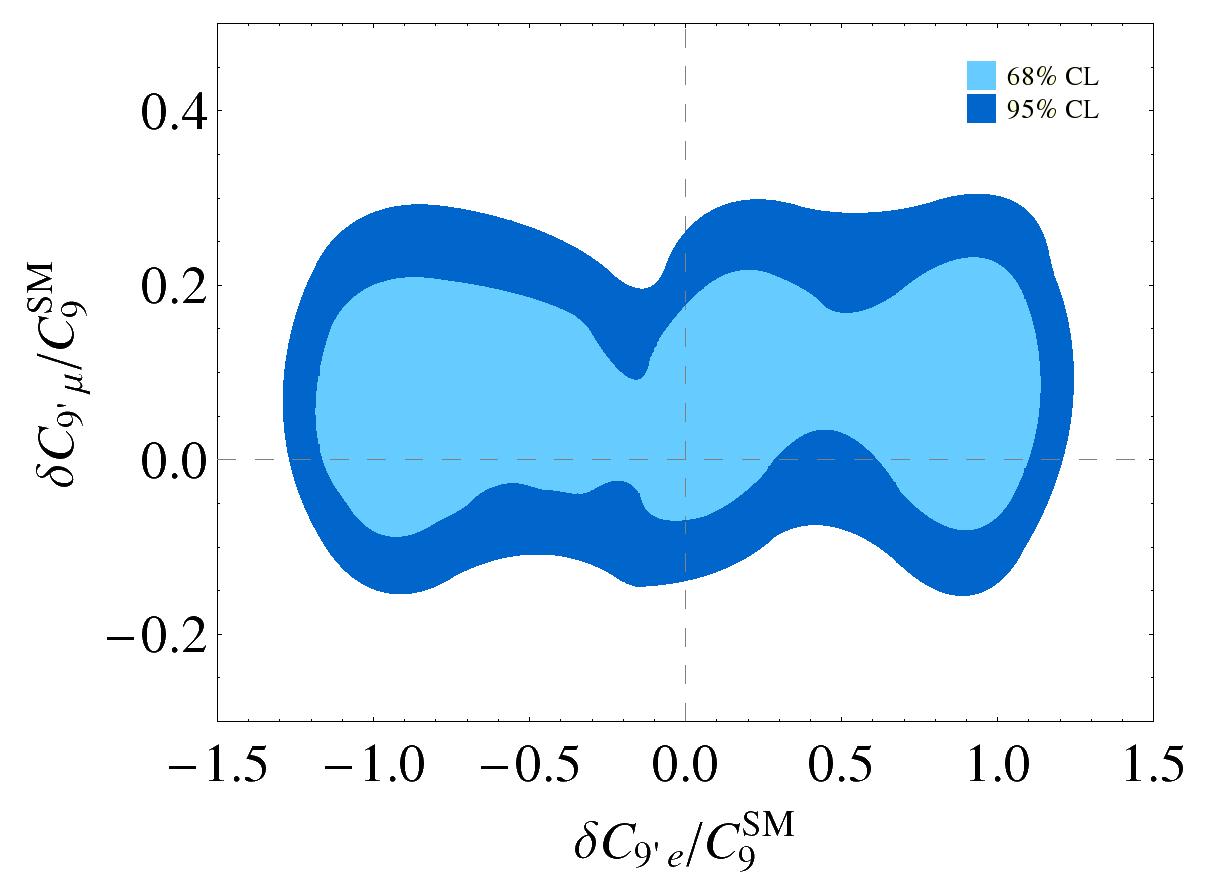}\\[5.mm]
\includegraphics[width=0.49\textwidth]{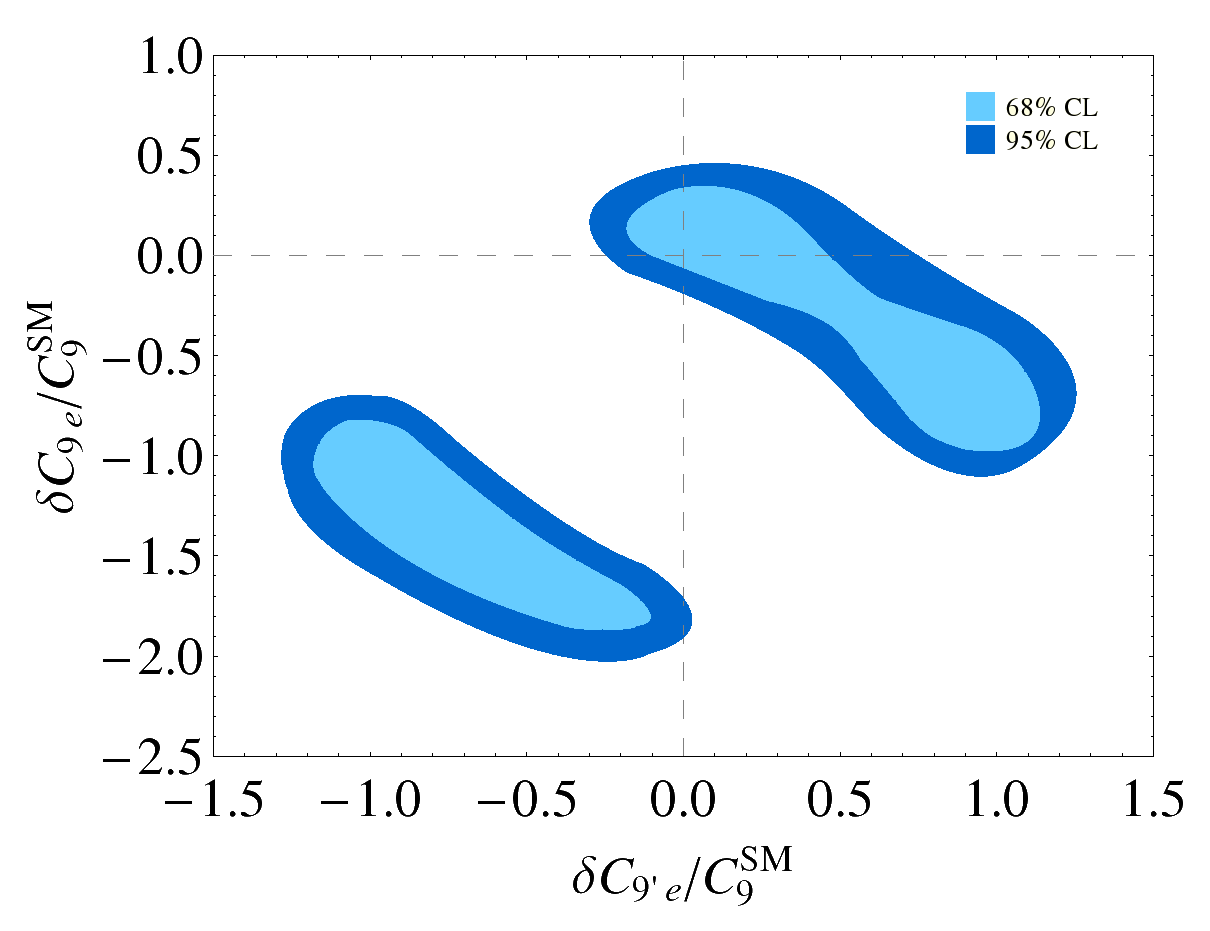}
\includegraphics[width=0.49\textwidth]{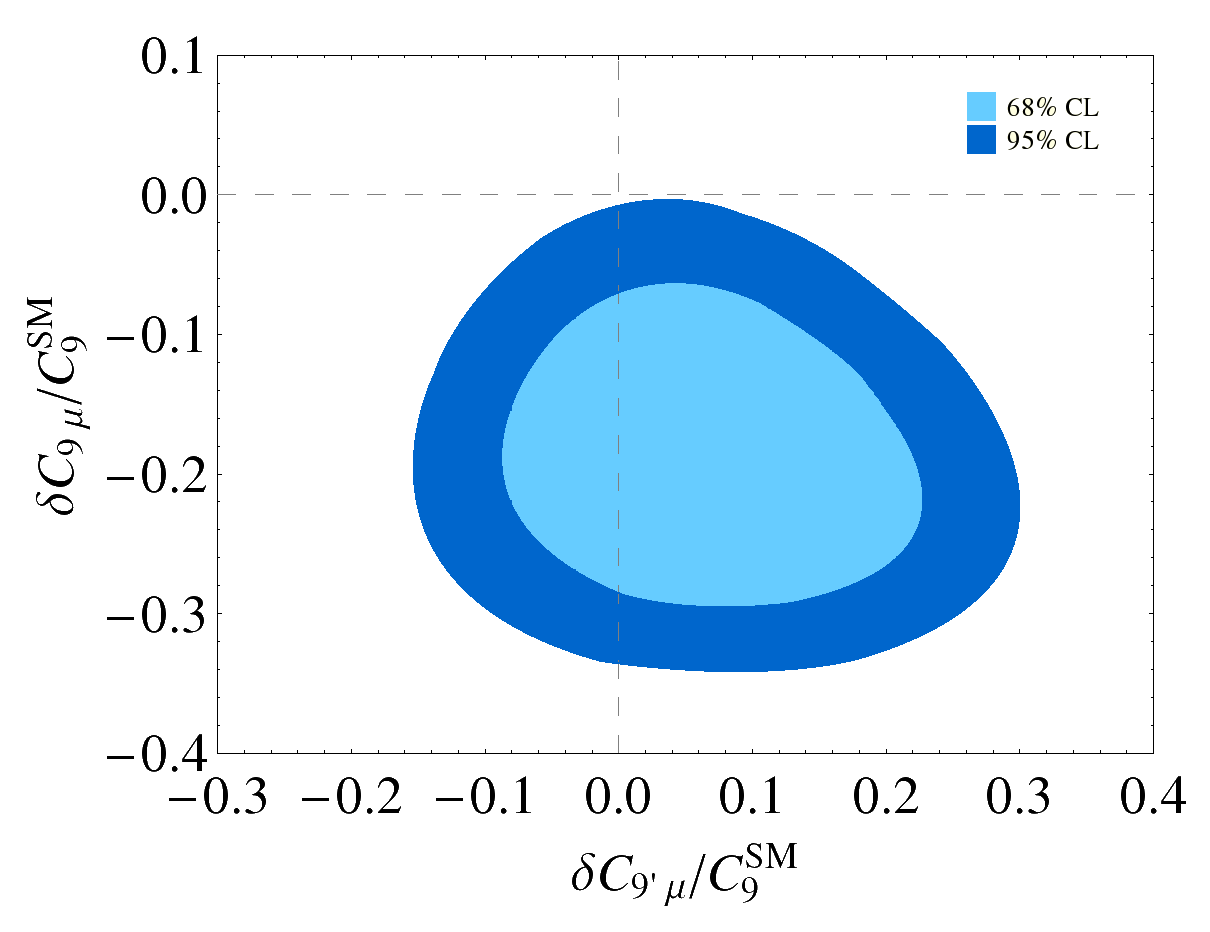}
\caption{Global fit results by using full form factors, with $\Delta \chi^2$ method. 
The (light) red contour in the upper left plot corresponds to the (1) 2$\sigma$ allowed region when new physics is considered in two operators only.
The gray line corresponds to the lepton flavour universality condition.
\label{fig:4-op-9e9m9pe9pm}}
\end{center}
\end{figure}

Assuming new physics to appear in the $O_9$ and $O_{9}^\prime$ operators with different contributions for the electron and muon sectors, 
we have fitted the \{$C_9^e, C_9^\mu, C_9^{\prime e}, C_9^{\prime\mu}$\} set of Wilson coefficients to the $b\to s$ data.
In this four operator fit the SM has 3.4$\sigma$ tension with the best fit point which has a $\chi^2$ of 113.3.
In Fig.~\ref{fig:4-op-9e9m9pe9pm}, a projection of the allowed regions at 1 and 2$\sigma$ on different Wilson coefficient planes are shown.
In the upper right plot of Fig.~\ref{fig:4-op-9e9m9pe9pm}, the projection of the 
1 and 2$\sigma$ regions in the \{$C_9^e, C_9^\mu$\} plane is given, where each point within the 
coloured region indicates that there exist at least some values of $C_9^{\prime e}$ and $C_9^{\prime\mu}$ for which the corresponding $C_9^e$ and $C_9^\mu$ value give a $\chi^2$ that is within the 2$\sigma$ region. 
In this plot, besides the projected four operator fit, we have overlaid the 1 and 2$\sigma$ contours of the two operator fit of \{$C_9^e, C_9^\mu$\} for comparison.
The comparison of the results shows that considering only the modification of two Wilson coefficients leads to much more restrictive results.
And while in the latter, lepton universality is in more than 2$\sigma$ tension with the best fit point, in the four operator fit lepton universality in $C_9$ is respected even within the 1$\sigma$ region. However,
in this case the lepton non-universality is appearing in $C_9^\prime$ which is not shown in the projected plane. 

\subsubsection{Global fit results for four operators \{$C_9^e, C_9^\mu, C_{10}^{e}, C_{10}^{\mu}$\}}
\begin{figure}[!t]
\begin{center}
\includegraphics[width=0.49\textwidth]{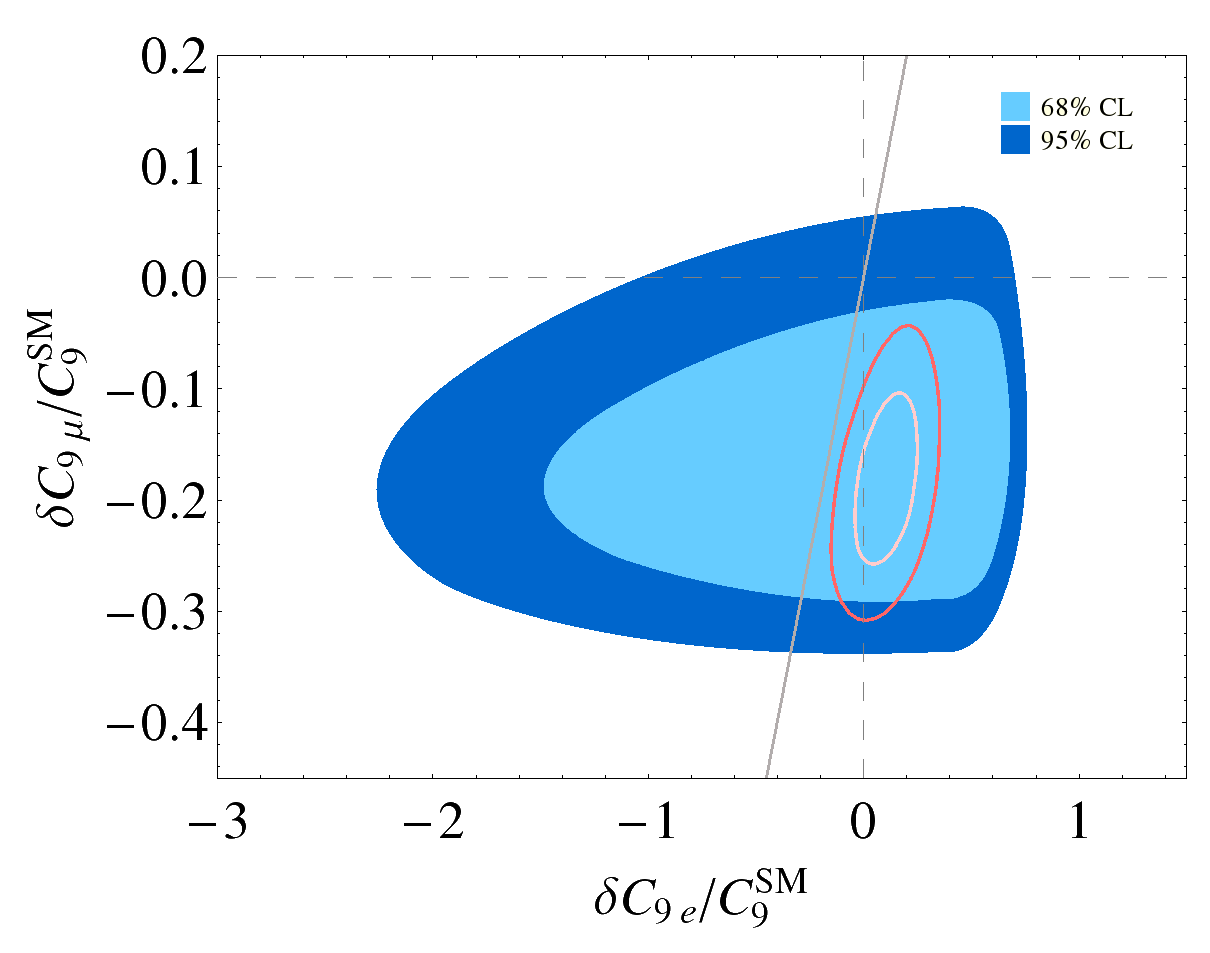}
\includegraphics[width=0.49\textwidth]{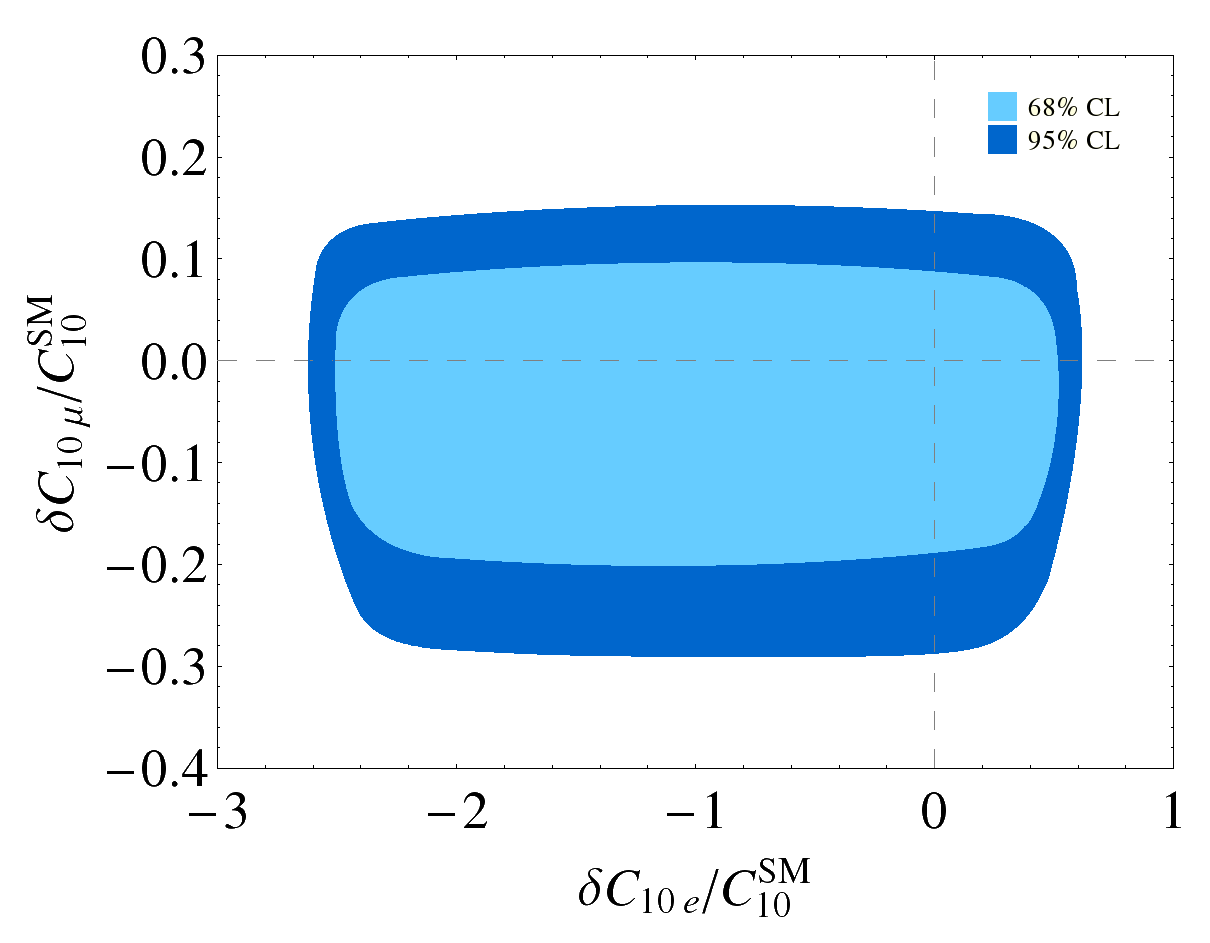}\\[5.mm]
\includegraphics[width=0.49\textwidth]{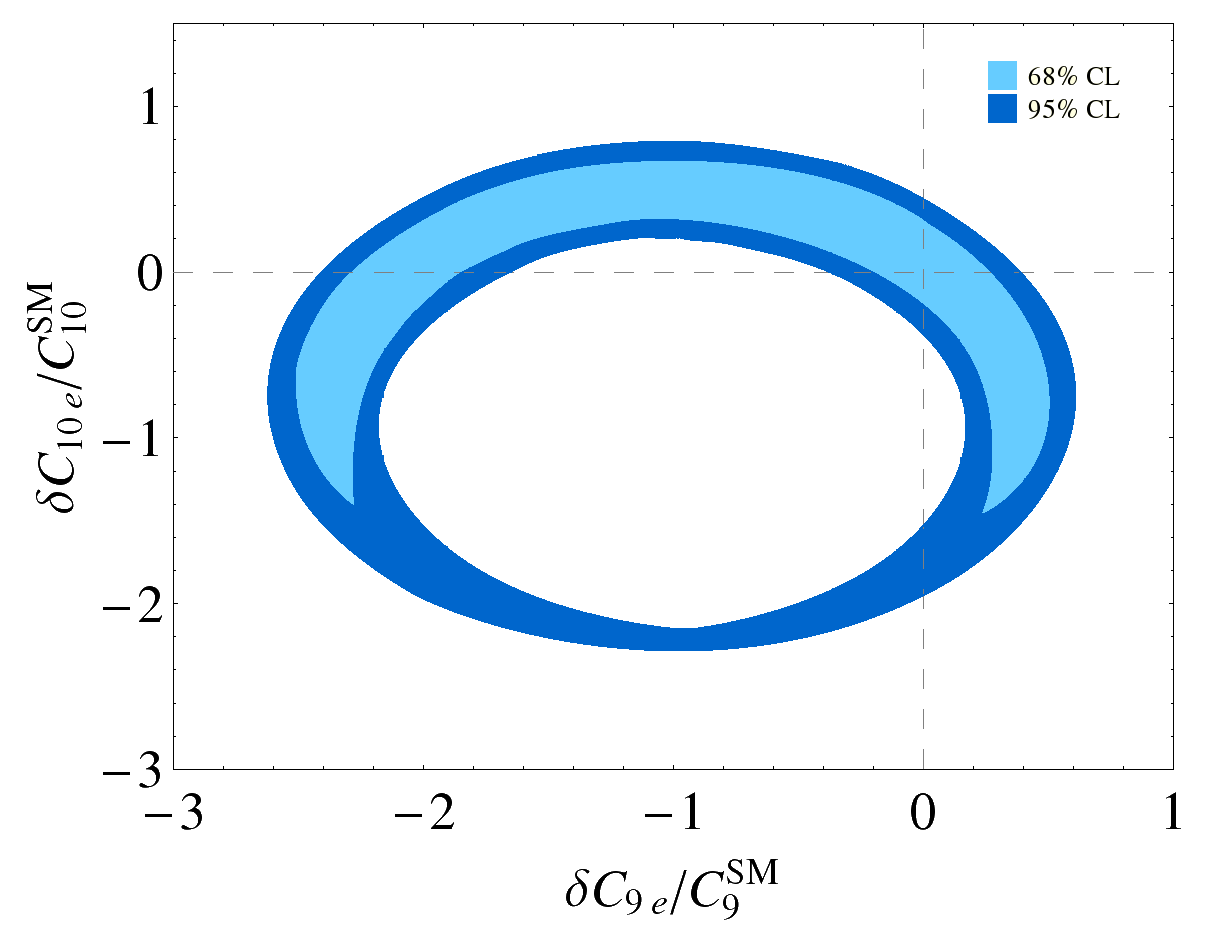}
\includegraphics[width=0.49\textwidth]{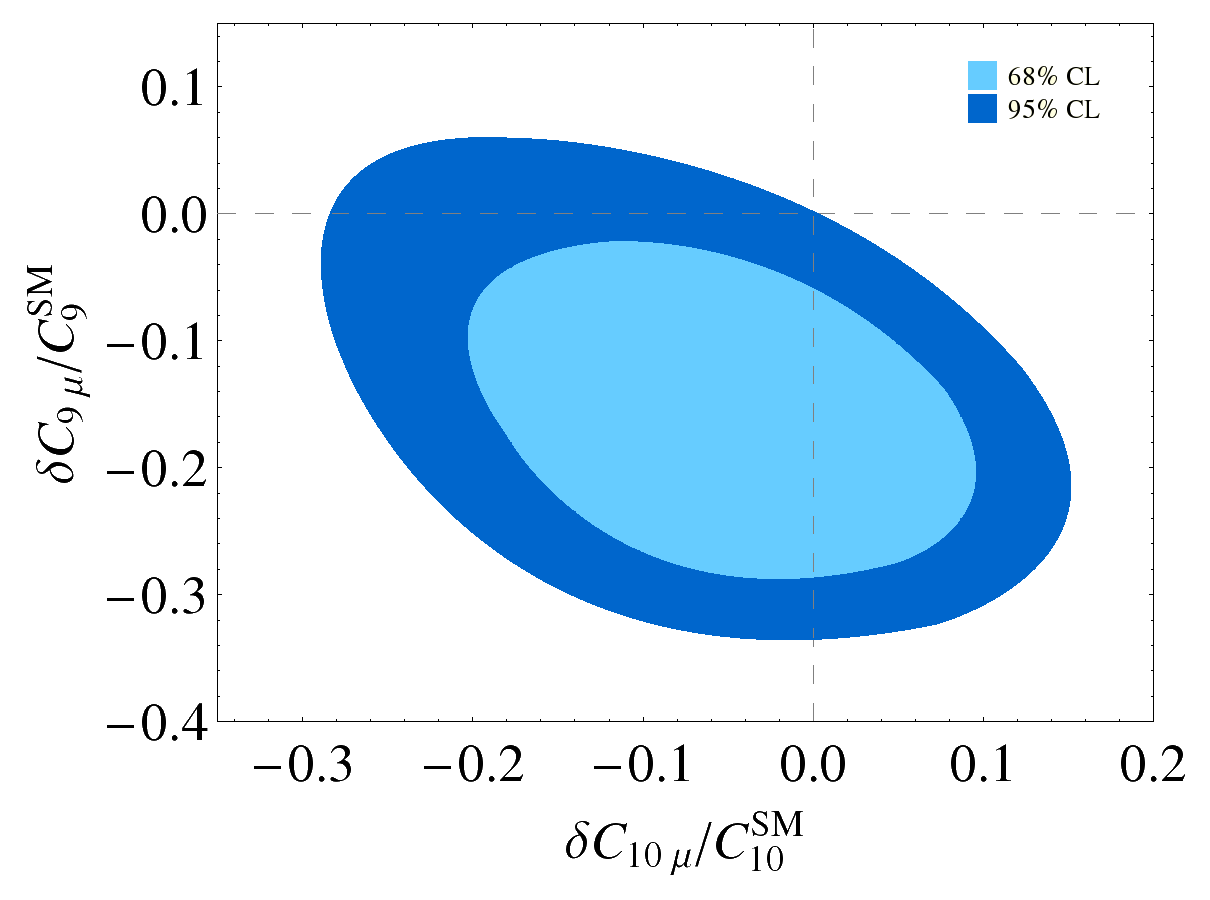}
\caption{Global fit results using full form factors, with $\Delta \chi^2$ method. 
The (light) red  contour in the upper left plot corresponds to the (1) 2$\sigma$ allowed region when new physics is considered in two operators only. 
The gray line corresponds to the lepton flavour universality condition.
\label{fig:4-op-9e9m10e10m}}
\end{center}
\end{figure}

We consider here the four operator fit assuming lepton non-universality in $O_9$ and $O_{10}$.
The best fit point of the \{$C_9^e, C_9^\mu, C_{10}^{e}, C_{10}^{\mu}$\} fit has a $\chi^2=114$,
which indicates that there is no improvement in the fit when replacing
the operator $O_9^\prime$ with $O_{10}$, even when considering different contributions for muons and electrons (the same result was also seen in the two operator fit with lepton universality).
In the \{$C_9^e, C_9^\mu, C_{10}^{e}, C_{10}^{\mu}$\} fit the SM value has a pull of 3.4$\sigma$ 
with respect to the best fit point, similar to the \{$C_9^e, C_9^\mu, C_9^{\prime e}, C_9^{\prime\mu}$\} fit.
Some of the possible two dimensional projections of the four operator fit are presented in Fig.~\ref{fig:4-op-9e9m10e10m}.
In the upper left plot, the 1 and 2$\sigma$ contours of the \{$C_9^e, C_9^\mu$\} two operator fit has also been shown.
A comparison between the allowed regions in the two and four operator fits shows that considering four operator fits considerably relaxes the constraints on the Wilson coefficients leaving room for more diverse new physics contributions which are otherwise overlooked.

\subsubsection{Global fit results for four operators \{$C_9, C_{10}, C_9^{\prime}, C_{10}^{\prime}$\}}
\begin{figure}[!h]
\begin{center}
\includegraphics[width=0.49\textwidth]{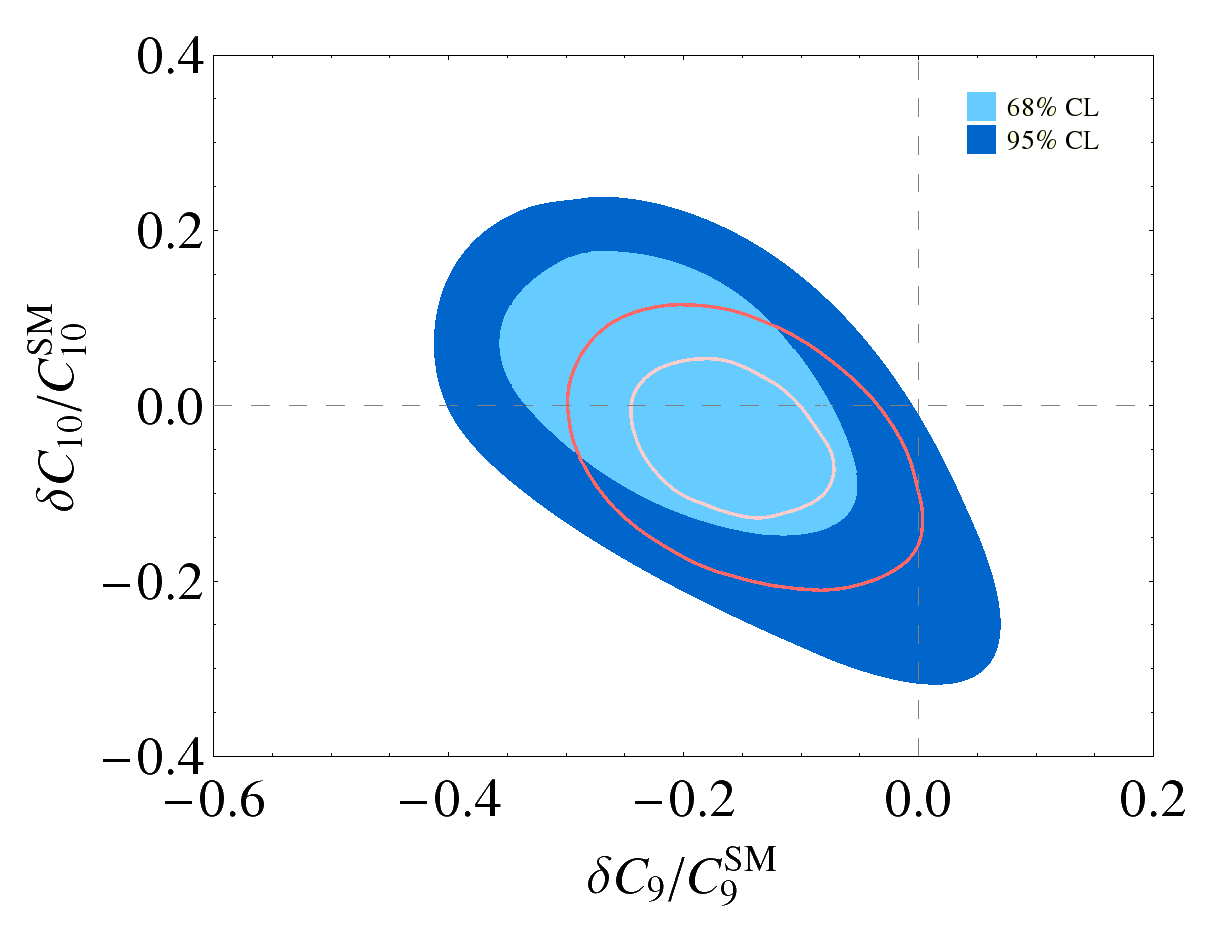}
\includegraphics[width=0.49\textwidth]{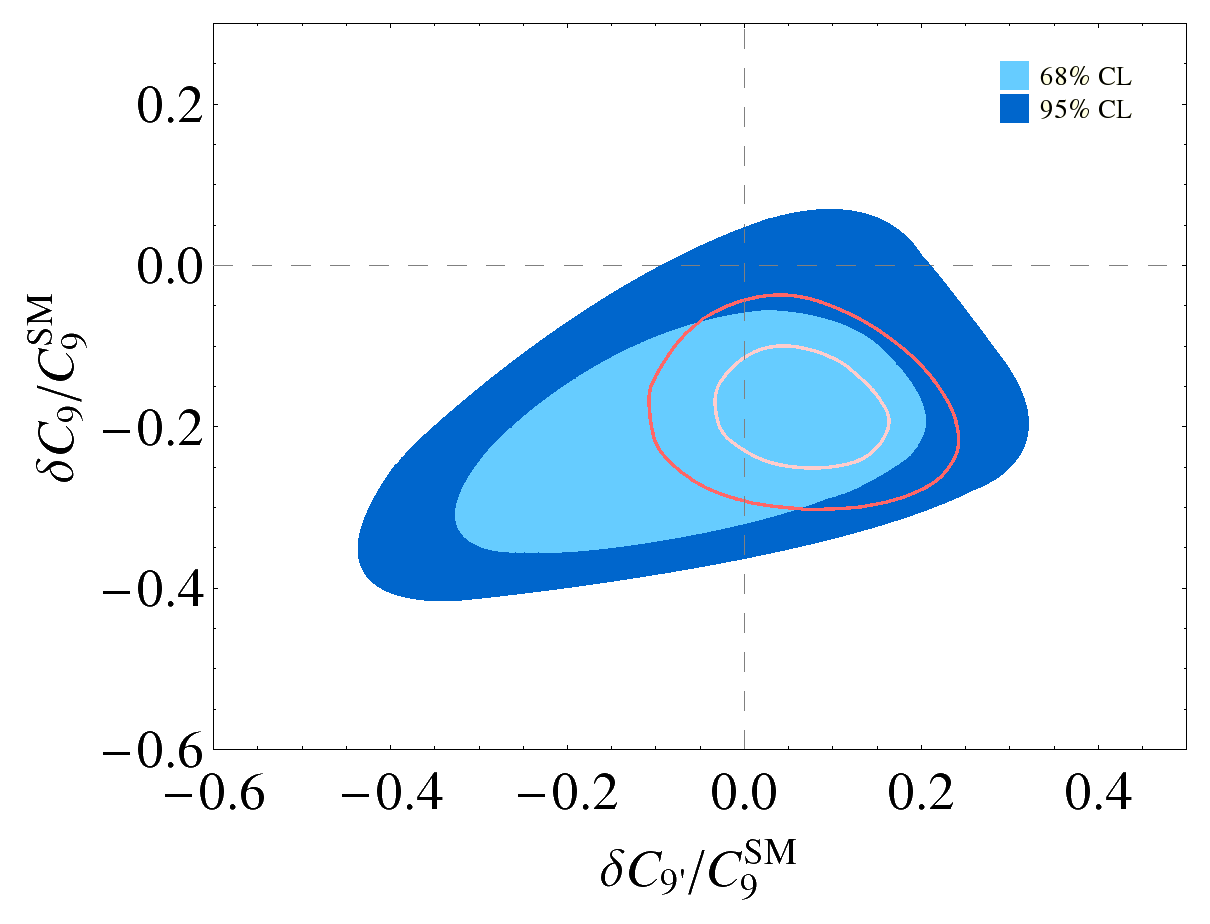}\\[5.mm]
\includegraphics[width=0.49\textwidth]{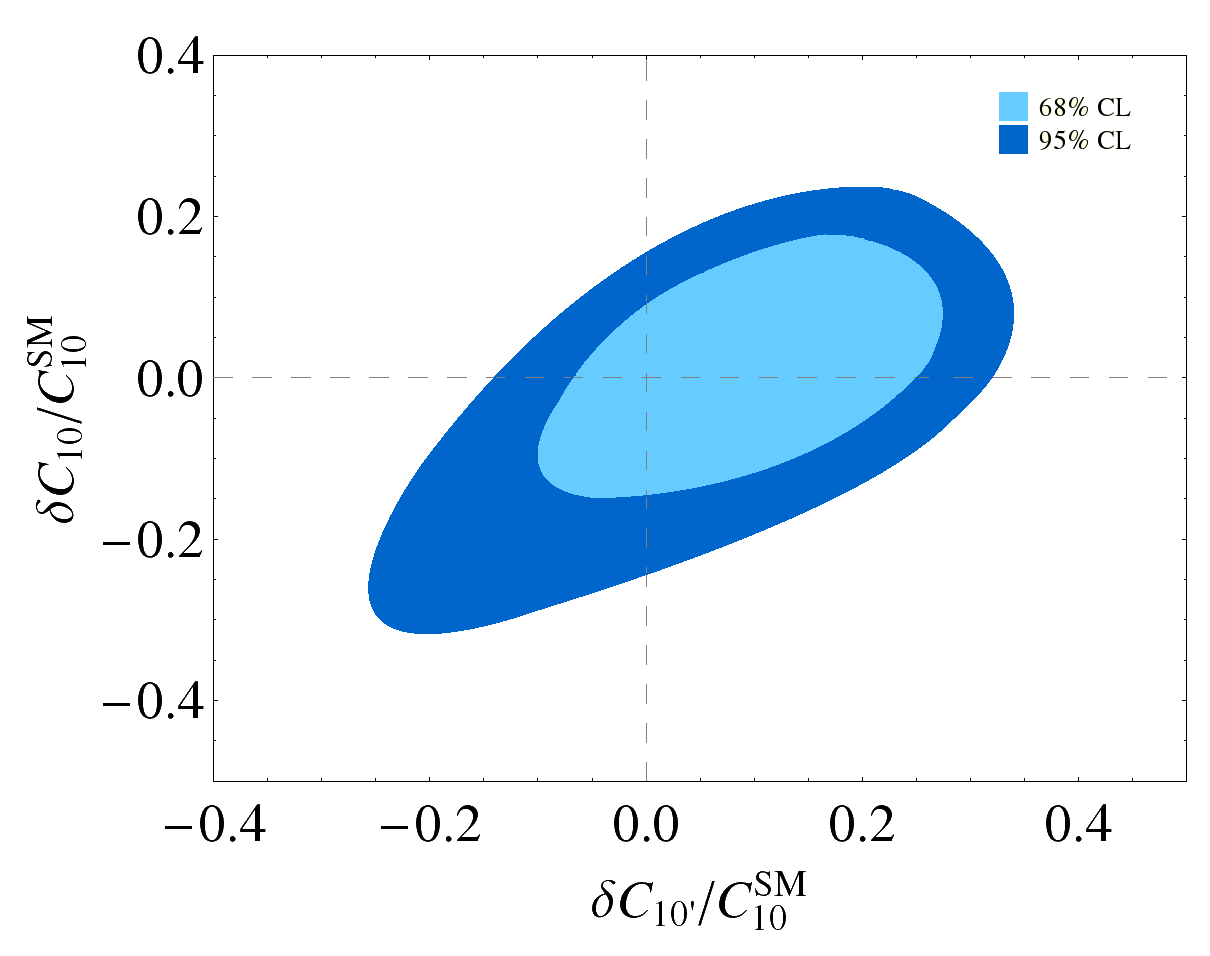}
\includegraphics[width=0.49\textwidth]{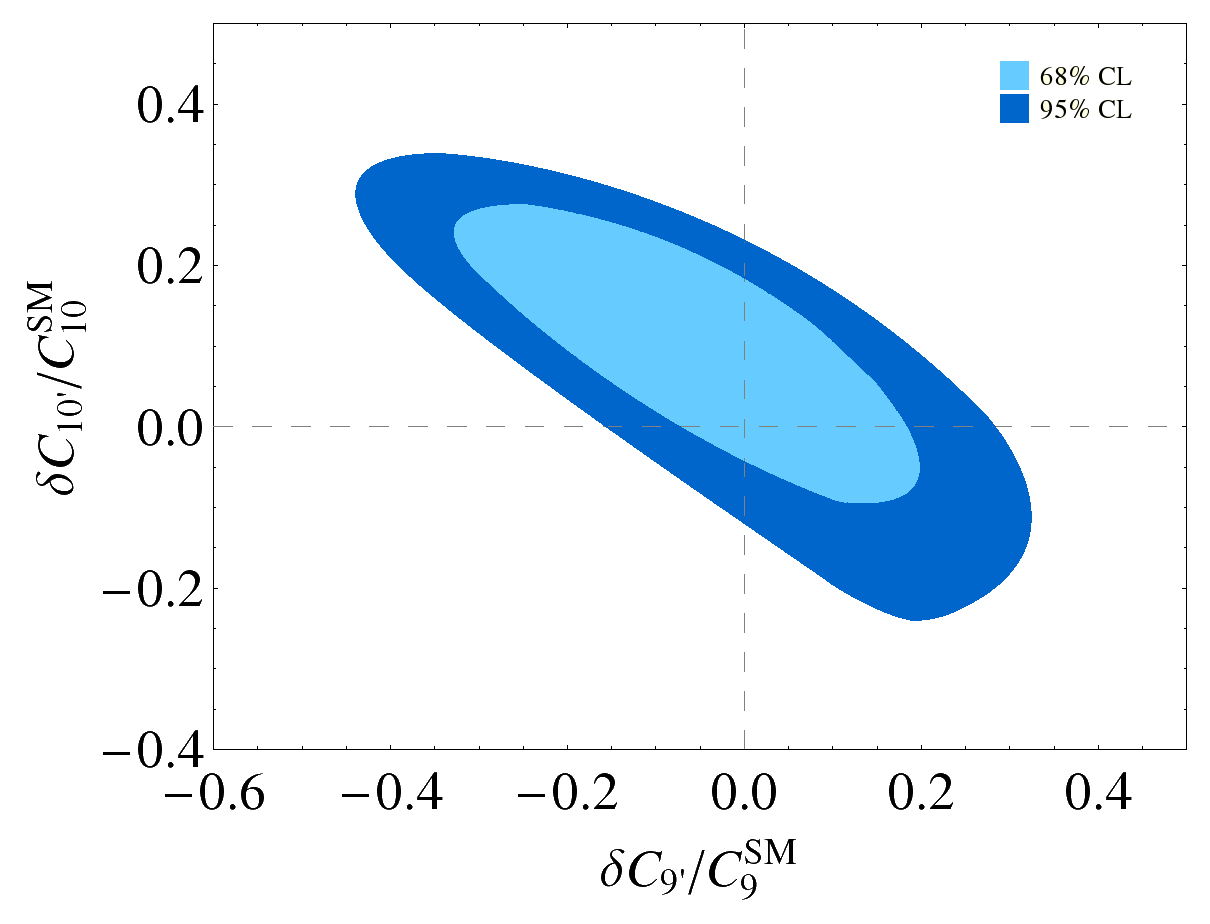}
\caption{Global fit results using full form factors, with $\Delta \chi^2$ method. 
The (light) red contour in the upper left plot corresponds to the (1) 2$\sigma$ allowed region when new physics is considered in two operators only.
\label{fig:4-op-910m9p10p}}
\end{center}
\end{figure}

In Fig.~\ref{fig:4-op-910m9p10p}, the projection of the \{$C_9, C_{10}, C_9^{\prime}, C_{10}^{\prime}$\} fit on different 2-dimensional planes are demonstrated.
This four operator fit has a best fit point with $\chi^2=121.6$ which indicates that the experimental measurements are better described assuming lepton non-universality as in the two previous subsections. 
In this case the SM value of the Wilson coefficients has a pull of 2.3$\sigma$ with the best fit point.
Including the primed operators with respect to the two operator fit for \{$C_9, C_{10}$\} (with $\chi^2=123.7$) does not 
improve the fit\footnote{In the four operator fit there are two less degrees of freedom 
with respect to the two operator fit.} (see also the upper left plot of Fig.~\ref{fig:4-op-910m9p10p}).
The two-operator fits are {overlaid} again in the projection plot of the four-operator fit. The comparison shows that
the bounds based on the two-operator fits are always stronger by construction.

\subsection{Global fit results  in MFV}
In this section we show the impact of the $b \to s$ data  within the framework of 
minimal flavour violation (MFV), see e.g.~\cite{Chivukula:1987py,Hall:1990ac,D'Ambrosio:2002ex,Hurth:2008jc,Hurth:2012jn}
and~\cite{Isidori:2012ts} for a recent review.
There are different definitions for the MFV framework. We follow the canonical one which is based on a symmetry principle introduced in Ref.~\cite{D'Ambrosio:2002ex}, which implies that in a MFV model all flavour-violating interactions can be traced back to the well-known structure of the Yukawa couplings and all CP-violating interactions are due to the physical phase in the CKM matrix. 

The specific hierarchy of the quark masses and CKM matrix elements implies that only a small number of operators are relevant in the MFV framework compared to a general model-independent analysis. 
Here, besides the operators present in the SM, especially $O_7,O_8,O_9,$ and $O_{10}$,  we also have to consider
the scalar-density operator with right-handed $b$-quarks ($O_l = e^2/16\pi^2 \times (\bar s_L b_R)(\bar \ell_R \ell_L)$) (see Ref.~\cite{Hurth:2012jn} for more details).

The MFV hypothesis represents an important benchmark in the sense that any measurement which is inconsistent with the general constraints and relations induced by the MFV hypothesis unambiguously indicates the existence of new flavour structures. 
Thus, any incompatibility of the $b \to s$ measurements with the MFV hypothesis would imply that new flavour structures are needed to explain the data. 

\begin{figure}[!t]
\begin{center}
\includegraphics[width=0.49\textwidth]{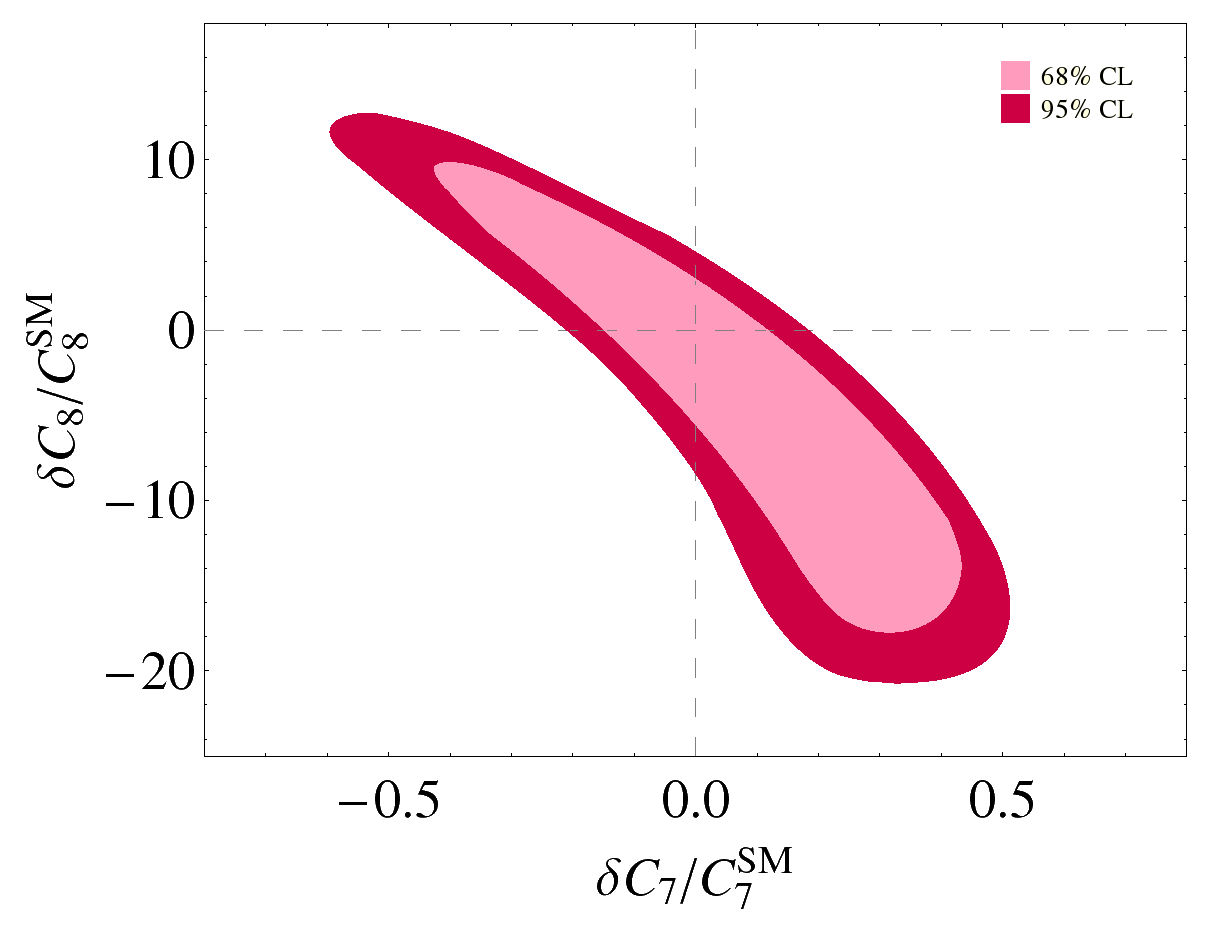}
\includegraphics[width=0.49\textwidth]{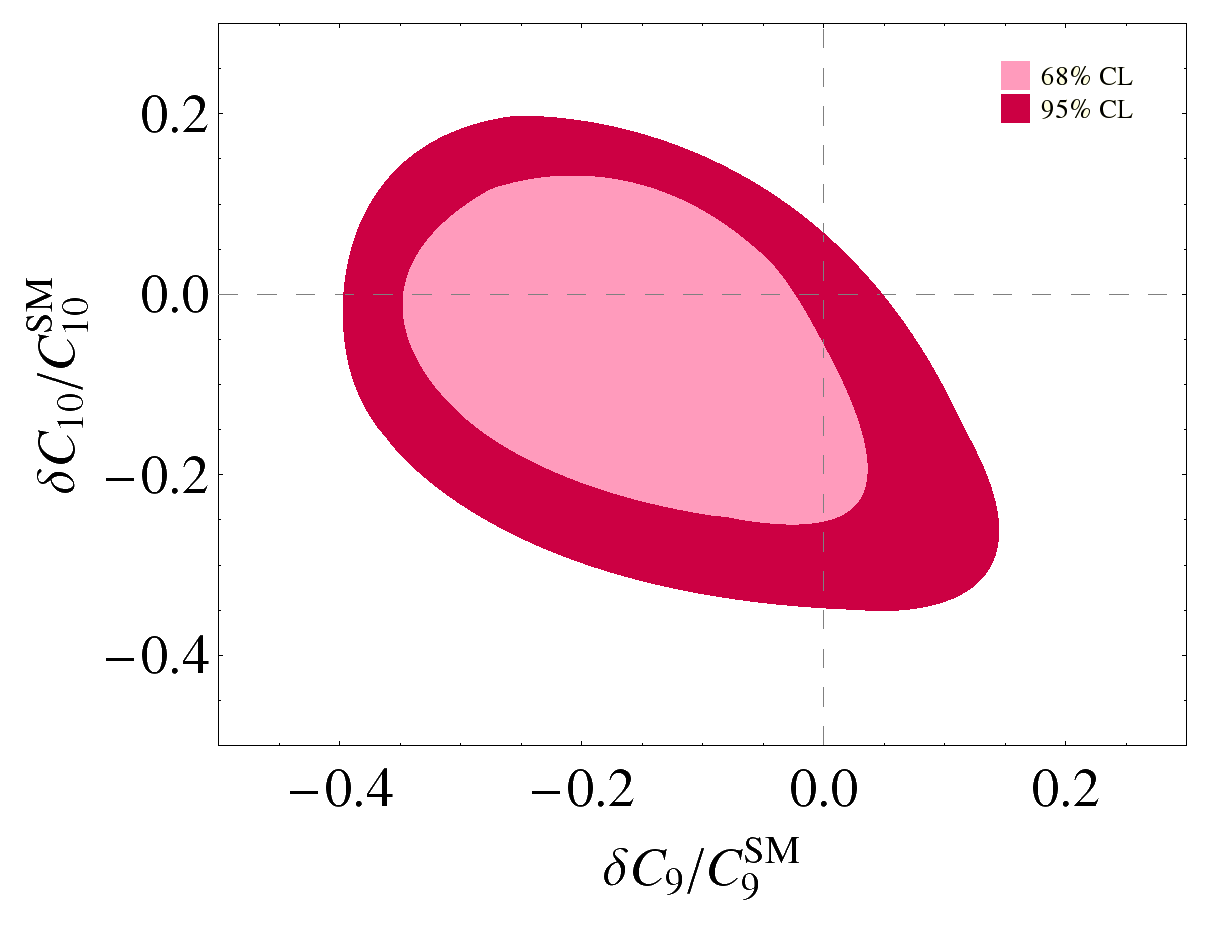}\\[5.mm]
\includegraphics[width=0.49\textwidth]{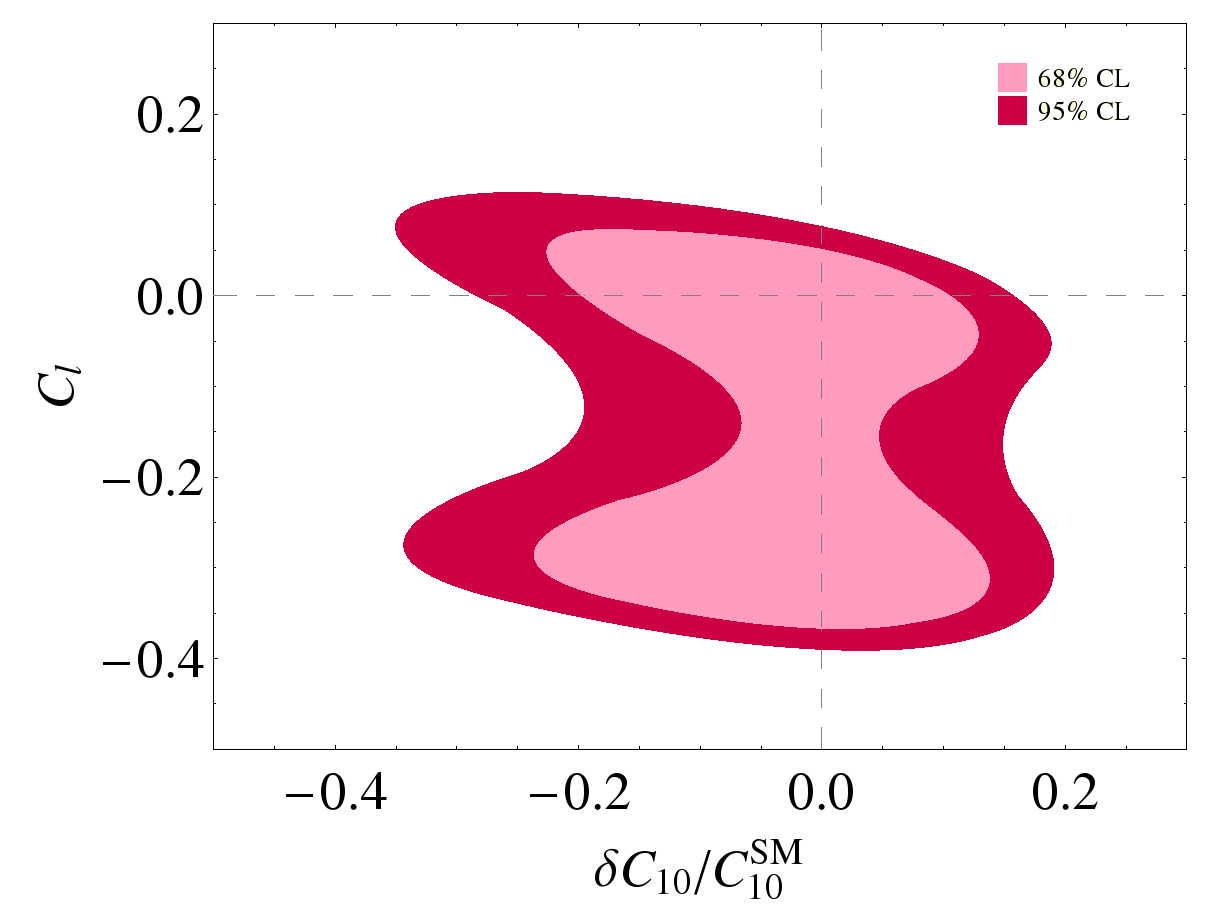}
\caption{ Global fit results for $\delta C_i$ in the MFV effective theory.\label{fig:MFV}}
\end{center}
\end{figure}

We have made a global fit within the MFV framework using the five operators listed above. In addition to the observables used in the previous fits, we include BR($B\to X_s\gamma$) and the isospin asymmetry of $B\to K^*\gamma$, which are sensitive to the $O_7$ and $O_8$ operators. The best fit point has a $\chi^2$ of 123.5 for 137 degrees of freedom which represents a good fit. The five operator fit within the MFV framework shows compatibility with the MFV hypothesis. 
In Fig.~\ref{fig:MFV}, the resulting bounds on the Wilson coefficients are shown.

%%%%%%%%%%%%%%%%%%%%%%%%%%%%%%%%%%%%%%%%%%%%%%
\subsection{$R_K$ and predictions for other ratios}

In the introduction, we have identified the observable $R_K$ as a possible key observable to clarify also the origin of the anomalies in the LHCb data. 
The ratio $R_K = {\rm BR}(B^+ \to K^+ \mu^+ \mu^-) / $ ${\rm BR}(B^+ \to K^+ e^+ e^-)$ in the low-$q^2$ region had been measured by LHCb using the full 3 fb$^{-1}$ of data, showing a $2.6\sigma$ deviation from the SM prediction~\cite{Aaij:2014ora}. This might be a sign for lepton non-universality. 
In contrast to the anomalies in the angular observables in $B \to K^{*} \mu^+\mu^-$, the ratio  $R_K$ is theoretically rather clean, in  particular it is unaffected by power corrections and also the electromagnetic corrections are under control. Because both tensions might be explained by new physics in the Wilson coefficients $C_9$, other ratios of observables  which may indicate lepton non-universality will be important crosschecks of the anomalies discussed in this paper. In Section~\ref{sec:without} we found that at present the $R_K$ ratio is the only driving force for lepton non-universality. 

In Ref.~\cite{Altmannshofer:2014rta}, the authors predict the central values for such ratios based on a specific choice of the Wilson coefficients, in particular on the assumption that the electron modes are SM like. In contrast, we make our predictions for such ratios based on the global fit considering 
two Wilson coefficients  $C_9^\mu$ and $C_9^e$. We find that in most cases the SM point is outside the 2$\sigma$ region of our indirect predictions reflecting the present 
tension in $R_K$ (see Table~\ref{tab:ratios}). Moreover, the ratio in the case of the $A_{FB}$ looks most promising from the theoretical point of view. 

Finally we note that we have shown previously~\cite{Hurth:2013ssa,Hurth:2014zja,Hurth:2014vma}, that the present data on inclusive and exclusive decays are compatible with each other
and there is no sign of lepton non-universality in the published data on the inclusive mode.

\begin{table}[!t]
\ra{0.90}
\rb{1.3mm}
\begin{center}
\setlength\extrarowheight{2pt}
\rowcolors{1}{}{light-gray}
\footnotesize{\begin{tabular}{c|c}
Observable &  \cen{95\% C.L. prediction} \\
\hline \hline  
${\rm BR}( B\to X_s \mu^+ \mu^- )/{\rm BR}( B\to X_s e^+ e^- )_{q^2 \in[1,6]  ({\rm GeV})^2}$                   &  $ [0.61, 0.93] $    \\ [1mm]
${\rm BR}( B\to X_s \mu^+ \mu^- )/{\rm BR}( B\to X_s e^+ e^- )_{q^2 > 14.2  ({\rm GeV})^2}$                     &  $ [0.68, 1.13] $    \\ [1mm]
${\rm BR}( B^0\to K^{*0} \mu^+ \mu^- )/\;{\rm BR}( B^0\to K^{*0} e^+ e^- )_{q^2 \in[1,6]  ({\rm GeV})^2}$       &  $ [0.65, 0.96] $    \\ [1mm]
$\av{F_L(B^0\to K^{*0} \mu^+ \mu^-)}/\av{F_L(B^0\to K^{*0} e^+ e^-)}_{q^2 \in[1,6]  ({\rm GeV})^2}$             &  $ [0.85, 0.96] $    \\ [1mm]
$\av{A_{F\!B}(B^0\to K^{*0} \mu^+ \mu^-)}/\av{A_{F\!B}(B^0\to K^{*0} e^+ e^-)}_{q^2 \in[4,6]  ({\rm GeV})^2}$   &  $ [-0.21, 0.71] $    \\ [1mm]
$\av{S_5(B^0\to K^{*0} \mu^+ \mu^-)}/\av{S_5(B^0\to K^{*0} e^+ e^-)}_{q^2 \in[4,6]  ({\rm GeV})^2}$             &  $ [0.53, 0.92] $    \\ [1mm]
${\rm BR}( B^0\to K^{*0} \mu^+ \mu^- )/\;{\rm BR}( B^0\to K^{*0} e^+ e^- )_{q^2 \in[15,19]  ({\rm GeV})^2}$     &  $ [0.58, 0.95] $    \\ [1mm]
$\av{F_L(B^0\to K^{*0} \mu^+ \mu^-)}/\av{F_L(B^0\to K^{*0} e^+ e^-)}_{q^2 \in[15,19]  ({\rm GeV})^2}$           &  $ [0.998, 0.999] $    \\ [1mm]
$\av{A_{F\!B}(B^0\to K^{*0} \mu^+ \mu^-)}/\av{A_{F\!B}(B^0\to K^{*0} e^+ e^-)}_{q^2 \in[15,19]  ({\rm GeV})^2}$ &  $ [0.87, 1.01] $    \\ [1mm]
$\av{S_5(B^0\to K^{*0} \mu^+ \mu^-)}/\av{S_5(B^0\to K^{*0} e^+ e^-)}_{q^2 \in[15,19]  ({\rm GeV})^2}$           &  $ [0.87, 1.01] $    \\ [1mm]
${\rm BR}( B^+\to K^{+} \mu^+ \mu^- )/\;{\rm BR}( B^+\to K^{+} e^+ e^- )_{q^2 \in[1,6]  ({\rm GeV})^2}$         &  $ [0.58, 0.95] $    \\ [1mm]
${\rm BR}( B^+\to K^{+} \mu^+ \mu^- )/\;{\rm BR}( B^+\to K^{+} e^+ e^- )_{q^2 \in[15,22]  ({\rm GeV})^2}$       &  $ [0.58, 0.95] $    \\ [1mm]
%%%%%%
\end{tabular} }
\caption{Predicted ratios of observables with muons in the final state to electrons in the final state, considering the two operator fit within the \{$C_9^\mu,C_9^e$\} set.\vspace*{-0.2cm}
\label{tab:ratios}}
\end{center} 
\end{table}

%%%%%%%%%%%%%%%%%%%%%%%%%%%%%%%%%%%%%%%%%%%%%%%%
\section{Conclusions}
The LHCb collaboration has recently presented new data on exclusive $b \to s \ell^+\ell^-$ penguin decays~\cite{Aaij:2015oid,Aaij:2015esa}. These data were eagerly awaited because of some tensions with the SM in the angular observables of $B \to K^* \mu^+\mu^-$.  

In view of these new data we have stressed again that there is a significant difference in the theoretical accuracy of the inclusive and exclusive $b \to s \ell^+\ell^-$ decays in the low-$q^2$ region.  
The theoretical description of power corrections exists in the inclusive case so that they can be at least estimated within the 
theoretical approach. On the contrary, no theoretical description of power corrections exists in the exclusive case in the framework of QCD factorisation and SCET, and power corrections can only be guesstimated. This issue makes it rather difficult or even impossible to separate new physics effects from such potentially large hadronic power corrections within these exclusive angular observables. 
Therefore, these tensions might stay unexplained until Belle II will clarify the situation by measuring the corresponding inclusive $b\to s \ell^+ \ell^-$ observables as we have demonstrated previously~\cite{Hurth:2013ssa,Hurth:2014zja,Hurth:2014vma}.

The present situation motivated a recent theory analysis in which the unknown power corrections were just fitted to the data using an ansatz with 18 additional real parameters in the fit~\cite{Ciuchini:2015qxb}. 
However, we have shown that this fit to the data needs very large power corrections from 20\% up to 50\%, and even larger, 
in the critical bins of the angular observables. The existence of such large hadronic corrections cannot be ruled out in principle, but they somehow question the validity of the QCD factorisation approach for such observables. 
 
We have analysed how the tensions in the present LHCb data depend on the input parameters like form factor calculations and corresponding correlations 
%and on the guesstimate of power corrections 
and have shown that they are rather insensitive to these inputs. Only quadrupling the form factor error makes a real difference.
We have also found that the standard guesstimate of non-factorisable power corrections from 5\% to 20\% at the amplitude level has no real impact on the theoretical predictions. As we have shown such variations lead to a 6\% error at the observable level for the three observables $S_3$, $S_4$ and $S_5$.
Only variations significantly larger than 60\% -- corresponding to errors larger that 20\% at the observable level -- have a real  impact. For example we have shown that for the three observables $S_3$, $S_4$ and $S_5$, a 150\% of the power correction error is needed in the critical bins to reproduce errors up to 50\% at the observable level which are comparable in size with the corrections needed in the aforementioned recent fit to the SM. 

In addition, we have explicitly shown that within a new physics analysis the observable $S_5$ is not the only observable that drives the new physics Wilson coefficient $\delta C_9$ to negative values, but also other observables play a similar role. 

We have noted that the other tensions in the LHCb data of $b \to s \ell\ell$ may play a crucial role in the near future, namely  
the observable $R_K$ with a $2.6\sigma$ deviation from the SM prediction~\cite{Aaij:2014ora} 
what might signal a first sign of lepton non-universality. In contrast to the anomaly in the rare decay $B \to K^{*} \mu^+\mu^-$ which is affected by power corrections, the ratio $R_K$ is theoretically rather clean and its tension with the SM cannot be explained by power corrections. Neither some missing 
electromagnetic corrections can  serve as explanation for the discrepancy from the SM. 

It might be just an accidental coincidence that the tensions in $R_K$ and in the angular observables can simultaneously be resolved by a negative new physics contribution
to the Wilson coefficient of the semileptonic operator $O_9$. If not, then the measurement of analogous ratios which may show signs of lepton non-universality will be also an important crosscheck of the anomalies in the angular observables and might even resolve the puzzle.  
Therefore, we have presented predictions for such ratios based on our global fits.

%%%%%%%%%%%%%%%%%%%%%%%%%%%%%%%%%%%%%%%%%%%%%%%%
\section*{Acknowledgement} 
TH thanks the CERN theory group for its hospitality during his regular visits to CERN where
part of this work was written.
The authors are grateful to J\'er\^ome Charles for his valuable comments, and to Roman Zwicky and Marcin Chrzaszcz for useful discussions.

%%%%%%%%%%%%%%%%%%%%%%%%%%%%%%%%%%%%%%%%%%%%%%%%

\appendix

\begin{table}
\parbox[t][4.5cm][t]{1.00\linewidth}{
\section{SM predictions and experimental values}\label{app:SMpredict}
The experimental values and SM predictions of the observables considered in this work have been given 
in Tables~\ref{tab:BRrest}-\ref{tab:BRBstophi}.
The angular observables in Tables~\ref{tab:BtoKstar}-\ref{tab:Bstophi} are all given according to the LHCb conventions.
}
\parbox[c][3cm][c]{0.90\linewidth}{
\ra{0.90}
\rb{1.3mm}
\begin{center}
\setlength\extrarowheight{2pt}
\rowcolors{1}{}{light-gray}
\scalebox{0.80}{\begin{tabular}{l|rrl}
\hline
Observable & \cen{SM prediction} & Measurement & \\ 
\hline \hline
%
%%%%%%
$10^4\times{\rm BR}(B \to X_s \gamma)$ & $ 3.40 \pm 0.19$ & $ 3.43 \pm 0.22 $ & \cite{Amhis:2012bh} \\ [1mm]
$10^2\times{\Delta_0}(B \to K^* \gamma)$ & $5.1 \pm 1.5$ & $ 5.2 \pm 2.6 $ & \cite{Amhis:2012bh} \\ [1mm] \hline
%%%%%%
$10^9\times{\rm BR}(B_s \to \mu^+ \mu^-)$ & $ 3.54 \pm 0.27 $ & $2.9 \pm 0.7$ & \cite{Aaij:2013aka,Chatrchyan:2013bka,CMS:2014xfa} \\ [1mm]
$10^{10}\times{\rm BR}(B_d \to \mu^+ \mu^-)$ & $ 1.07 \pm 0.27 $ & $3.6 \pm 1.6$ &  \cite{Aaij:2013aka,Chatrchyan:2013bka,CMS:2014xfa} \\ [1mm] \hline
$R_{K\:{q^2 \in [1.0,6.0]({\rm GeV})^2}}$ & $1.0006 \pm 0.0004$ & $0.745\pm{0.097}$ & \cite{Aaij:2014ora} \\ [1mm] \hline
%%%%%%
$10^{6}\times{\rm BR}\left( B\to X_s e^+ e^- \right)_{q^2 \in[1,6]  ({\rm GeV})^2}$     & $ 1.73\err{0.12}{0.12} $ &  $1.93 \pm{0.55}$ & \cite{Lees:2013nxa}  \\ [1mm]
$10^{6}\times{\rm BR}\left( B\to X_s e^+ e^- \right)_{q^2 > 14.2 ({\rm GeV})^2}$        & $ 0.20\err{0.06}{0.06} $ &  $0.56 \pm{0.19}$ & \cite{Lees:2013nxa} \\ [1mm] \hline 
$10^{6}\times{\rm BR}\left( B\to X_s \mu^+ \mu^- \right)_{q^2 \in[1,6]  ({\rm GeV})^2}$ & $ 1.67\err{0.12}{0.12} $ &  $0.66 \pm{0.88}$ & \cite{Lees:2013nxa}  \\ [1mm]
$10^{6}\times{\rm BR}\left( B\to X_s \mu^+ \mu^- \right)_{q^2 > 14.2 ({\rm GeV})^2}$    & $ 0.23\err{0.07}{0.06} $ &  $0.60 \pm{0.31}$ & \cite{Lees:2013nxa}  \\ [1mm] 
%%%%%%
%%%%%%
\end{tabular}}
\caption{The SM predictions and experimental values.\label{tab:BRrest}} 
\end{center} 
}\\
\parbox[b][16cm][b]{0.45\linewidth}{
\ra{0.90}
\rb{1.3mm}
\begin{center}
\rowcolors{2}{light-gray}{}
\footnotesize{\begin{tabular}[b]{l||l|l}
\toprule[.6pt]
\multicolumn{3}{l}{$B\to K  \mu^+\mu^-$  differential branching ratios}\\
\hline
\toprule[.2pt]
\rowcolor{white}Bin (GeV$^2$) & SM prediction & \cen{Measurement} \\  
\hline \hline
%
%%%%%%
%\rowcolor{yellow}
\multicolumn{3}{c}{$10^9\times \av{dBR/dq^2}(B^0\to K^0 \mu^+ \mu^-)$} \\
%%%%%%
$[0.1-2.0]$   &  $ 26.6 \pm{5.7}$ & $12.2 ^{+5.9}_{-5.2} \pm 0.6$ \\ 
$[2.0-4.0]$   &  $ 27.0 \pm{6.1}$ & $18.7 ^{+5.5}_{-4.9} \pm 0.9 $ \\ 
$[4.0-6.0]$   &  $ 27.8 \pm{7.3}$ & $17.3 ^{+5.3}_{-4.8} \pm 0.9 $ \\ 
$[6.0-8.0]$   &  $ 28.2 \pm{8.4}$ & $27.0 ^{+5.8}_{-5.3} \pm 1.4 $ \\ 
% $[11.0-12.5]$ & $ 25.6 \pm{4.6}$ & $12.7 ^{+4.5}_{-4.0} \pm 0.6 $ \\ 
$[15.0-17.0]$ & $ 19.2 \pm{2.2}$ & $14.3 ^{+3.5}_{-3.2} \pm 0.7 $ \\ 
$[17.0-22.0]$ & $ 11.0 \pm{1.2}$ & $\phantom{0}7.8 ^{+1.7}_{-1.5} \pm 0.4$ \\ \hline
$[1.1-6.0]$   & $ 27.1 \pm{6.4}$ & $18.7 ^{+3.5}_{-3.2} \pm 0.9$ \\ 
$[15.0-22.0]$ & $ 13.3 \pm{1.5}$ & $ \phantom{0}9.5 ^{+1.6}_{-1.5} \pm 0.5 $ \\ \hline
%%%%%%% 
\rowcolor{white} & &    \\ [-12pt] %\hline
\midrule[.5pt]
%%%%%%
%\rowcolor{yellow}
\multicolumn{3}{c}{$10^9\times \av{dBR/dq^2}(B^+\to K^+ \mu^+ \mu^-)$} \\
%%%%%%
$[0.1-0.98]$    & $ 28.4 \pm{6.2}$ & $33.2 \pm 1.8 \pm 1.7$ \\ 
$[1.1-2.0]$     & $ 29.3 \pm{6.3}$ & $23.3 \pm 1.5 \pm 1.2$ \\ 
$[2.0-3.0]$     & $ 29.5 \pm{6.5}$ & $28.2 \pm 1.6 \pm 1.4$ \\ 
$[3.0-4.0]$     & $ 29.7 \pm{6.9}$ & $25.4 \pm 1.5 \pm 1.3$ \\ 
$[4.0-5.0]$     & $ 29.9 \pm{7.5}$ & $22.1 \pm 1.4 \pm 1.1$ \\ 
$[5.0-6.0]$     & $ 30.0 \pm{8.3}$ & $23.1 \pm 1.4 \pm 1.2$ \\ 
$[6.0-7.0]$     & $ 30.3 \pm{9.2}$ & $24.5 \pm 1.4 \pm 1.2$ \\ 
$[7.0-8.0]$     & $ 30.8 \pm{9.0}$ & $23.1 \pm 1.4 \pm 1.2$ \\ 
$[15.0-16.0]$   & $ 21.7 \pm{2.5}$ & $16.1 \pm 1.0 \pm 0.8$ \\ 
$[16.0-17.0]$   & $ 19.7 \pm{2.2}$ & $16.4 \pm 1.0 \pm 0.8$ \\ 
$[17.0-18.0]$   & $ 17.5 \pm{1.9}$ & $20.6 \pm 1.1 \pm 1.0$ \\ 
$[18.0-19.0]$   & $ 15.0 \pm{1.6}$ & $13.7 \pm 1.0 \pm 0.7$ \\ 
$[19.0-20.0]$   & $ 12.3 \pm{1.3}$ & $\phantom{0}7.4 \pm 0.8 \pm 0.4$ \\ 
$[20.0-21.0]$   & $ \phantom{0}9.3 \pm{1.0}$ & $\phantom{0}5.9 \pm 0.7 \pm 0.3$ \\ 
$[21.0-22.0]$   & $ \phantom{0}5.9 \pm{0.8}$ & $\phantom{0}4.3 \pm 0.7 \pm 0.2$ \\\hline 
$[1.1-6.0]$   & $ 29.7 \pm{7.0}$ & $24.2 \pm 0.7 \pm 1.2$ \\ 
$[15.0-22.0]$ & $ 14.4 \pm{1.6}$ & $12.1 \pm 0.4 \pm 0.6$ \\ 
%%%%%%% 
\end{tabular}}
\caption{SM predictions and experimental values for the differential branching ratio of the $B\to K  \mu^+\mu^-$ decay.
The uncertainties of the experimental values~\cite{Aaij:2014pli} are (from left to right) statistical and systematic.\label{tab:BRBtoKstar}}
\end{center} 
}
\hfill
\parbox[b][13cm][b]{.45\linewidth}{
%%%%%%
%%%%%%%%B->K mu mu%%%%%%%%%%%%%%%
% \begin{table}
\ra{0.90}
\rb{1.3mm}
\begin{center}
\rowcolors{2}{light-gray}{}
\footnotesize{\begin{tabular}[b]{l||l|l}
\toprule[.6pt]
\multicolumn{3}{l}{$B\to K^{*}  \mu^+\mu^-$ differential branching ratios} \\
\hline
\toprule[.2pt]
\rowcolor{white}Bin (GeV$^2$) & SM prediction & \cen{Measurement} \\  
\hline \hline
%
%%%%%%
%\rowcolor{yellow}
\multicolumn{3}{c}{$10^9\times \av{dBR/dq^2}(B^+\to K^{*+} \mu^+ \mu^-)$} \\
%%%%%%
$[0.1-2.0]$    & $ 83.1 \pm{10.9}$ & $59.2 ^{+14.4}_{-13.0} \pm 4.0$  \\ 
$[2.0-4.0]$    & $ 46.1 \pm{7.1}$  & $55.9 ^{+15.9}_{-14.4} \pm 3.8 $ \\ 
$[4.0-6.0]$    & $ 50.2 \pm{7.0}$  & $24.9 ^{+11.0}_{-9.6}  \pm 1.7 $ \\ 
$[6.0-8.0]$    & $ 57.0 \pm{7.0}$  & $33.0 ^{+11.3}_{-10.0} \pm 2.3 $ \\ 
$[15.0-17.0]$  & $ 67.0 \pm{7.9}$  & $64.4 ^{+12.9}_{-11.5} \pm 4.4 $ \\ 
$[17.0-22.0]$  & $ 46.7 \pm{6.0}$  & $11.6 ^{9.1}_{-7.6}    \pm 0.8$  \\ \hline
%%%%%%% 
\rowcolor{white} & &    \\ [-12pt] %\hline
\midrule[.5pt]
%%%%%%
%\rowcolor{yellow}
\multicolumn{3}{c}{$10^7\times \av{dBR/dq^2}(B^0\to K^{*0} \mu^+ \mu^-)$} \\
%%%%%%
$[0.10-2.00]$    & $ 0.81 \pm{0.10}$ & $0.60 \pm{0.10} $ \\ 
$[2.00-4.30]$    & $ 0.43 \pm{0.07}$ & $0.30 \pm{0.05} $ \\ 
$[4.30-8.68]$    & $ 0.51 \pm{0.07}$ & $0.49 \pm{0.08} $ \\ 
$[14.18-16.00]$  & $ 0.65 \pm{0.08}$ & $0.56 \pm{0.10} $ \\ 
$[16.00-19.0]$   & $ 0.48 \pm{0.06}$ & $0.41 \pm{0.07} $ \\
%%%%%%% 
\end{tabular}}
\caption{SM predictions and experimental values for the differential branching ratio of the $B\to K^*  \mu^+\mu^-$ decay.
The experimental error of the $B^+\to K^{*+}  \mu^+\mu^-$ 
decay~\cite{Aaij:2014pli} are (from left to right) statistical and systematic.
The experimental error of the $B^0\to K^{*0}  \mu^+\mu^-$ decay~\cite{Aaij:2013iag} have been added in quadrature, 
taking the largest side error in case of 
non-symmetrical uncertainties.
\label{tab:BRBtoK} }
\end{center} 
}
\end{table}
%%%%%%%%%%%%
%%%%%%%%%%%%
\clearpage

\begin{table}
\ra{0.90}
\rb{1.3mm}
\begin{center}
\setlength\extrarowheight{2pt}
\rowcolors{1}{}{light-gray}
\footnotesize{\begin{tabular}{lrcc||lrc}
\toprule[.6pt]
\multicolumn{7}{l}{$B^0\to K^{*0}\mu^+\mu^-$ angular observables (measurement obtained by the unbinned maximum likelihood fit)}\\
\hline
\toprule[.2pt]
\rowcolor{white}Observable & \cen{SM prediction} & Measurement        &  & Observable & \cen{SM prediction} & Measurement \\ 
\hline \hline
%
%%%%%%
%\rowcolor{yellow}
\multicolumn{3}{c}{$q^2 \in [\,0.10\,,\,0.98\,]\,{\rm GeV}^2 $}             & & \multicolumn{3}{c}{$q^2 \in [\,1.1\,,\,2.5\,]\,{\rm GeV}^2 $ }    \\  
%%%%%%
$\av{F_L}$    & $ 0.252   \pm{0.037} $ &  $\phantom{-}0.263\,{}^{+0.045}_{-0.044} \pm 0.017$ & & $\av{F_L}$    & $ 0.728  \pm{0.036} $ &  $\phantom{-}0.660\,{}^{+0.083}_{-0.077} \pm 0.022$ \\ 
$\av{A_{FB}}$ & $ -0.088  \pm{0.008} $ &  $-0.003\,{}^{+0.058}_{-0.057} \pm 0.009$           & & $\av{A_{FB}}$ & $ -0.153 \pm{0.023} $ &  $-0.191\,{}^{+0.068}_{-0.080} \pm 0.012$           \\ 
$\av{S_3}$    & $ 0.007   \pm{0.003} $ &  $-0.036\,{}^{+0.063}_{-0.063} \pm 0.005$           & & $\av{S_3}$    & $ 0.003  \pm{0.002} $ &  $-0.077\,{}^{+0.087}_{-0.105} \pm 0.005$           \\ 
$\av{S_4}$    & $ 0.097   \pm{0.005} $ &  $\phantom{-}0.082\,{}^{+0.068}_{-0.069} \pm 0.009$ & & $\av{S_4}$    & $ 0.009  \pm{0.009} $ &  $-0.077\,{}^{+0.111}_{-0.113} \pm 0.005$           \\ 
$\av{S_5}$    & $ 0.241   \pm{0.010} $ &  $\phantom{-}0.170\,{}^{+0.059}_{-0.058} \pm 0.018$ & & $\av{S_5}$    & $ 0.101  \pm{0.015} $ &  $\phantom{-}0.137\,{}^{+0.099}_{-0.094} \pm 0.009$ \\ 
$\av{S_7}$    & $ 0.021   \pm{0.006} $ &  $\phantom{-}0.015\,{}^{+0.059}_{-0.059} \pm 0.006$ & & $\av{S_7}$    & $ 0.033  \pm{0.008} $ &  $-0.219\,{}^{+0.094}_{-0.104} \pm 0.004$           \\ 
$\av{S_8}$    & $ 0.004   \pm{0.002} $ &  $\phantom{-}0.079\,{}^{+0.076}_{-0.075} \pm 0.007$ & & $\av{S_8}$    & $ 0.011  \pm{0.004} $ &  $-0.098\,{}^{+0.108}_{-0.123} \pm 0.005$           \\ 
$\av{S_9}$    & $ -0.001  \pm{0.001} $ &  $-0.083\,{}^{+0.058}_{-0.057} \pm 0.004$           & & $\av{S_9}$    & $ -0.001 \pm{0.001} $ &  $-0.119\,{}^{+0.087}_{-0.104} \pm 0.005$           \\ 
%%%%%%% 
\rowcolor{white} & &    \\ [-12pt] %\hline
\midrule[.5pt]
%%%%%%
%\rowcolor{yellow}
\multicolumn{3}{c}{$q^2 \in [\,2.5\,,\,4.0\,]\,{\rm GeV}^2 $ }          & & \multicolumn{3}{c}{$q^2 \in [\,4.0\,,\,6.0\,]\,{\rm GeV}^2 $ }   \\  
%%%%%%
$\av{F_L}$    & $ 0.811   \pm{0.027} $ &  $\phantom{-}0.876\,{}^{+0.109}_{-0.097} \pm 0.017$ & & $\av{F_L}$    & $ 0.742  \pm{0.035} $ &  $\phantom{-}0.611\,{}^{+0.052}_{-0.053} \pm 0.017$    \\ 
$\av{A_{FB}}$ & $ -0.047  \pm{0.011} $ &  $-0.118\,{}^{+0.082}_{-0.090} \pm 0.007$           & & $\av{A_{FB}}$ & $ 0.095  \pm{0.017} $ &  $\phantom{-}0.025\,{}^{+0.051}_{-0.052} \pm 0.004$    \\ 
$\av{S_3}$    & $ -0.010  \pm{0.003} $ &  $\phantom{-}0.035\,{}^{+0.098}_{-0.089} \pm 0.007$ & & $\av{S_3}$    & $ -0.026 \pm{0.007} $ &  $\phantom{-}0.035\,{}^{+0.069}_{-0.068} \pm 0.007$    \\ 
$\av{S_4}$    & $ -0.126  \pm{0.013} $ &  $-0.234\,{}^{+0.127}_{-0.144} \pm 0.006$           & & $\av{S_4}$    & $ -0.213 \pm{0.013} $ &  $-0.219\,{}^{+0.086}_{-0.084} \pm 0.008$              \\ 
$\av{S_5}$    & $ -0.148  \pm{0.017} $ &  $-0.022\,{}^{+0.110}_{-0.103} \pm 0.008$           & & $\av{S_5}$    & $ -0.311 \pm{0.017} $ &  $-0.146\,{}^{+0.077}_{-0.078} \pm 0.011$              \\ 
$\av{S_7}$    & $ 0.024   \pm{0.006} $ &  $\phantom{-}0.068\,{}^{+0.120}_{-0.112} \pm 0.005$ & & $\av{S_7}$    & $ 0.016  \pm{0.004} $ &  $-0.016\,{}^{+0.081}_{-0.080} \pm 0.004$              \\ 
$\av{S_8}$    & $ 0.010   \pm{0.003} $ &  $\phantom{-}0.030\,{}^{+0.129}_{-0.131} \pm 0.006$ & & $\av{S_8}$    & $ 0.008  \pm{0.002} $ &  $\phantom{-}0.167\,{}^{+0.094}_{-0.091} \pm 0.004$    \\ 
$\av{S_9}$    & $ -0.001  \pm{0.001} $ &  $-0.092\,{}^{+0.105}_{-0.125} \pm 0.007$           & & $\av{S_9}$    & $ -0.001 \pm{0.001} $ &  $-0.032\,{}^{+0.071}_{-0.071} \pm 0.004$              \\ 
%%%%%
\rowcolor{white} & &   \\ [-12pt] %\hline
\midrule[.5pt]
%%%%%%
%\rowcolor{yellow}
\multicolumn{3}{c}{$q^2 \in [\,6.0\,,\,8.0\,]\,{\rm GeV}^2 $}           & & \multicolumn{3}{c}{$q^2 \in [\,15.0\,,\,17.0\,]\,{\rm GeV}^2 $}    \\  
%%%%%%
$\av{F_L}$    & $ 0.627  \pm{0.041} $ &  $\phantom{-}0.579\,{}^{+0.046}_{-0.046} \pm 0.015$   & & $\av{F_L}$    & $ 0.340  \pm{0.039} $ &  $\phantom{-}0.349\,{}^{+0.039}_{-0.039} \pm 0.009$   \\ 
$\av{A_{FB}}$ & $ 0.230  \pm{0.026} $ &  $\phantom{-}0.152\,{}^{+0.041}_{-0.040} \pm 0.008$   & & $\av{A_{FB}}$ & $ 0.409  \pm{0.026} $ &  $\phantom{-}0.411\,{}^{+0.041}_{-0.037} \pm 0.008$   \\ 
$\av{S_3}$    & $ -0.045 \pm{0.010} $ &  $-0.042\,{}^{+0.058}_{-0.059} \pm 0.011$             & & $\av{S_3}$    & $ -0.181 \pm{0.024} $ &  $-0.142\,{}^{+0.044}_{-0.049} \pm 0.007$             \\ 
$\av{S_4}$    & $ -0.261 \pm{0.009} $ &  $-0.296\,{}^{+0.063}_{-0.067} \pm 0.011$             & & $\av{S_4}$    & $ -0.294 \pm{0.008} $ &  $-0.321\,{}^{+0.055}_{-0.074} \pm 0.007$             \\ 
$\av{S_5}$    & $ -0.392 \pm{0.013} $ &  $-0.249\,{}^{+0.059}_{-0.060} \pm 0.012$             & & $\av{S_5}$    & $ -0.315 \pm{0.024} $ &  $-0.316\,{}^{+0.051}_{-0.057} \pm 0.009$             \\ 
$\av{S_7}$    & $ 0.008  \pm{0.004} $ &  $-0.047\,{}^{+0.068}_{-0.066} \pm 0.003$             & & $\av{S_7}$    & $ 0.000  \pm{0.034} $ &  $\phantom{-}0.061\,{}^{+0.058}_{-0.058} \pm 0.005$   \\ 
$\av{S_8}$    & $ 0.005  \pm{0.002} $ &  $-0.085\,{}^{+0.072}_{-0.070} \pm 0.006$             & & $\av{S_8}$    & $ 0.000  \pm{0.009} $ &  $\phantom{-}0.003\,{}^{+0.061}_{-0.061} \pm 0.003$   \\ 
$\av{S_9}$    & $ -0.001 \pm{0.002} $ &  $-0.024\,{}^{+0.059}_{-0.060} \pm 0.005$             & & $\av{S_9}$    & $ 0.000  \pm{0.016} $ &  $-0.019\,{}^{+0.054}_{-0.056} \pm 0.004$             \\ 
%%%%%
\rowcolor{white} & &   \\ [-12pt] %\hline
\midrule[.5pt]
%%%%%%
%\rowcolor{yellow} 
\multicolumn{3}{c}{$q^2 \in [\,17.0\,,\,19.0\,]\,{\rm GeV}^2 $} \\ 
%%%%%%
$\av{F_L}$    & $ 0.322  \pm{0.042} $ &  $\phantom{-}0.354\,{}^{+0.049}_{-0.048} \pm 0.025$   \\ 
$\av{A_{FB}}$ & $ 0.321  \pm{0.024} $ &  $\phantom{-}0.305\,{}^{+0.049}_{-0.048} \pm 0.013$   \\ 
$\av{S_3}$    & $ -0.257 \pm{0.024} $ &  $-0.188\,{}^{+0.074}_{-0.084} \pm 0.017$             \\ 
$\av{S_4}$    & $ -0.309 \pm{0.010} $ &  $-0.266\,{}^{+0.063}_{-0.072} \pm 0.010$             \\ 
$\av{S_5}$    & $ -0.224 \pm{0.022} $ &  $-0.323\,{}^{+0.063}_{-0.072} \pm 0.009$             \\ 
$\av{S_7}$    & $ 0.000  \pm{0.036} $ &  $\phantom{-}0.044\,{}^{+0.073}_{-0.072} \pm 0.013$   \\ 
$\av{S_8}$    & $ 0.000  \pm{0.007} $ &  $\phantom{-}0.013\,{}^{+0.071}_{-0.070} \pm 0.005$   \\ 
$\av{S_9}$    & $ 0.000  \pm{0.013} $ &  $-0.094\,{}^{+0.065}_{-0.067} \pm 0.004$             \\ 
%%%%%
\end{tabular} }
\caption{The SM predictions and experimental values of the $B\to K^* \mu^+ \mu^-$ angular observables, evaluated by
the unbinned maximum likelihood fit.
The experimental values~\cite{Aaij:2015oid}, are (from left to right) statistical and systematic. 
\label{tab:BtoKstar}}
\end{center} 
\end{table}  
%%%%%%%%%%%%%%%%%%%%%%%%%%%%%%%%%%%%%%%%%%%%%%%%%%%%%%%%%%%%%%%%%

\begin{table}
\ra{0.90}
\rb{1.3mm}
\begin{center}
\setlength\extrarowheight{2pt}
\rowcolors{1}{}{light-gray}
\scalebox{0.67}{\begin{tabular}{l|rcc||l|rcc||l|rc}
\toprule[.6pt]
\multicolumn{11}{l}{$B^0\to K^{*0}\mu^+\mu^-$ angular observables (measurement obtained by the method of moments)}\\
\hline
\toprule[.2pt]
\rowcolor{white}Obs. & \cen{SM} & Measurement        &  & Obs. & \cen{SM} & Measurement &  & Obs. & \cen{SM} & Measurement\\ 
\hline \hline
%
%%%%%%
%\rowcolor{yellow}
\multicolumn{3}{c}{$q^2 \in [\,0.10\,,\,0.98\,]\,{\rm GeV}^2 $}            & & \multicolumn{3}{c}{$q^2 \in [\,1.1\,,\,2.0\,]\,{\rm GeV}^2 $ }  & & \multicolumn{3}{c}{$q^2 \in [\,2.0\,,\,3.0\,]\,{\rm GeV}^2 $ }  \\  
%%%%%%
$\av{F_L}$    & $ 0.252  \pm{0.037} $ & $\phantom{-}0.242\,{}^{+0.058}_{-0.056} \pm 0.026$  & & $\av{F_L}$    & $ 0.695  \pm{0.039} $ & $\phantom{-}0.768\,{}^{+0.141}_{-0.130} \pm 0.025 $ & & $\av{F_L}$    & $ 0.805  \pm{0.028} $ & $\phantom{-}0.690\,{}^{+0.113}_{-0.082} \pm 0.023 $  \\ 
$\av{A_{FB}}$ & $ -0.088 \pm{0.008} $ & $-0.138\,{}^{+0.095}_{-0.092} \pm 0.072 $           & & $\av{A_{FB}}$ & $ -0.164 \pm{0.024} $ & $-0.333\,{}^{+0.115}_{-0.130} \pm 0.012 $           & & $\av{A_{FB}}$ & $ -0.110 \pm{0.019} $ & $-0.158\,{}^{+0.080}_{-0.090} \pm 0.008 $            \\ 
$\av{S_3}$    & $ 0.007  \pm{0.003} $ & $-0.014\,{}^{+0.059}_{-0.060} \pm 0.008 $           & & $\av{S_3}$    & $ 0.005  \pm{0.002} $ & $\phantom{-}0.065\,{}^{+0.137}_{-0.127} \pm 0.007 $ & & $\av{S_3}$    & $ -0.003 \pm{0.002} $ & $\phantom{-}0.006\,{}^{+0.100}_{-0.100} \pm 0.007 $  \\ 
$\av{S_4}$    & $ 0.097  \pm{0.005} $ & $\phantom{-}0.039\,{}^{+0.091}_{-0.090} \pm 0.015 $ & & $\av{S_4}$    & $ 0.035  \pm{0.008} $ & $\phantom{-}0.127\,{}^{+0.190}_{-0.180} \pm 0.027 $ & & $\av{S_4}$    & $ -0.066 \pm{0.011} $ & $-0.339\,{}^{+0.115}_{-0.140} \pm 0.041 $            \\ 
$\av{S_5}$    & $ 0.241  \pm{0.010} $ & $\phantom{-}0.129\,{}^{+0.068}_{-0.066} \pm 0.011 $ & & $\av{S_5}$    & $ 0.148  \pm{0.014} $ & $\phantom{-}0.286\,{}^{+0.168}_{-0.172} \pm 0.009 $ & & $\av{S_5}$    & $ -0.037 \pm{0.015} $ & $\phantom{-}0.206\,{}^{+0.131}_{-0.115} \pm 0.009 $  \\ 
$\av{S_7}$    & $ 0.021  \pm{0.006} $ & $\phantom{-}0.038\,{}^{+0.063}_{-0.062} \pm 0.009 $ & & $\av{S_7}$    & $ 0.034  \pm{0.008} $ & $-0.293\,{}^{+0.180}_{-0.176} \pm 0.005 $           & & $\av{S_7}$    & $ 0.029  \pm{0.007} $ & $-0.252\,{}^{+0.127}_{-0.151} \pm 0.002 $            \\ 
$\av{S_8}$    & $ -0.004 \pm{0.002} $ & $\phantom{-}0.063\,{}^{+0.079}_{-0.080} \pm 0.009 $ & & $\av{S_8}$    & $ 0.011  \pm{0.004} $ & $-0.114\,{}^{+0.185}_{-0.196} \pm 0.006 $           & & $\av{S_8}$    & $ 0.011  \pm{0.004} $ & $-0.176\,{}^{+0.149}_{-0.165} \pm 0.006 $            \\ 
$\av{S_9}$    & $ -0.001 \pm{0.001} $ & $-0.113\,{}^{+0.061}_{-0.063} \pm 0.004 $           & & $\av{S_9}$    & $ -0.001 \pm{0.001} $ & $-0.110\,{}^{+0.140}_{-0.138} \pm 0.001 $           & & $\av{S_9}$    & $ -0.001 \pm{0.001} $ & $-0.000\,{}^{+0.100}_{-0.102} \pm 0.003 $            \\ 
%%%%%%% 
\rowcolor{white} & &    \\ [-12pt] %\hline
\midrule[.5pt]
%%%%%%
%\rowcolor{yellow}
\multicolumn{3}{c}{$q^2 \in [\,3.0\,,\,4.0\,]\,{\rm GeV}^2 $}              & & \multicolumn{3}{c}{$q^2 \in [\,4.0\,,\,5.0\,]\,{\rm GeV}^2 $ }                              & & \multicolumn{3}{c}{$q^2 \in [\,5.0\,,\,6.0\,]\,{\rm GeV}^2 $ }  \\  
%%%%%%
$\av{F_L}$    & $ 0.808  \pm{0.028} $ & $\phantom{-}0.873\,{}^{+0.154}_{-0.105} \pm 0.023 $ & & $\av{F_L}$    & $ 0.769  \pm{0.032} $ & $\phantom{-}0.899\,{}^{+0.106}_{-0.104} \pm 0.023 $ & & $\av{F_L}$    & $ 0.717  \pm{0.037} $ & $\phantom{-}0.644\,{}^{+0.130}_{-0.121} \pm 0.025 $  \\ 
$\av{A_{FB}}$ & $ -0.026 \pm{0.010} $ & $-0.041\,{}^{+0.091}_{-0.091} \pm 0.002 $           & & $\av{A_{FB}}$ & $ 0.056  \pm{0.013} $ & $\phantom{-}0.052\,{}^{+0.080}_{-0.080} \pm 0.004 $ & & $\av{A_{FB}}$ & $ 0.130  \pm{0.020} $ & $\phantom{-}0.057\,{}^{+0.094}_{-0.090} \pm 0.006 $  \\ 
$\av{S_3}$    & $ -0.012 \pm{0.004} $ & $\phantom{-}0.078\,{}^{+0.131}_{-0.122} \pm 0.008 $ & & $\av{S_3}$    & $ -0.022 \pm{0.006} $ & $\phantom{-}0.200\,{}^{+0.101}_{-0.097} \pm 0.007 $ & & $\av{S_3}$    & $ -0.031 \pm{0.008} $ & $-0.122\,{}^{+0.119}_{-0.126} \pm 0.009 $            \\ 
$\av{S_4}$    & $ -0.144 \pm{0.014} $ & $-0.046\,{}^{+0.193}_{-0.196} \pm 0.046 $           & & $\av{S_4}$    & $ -0.195 \pm{0.014} $ & $-0.148\,{}^{+0.154}_{-0.154} \pm 0.047 $           & & $\av{S_4}$    & $ -0.230 \pm{0.013} $ & $-0.273\,{}^{+0.174}_{-0.184} \pm 0.048 $            \\ 
$\av{S_5}$    & $ -0.182 \pm{0.018} $ & $-0.110\,{}^{+0.163}_{-0.169} \pm 0.004 $           & & $\av{S_5}$    & $ -0.278 \pm{0.018} $ & $-0.306\,{}^{+0.138}_{-0.141} \pm 0.004 $           & & $\av{S_5}$    & $ -0.340 \pm{0.017} $ & $-0.095\,{}^{+0.137}_{-0.142} \pm 0.004 $            \\ 
$\av{S_7}$    & $ 0.023  \pm{0.006} $ & $\phantom{-}0.171\,{}^{+0.175}_{-0.158} \pm 0.002 $ & & $\av{S_7}$    & $ 0.019  \pm{0.005} $ & $-0.082\,{}^{+0.129}_{-0.128} \pm 0.001 $           & & $\av{S_7}$    & $ 0.014  \pm{0.004} $ & $\phantom{-}0.038\,{}^{+0.135}_{-0.135} \pm 0.002 $  \\ 
$\av{S_8}$    & $ 0.009  \pm{0.003} $ & $\phantom{-}0.097\,{}^{+0.189}_{-0.184} \pm 0.002 $ & & $\av{S_8}$    & $ 0.008  \pm{0.002} $ & $\phantom{-}0.107\,{}^{+0.144}_{-0.146} \pm 0.003 $ & & $\av{S_8}$    & $ 0.007  \pm{0.002} $ & $-0.037\,{}^{+0.160}_{-0.159} \pm 0.003 $            \\ 
$\av{S_9}$    & $ -0.001 \pm{0.001} $ & $-0.203\,{}^{+0.112}_{-0.132} \pm 0.002 $           & & $\av{S_9}$    & $ -0.001 \pm{0.001} $ & $\phantom{-}0.181\,{}^{+0.105}_{-0.099} \pm 0.001 $ & & $\av{S_9}$    & $ -0.001 \pm{0.001} $ & $-0.080\,{}^{+0.117}_{-0.120} \pm 0.001 $            \\ 
%%%%%%% 
\rowcolor{white} & &    \\ [-12pt] %\hline
\midrule[.5pt]
%%%%%%
%\rowcolor{yellow}
\multicolumn{3}{c}{$q^2 \in [\,6.0\,,\,7.0\,]\,{\rm GeV}^2 $}              & & \multicolumn{3}{c}{$q^2 \in [\,7.0\,,\,8.0\,]\,{\rm GeV}^2 $ }  & & \multicolumn{3}{c}{$q^2 \in [\,15.0\,,\,16.0\,]\,{\rm GeV}^2 $ }  \\  
%%%%%%
$\av{F_L}$    & $ 0.658  \pm{0.040} $ & $\phantom{-}0.644\,{}^{+0.089}_{-0.084} \pm 0.025 $ & & $\av{F_L}$    & $ 0.598  \pm{0.041} $ & $\phantom{-}0.609\,{}^{+0.103}_{-0.082} \pm 0.025 $ & & $\av{F_L}$    & $ 0.345  \pm{0.039} $ & $\phantom{-}0.385\,{}^{+0.067}_{-0.066} \pm 0.013 $  \\ 
$\av{A_{FB}}$ & $ 0.198  \pm{0.025} $ & $\phantom{-}0.058\,{}^{+0.064}_{-0.063} \pm 0.009 $ & & $\av{A_{FB}}$ & $ 0.260  \pm{0.027} $ & $\phantom{-}0.241\,{}^{+0.080}_{-0.062} \pm 0.012 $ & & $\av{A_{FB}}$ & $ 0.418  \pm{0.026} $ & $\phantom{-}0.396\,{}^{+0.068}_{-0.047} \pm 0.009 $  \\ 
$\av{S_3}$    & $ -0.040 \pm{0.010} $ & $-0.069\,{}^{+0.089}_{-0.091} \pm 0.004 $           & & $\av{S_3}$    & $ -0.050 \pm{0.011} $ & $-0.054\,{}^{+0.097}_{-0.099} \pm 0.005 $           & & $\av{S_3}$    & $ -0.167 \pm{0.024} $ & $-0.060\,{}^{+0.085}_{-0.088} \pm 0.006 $            \\ 
$\av{S_4}$    & $ -0.253 \pm{0.011} $ & $-0.311\,{}^{+0.111}_{-0.118} \pm 0.052 $           & & $\av{S_4}$    & $ -0.269 \pm{0.008} $ & $-0.236\,{}^{+0.116}_{-0.136} \pm 0.058 $           & & $\av{S_4}$    & $ -0.292 \pm{0.008} $ & $-0.321\,{}^{+0.082}_{-0.099} \pm 0.007 $            \\ 
$\av{S_5}$    & $ -0.380 \pm{0.014} $ & $-0.339\,{}^{+0.108}_{-0.114} \pm 0.008 $           & & $\av{S_5}$    & $ -0.404 \pm{0.012} $ & $-0.386\,{}^{+0.105}_{-0.135} \pm 0.007 $           & & $\av{S_5}$    & $ -0.330 \pm{0.024} $ & $-0.360\,{}^{+0.074}_{-0.092} \pm 0.006 $            \\ 
$\av{S_7}$    & $ 0.010  \pm{0.003} $ & $\phantom{-}0.009\,{}^{+0.123}_{-0.124} \pm 0.004 $ & & $\av{S_7}$    & $ 0.007  \pm{0.006} $ & $-0.094\,{}^{+0.123}_{-0.130} \pm 0.003 $           & & $\av{S_7}$    & $ 0.000  \pm{0.033} $ & $\phantom{-}0.040\,{}^{+0.092}_{-0.089} \pm 0.002 $  \\ 
$\av{S_8}$    & $ 0.005  \pm{0.002} $ & $\phantom{-}0.080\,{}^{+0.131}_{-0.129} \pm 0.002 $ & & $\av{S_8}$    & $ 0.004  \pm{0.003} $ & $-0.295\,{}^{+0.119}_{-0.139} \pm 0.002 $           & & $\av{S_8}$    & $ 0.000  \pm{0.010} $ & $-0.057\,{}^{+0.093}_{-0.095} \pm 0.005 $            \\ 
$\av{S_9}$    & $ -0.001 \pm{0.001} $ & $\phantom{-}0.061\,{}^{+0.091}_{-0.091} \pm 0.001 $ & & $\av{S_9}$    & $ -0.001 \pm{0.002} $ & $\phantom{-}0.030\,{}^{+0.100}_{-0.098} \pm 0.001 $ & & $\av{S_9}$    & $ 0.000  \pm{0.016} $ & $-0.054\,{}^{+0.083}_{-0.087} \pm 0.005 $            \\ 
%%%%%%% 
\rowcolor{white} & &    \\ [-12pt] %\hline
\midrule[.5pt]
%%%%%%
%\rowcolor{yellow}
\multicolumn{3}{c}{$q^2 \in [\,16.0\,,\,17.0\,]\,{\rm GeV}^2 $}            & & \multicolumn{3}{c}{$q^2 \in [\,17.0\,,\,18.0\,]\,{\rm GeV}^2 $ }  & & \multicolumn{3}{c}{$q^2 \in [\,18.0\,,\,19.0\,]\,{\rm GeV}^2 $ }  \\  
%%%%%%
$\av{F_L}$    & $ 0.333  \pm{0.040} $ & $\phantom{-}0.295\,{}^{+0.058}_{-0.062} \pm 0.013 $ & & $\av{F_L}$    & $ 0.324  \pm{0.041} $ & $\phantom{-}0.363\,{}^{+0.073}_{-0.072} \pm 0.017 $ & & $\av{F_L}$    & $ 0.319  \pm{0.044} $ & $\phantom{-}0.421\,{}^{+0.100}_{-0.100} \pm 0.013 $  \\ 
$\av{A_{FB}}$ & $ 0.399  \pm{0.026} $ & $\phantom{-}0.451\,{}^{+0.071}_{-0.048} \pm 0.007 $ & & $\av{A_{FB}}$ & $ 0.358  \pm{0.025} $ & $\phantom{-}0.274\,{}^{+0.069}_{-0.061} \pm 0.008 $ & & $\av{A_{FB}}$ & $ 0.267  \pm{0.022} $ & $\phantom{-}0.354\,{}^{+0.111}_{-0.099} \pm 0.012 $  \\ 
$\av{S_3}$    & $ -0.197 \pm{0.024} $ & $-0.250\,{}^{+0.079}_{-0.092} \pm 0.007 $           & & $\av{S_3}$    & $ -0.236 \pm{0.024} $ & $-0.099\,{}^{+0.091}_{-0.092} \pm 0.011 $           & & $\av{S_3}$    & $ -0.287 \pm{0.024} $ & $-0.131\,{}^{+0.128}_{-0.130} \pm 0.012 $            \\ 
$\av{S_4}$    & $ -0.297 \pm{0.009} $ & $-0.246\,{}^{+0.083}_{-0.096} \pm 0.029 $           & & $\av{S_4}$    & $ -0.305 \pm{0.009} $ & $-0.229\,{}^{+0.090}_{-0.096} \pm 0.045 $           & & $\av{S_4}$    & $ -0.316 \pm{0.011} $ & $-0.607\,{}^{+0.153}_{-0.170} \pm 0.059 $            \\ 
$\av{S_5}$    & $ -0.299 \pm{0.024} $ & $-0.254\,{}^{+0.069}_{-0.081} \pm 0.010 $           & & $\av{S_5}$    & $ -0.254 \pm{0.024} $ & $-0.305\,{}^{+0.081}_{-0.088} \pm 0.015 $           & & $\av{S_5}$    & $ -0.180 \pm{0.020} $ & $-0.534\,{}^{+0.131}_{-0.150} \pm 0.015 $            \\ 
$\av{S_7}$    & $ 0.000  \pm{0.034} $ & $\phantom{-}0.144\,{}^{+0.091}_{-0.085} \pm 0.005 $ & & $\av{S_7}$    & $ 0.000  \pm{0.035} $ & $\phantom{-}0.022\,{}^{+0.094}_{-0.093} \pm 0.011 $ & & $\av{S_7}$    & $ 0.000  \pm{0.036} $ & $\phantom{-}0.058\,{}^{+0.123}_{-0.124} \pm 0.006 $  \\ 
$\av{S_8}$    & $ 0.000  \pm{0.009} $ & $\phantom{-}0.055\,{}^{+0.090}_{-0.088} \pm 0.005 $ & & $\av{S_8}$    & $ 0.000  \pm{0.007} $ & $-0.007\,{}^{+0.098}_{-0.098} \pm 0.001 $           & & $\av{S_8}$    & $ 0.000  \pm{0.005} $ & $\phantom{-}0.149\,{}^{+0.139}_{-0.138} \pm 0.010 $  \\ 
$\av{S_9}$    & $ 0.000  \pm{0.016} $ & $-0.014\,{}^{+0.084}_{-0.086} \pm 0.004 $           & & $\av{S_9}$    & $ 0.000  \pm{0.014} $ & $-0.090\,{}^{+0.092}_{-0.095} \pm 0.002 $           & & $\av{S_9}$    & $ 0.000  \pm{0.010} $ & $-0.079\,{}^{+0.122}_{-0.121} \pm 0.007 $            \\ 
%%%%%%% 
\end{tabular} }
\caption{The SM predictions and experimental values of the $B\to K^* \mu^+ \mu^-$ angular observables, evaluated 
by the method of moments. 
The experimental values~\cite{Aaij:2015oid}, are (from left to right) statistical and systematic.
\label{tab:BtoKstarMoment}}
\end{center} 
\end{table}

%%%%%%%%Bs->phi mu mu%%%%%%%%%%%%%%%
\begin{table}
\ra{0.90}
\rb{1.3mm}
\begin{center}
\setlength\extrarowheight{2pt}
\rowcolors{1}{}{light-gray}
\footnotesize{\begin{tabular}{lrcc||lrc}
\toprule[.6pt]
\multicolumn{7}{l}{$B_s\to \phi \,\mu^+\mu^-$ angular observables}\\
\hline
\toprule[.2pt]
\rowcolor{white}Observable & \cen{SM prediction} & Measurement        &  & Observable & \cen{SM prediction} & Measurement \\ 
\hline \hline
%
%%%%%%
%\rowcolor{yellow}
\multicolumn{3}{c}{$q^2 \in [\,0.1\,,\,2.0\,]\,{\rm GeV}^2 $}         & & \multicolumn{3}{c}{$q^2 \in [\,2.0\,,\,5.0\,]\,{\rm GeV}^2 $ }    \\  
%%%%%%
$\av{F_L}$    & $ 0.399 \pm{0.035} $ & $\phantom{-} 0.20 ^{+0.08}_{-0.09}\pm 0.02 $      & & $\av{F_L}$    & $ 0.796  \pm{0.021} $ & $\phantom{-} 0.68 ^{+ 0.16 }_{ -0.13 } \pm 0.03 $  \\ 
$\av{S_3}$    & $ 0.007 \pm{0.004} $ & $ -0.05 ^{+ 0.13 }_{ -0.13 } \pm 0.01 $           & & $\av{S_3}$    & $ -0.014 \pm{0.005} $ & $ -0.06 ^{+ 0.19 }_{ -0.23 } \pm 0.01 $            \\ 
$\av{S_4}$    & $ 0.080 \pm{0.004} $ & $\phantom{-} 0.27 ^{+ 0.28 }_{ -0.18 } \pm 0.01 $ & & $\av{S_4}$    & $ -0.139 \pm{0.012} $ & $ -0.47 ^{+ 0.30 }_{ -0.44 } \pm 0.01 $            \\ 
$\av{S_7}$    & $ 0.029 \pm{0.007} $ & $\phantom{-} 0.04^{+ 0.12 }_{ -0.12 } \pm 0.00 $  & & $\av{S_7}$    & $ 0.026  \pm{0.007} $ & $-0.03^{+ 0.18 }_{ -0.23 } \pm 0.01 $              \\ 
%%%%%%% 
\rowcolor{white} & &    \\ [-12pt] %\hline
\midrule[.5pt]
%%%%%%
%\rowcolor{yellow}
\multicolumn{3}{c}{$q^2 \in [\,5.0\,,\,8.0\,]\,{\rm GeV}^2 $ }          & & \multicolumn{3}{c}{$q^2 \in [\,15.0\,,\,17.0\,]\,{\rm GeV}^2 $ }   \\  
%%%%%%
$\av{F_L}$    & $ 0.656  \pm{0.046} $ & $\phantom{-} 0.54 ^{+ 0.10 }_{ -0.09 } \pm 0.02 $  & & $\av{F_L}$    & $ 0.337  \pm{0.037} $ & $\phantom{-} 0.23 ^{+ 0.09 }_{ -0.08 } \pm 0.02 $   \\ 
$\av{S_3}$    & $ -0.047 \pm{0.031} $ & $ -0.10 ^{+ 0.20 }_{ -0.29 } \pm 0.01 $            & & $\av{S_3}$    & $ -0.199 \pm{0.031} $ & $ -0.06 ^{+ 0.16 }_{ -0.19 } \pm 0.01 $             \\ 
$\av{S_4}$    & $ -0.255 \pm{0.016} $ & $ -0.10 ^{+ 0.15 }_{ -0.18 } \pm 0.01 $            & & $\av{S_4}$    & $ -0.299 \pm{0.013} $ & $ -0.03 ^{+ 0.15 }_{ -0.15 } \pm 0.01 $             \\ 
$\av{S_7}$    & $ 0.012  \pm{0.004} $ & $\phantom{-} 0.04^{+ 0.16 }_{ -0.20 } \pm 0.01 $   & & $\av{S_7}$    & $ 0.000  \pm{0.023} $ & $\phantom{-} 0. 12^{+ 0.16 }_{ -0.13 } \pm 0.01 $   \\ 
%%%%%
\rowcolor{white} & &   \\ [-12pt] %\hline
\midrule[.5pt]
%%%%%%
%\rowcolor{yellow}
\multicolumn{3}{c}{$q^2 \in [\,17.0\,,\,19.0\,]\,{\rm GeV}^2 $} \\ 
%%%%%%
$\av{F_L}$    & $ 0.321  \pm{0.037} $ & $\phantom{-} 0.40 ^{+ 0.13 }_{ -0.15 } \pm 0.02 $  \\ 
$\av{S_3}$    & $ -0.274 \pm{0.023} $ & $ -0.07 ^{+ 0.23 }_{ -0.27 } \pm 0.02 $            \\ 
$\av{S_4}$    & $ -0.313 \pm{0.008} $ & $ -0.39 ^{+ 0.25 }_{ -0.34 } \pm 0.02 $            \\ 
$\av{S_7}$    & $ 0.000  \pm{0.034} $ & $\phantom{-} 0.20^{+ 0.29 }_{ -0.22 } \pm 0.01 $   \\ 
%%%%%
\end{tabular} }
\caption{SM predictions and experimental values of the $B_s\to \phi\; \mu^+\mu^-$ observables.
The uncertainties of the experimental values~\cite{Aaij:2015esa} are (from left to right) statistical and systematic. 
\label{tab:Bstophi}}
\end{center} 
\end{table}  
%%%%%%%%dBR/dq2 of Bs->Phi mu mu%%%%%%%%%%%%%%%
\begin{table}
\ra{0.90}
\rb{1.3mm}
\begin{center}
\rowcolors{2}{light-gray}{}
\footnotesize{\begin{tabular}[b]{l||l|l}
\toprule[.6pt]
\multicolumn{3}{l}{$B_s\to \phi  \mu^+\mu^-$ differential branching ratio} \\
\hline
\toprule[.2pt]
\rowcolor{white}Bin (GeV$^2$) & SM prediction & \cen{Measurement} \\  
\hline \hline
%
%%%%%%
%\rowcolor{yellow}
\multicolumn{3}{c}{$10^8\times \av{dBR/dq^2}(B_s \to \phi \mu^+ \mu^-)$} \\
%%%%%%
$[0.1-2.0]$    & $ 8.410 \pm{1.130}$  & $5.85 ^{+0.73}_{-0.69} \pm{0.14} \pm{0.44}$ \\ 
$[2.0-5.0]$    & $ 4.452 \pm{0.553}$  & $2.56 ^{+0.42}_{-0.39} \pm{0.06} \pm{0.19}$ \\ 
$[5.0-8.0]$    & $ 5.211 \pm{0.689}$  & $3.21 ^{+0.44}_{-0.42} \pm{0.08} \pm{0.24}$ \\ 
% $[11.0-12.5]$  & $ 6.461 \pm{0.774}$  & $4.71 ^{+0.69}_{-0.65} \pm{0.15} \pm{0.36}$ \\ 
$[15.0-17.0]$  & $ 5.935 \pm{0.692}$  & $4.52 ^{+0.57}_{-0.54} \pm{0.12} \pm{0.34}$ \\ 
$[17.0-19.0]$  & $ 3.722 \pm{0.467}$  & $3.96 ^{+0.57}_{-0.54} \pm{0.14} \pm{0.30}$ 
\end{tabular}}
\caption{SM predictions and experimental values for the differential branching ratio of the $B_s\to \phi  \mu^+\mu^-$ decay.
The experimental error of the $B_s\to \phi  \mu^+\mu^-$ 
decay~\cite{Aaij:2015esa} are (from left to right) statistical and systematic and the uncertainty on the
branching fraction of the normalisation mode $B_s^0 \to J/\psi\phi$.
\label{tab:BRBstophi} }
\end{center} 
\end{table}
%%%%%%%%%%

%%%%%%%%B->K mu mu comparison%%%%%%%%%%%%%%%
\begin{table}
\ra{0.90}
\rb{1.3mm}
\begin{center}
\rowcolors{2}{light-gray}{}
\footnotesize{\begin{tabular}[b]{l||c|c}
\toprule[.6pt]
\multicolumn{3}{l}{$B\to K  \mu^+\mu^-$  differential branching ratio}\\
\hline
\toprule[.2pt]
%\rowcolor{white}Observable & \cen{Soft FF (10\%)} &  \cen{Soft FF (20\%)} & \cen{Full FF (5\%)} & \cen{Full FF (10\%)} \\ 
\rowcolor{white}Bin (GeV$^2$) & SM prediction (SuperIso) & SM prediction (Khodjamirian et al.) \\  
\hline \hline
%
%%%%%%
%\rowcolor{yellow}
\multicolumn{3}{c}{$10^7\times \av{BR}(B^0\to K^0 \mu^+ \mu^-)$} \\
%%%%%%
$[0.05-2.00]$   &  $ 0.52 \pm{0.11}$ & $0.71 ^{+0.22}_{-0.08} $ \\ 
$[2.00-4.30]$   &  $ 0.62 \pm{0.14}$ & $0.80 ^{+0.27}_{-0.11} $ \\ 
$[4.30-8.68]$   &  $ 1.23 \pm{0.34}$ & $1.39 ^{+0.53}_{-0.22} $ \\ 
$[1.00-6.00]$   &  $ 1.36 \pm{0.32}$ & $1.76 ^{+0.60}_{-0.23} $ 
%%%%%%% 
\end{tabular}}
\caption{Comparison of the SM predictions for  BR($B^0\to K^0 \mu^+ \mu^-$) from SuperIso using the full FF approach and assuming 10\% power correction with the result of Table 4 of Ref.~\cite{Khodjamirian:2012rm}.
}
\end{center} 
\end{table}

\clearpage

\section{Form factors}\label{app:FF}
The methods used for obtaining form factor results depend on the recoil energy of the outgoing light meson.
At high-$q^2$ which corresponds to the low recoil region, the $B \to K^*$ and $B_s \to \phi$ form factor results are available from unquenched lattice calculations~\cite{Horgan:2013hoa,Horgan:2015vla},
while for the low-$q^2$ region the form factors can be taken from LCSR
calculations which are available  from Ref.\cite{Khodjamirian:2010vf} (KMPW) as well as from the Ref.\cite{Straub:2015ica} (BSZ).
The theoretical errors associated with the KMPW form factors are larger than the ones from BSZ.
The larger error of the KMPW form factor is mostly by construction and due to the different 
choices of distribution amplitudes where they employ the 
$B$-meson distribution for which less information is available compared to the $K^*$ meson which has been used for the BSZ form factors.
Moreover, for the BSZ form factors, interpolation with lattice data and additional correlations between LCSR intrinsic parameters have been used which also have some effect in reducing the theoretical uncertainty. 

The form factor uncertainties are correlated through
hadronic inputs as well as kinematic relations at the endpoint $q^2$=0. As claimed in Ref.\cite{Straub:2015ica} there are additional correlations due to intrinsic LCSR parameters.
While in the soft FF approach due the relations among form factors at high recoil energy 
the number of independent form factors reduces from seven to two form factors, and most of the latter correlations are included by construction, in the full FF approaches analogous implications  can be derived directly within the  LCSR results via these additional correlations mentioned above.

Unfortunately the form factor correlations have not been given for the KMPW results, however, they have been 
provided in Ref.\cite{Straub:2015ica} for the BSZ form factors.

\begin{figure}[!t]
\begin{center}
\includegraphics[width=6.cm]{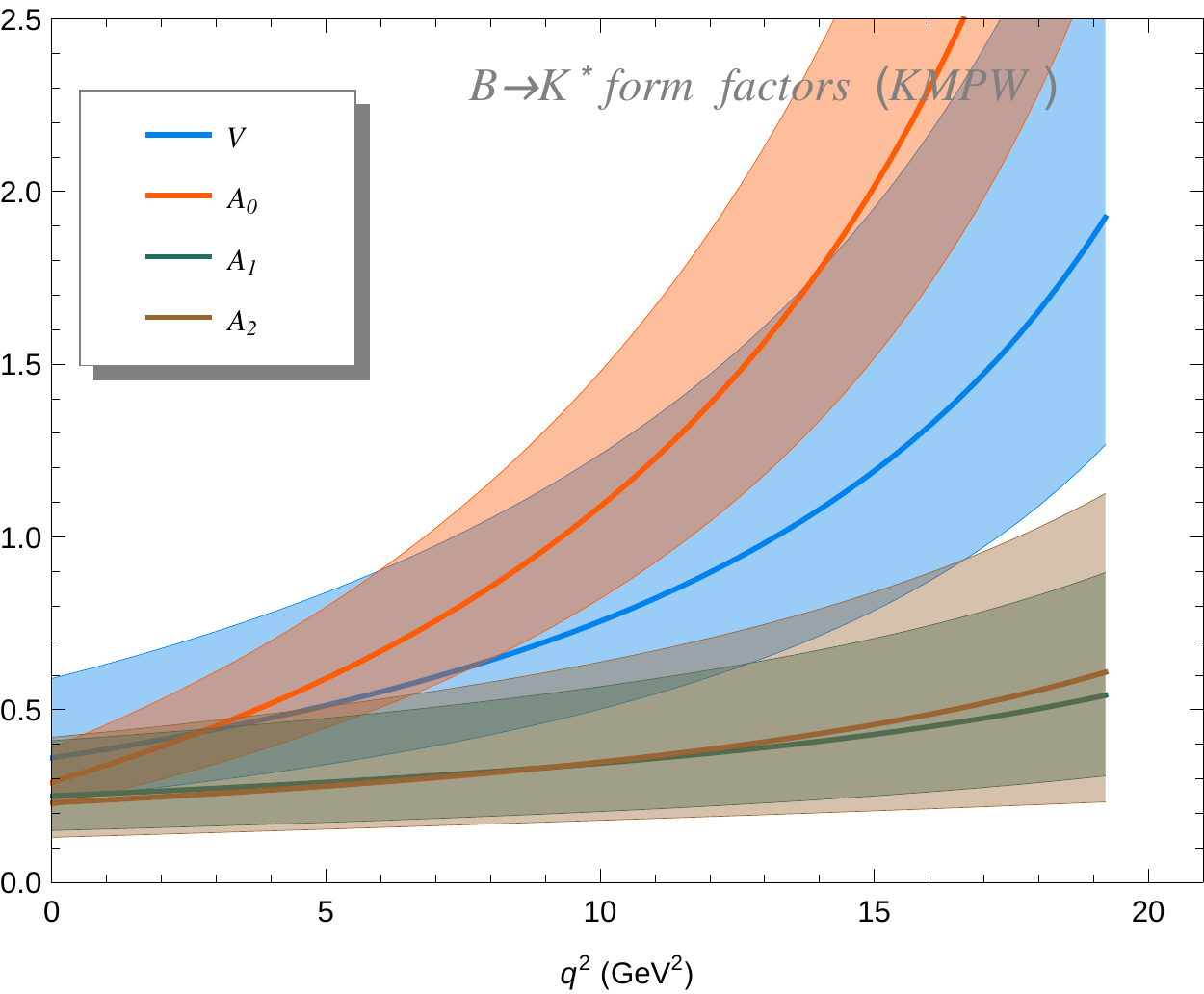}\qquad\includegraphics[width=6.cm]{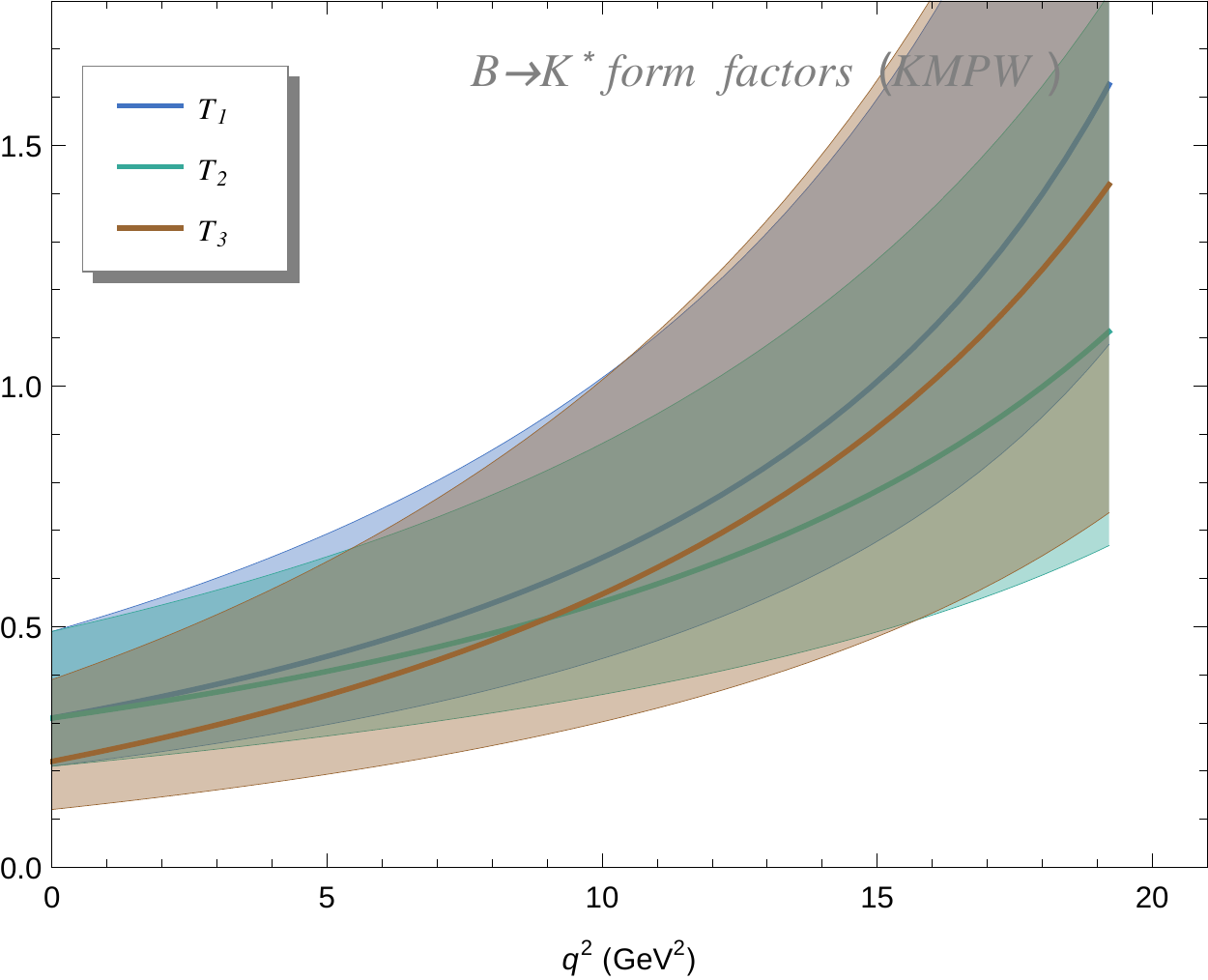}\\[5.mm]
\includegraphics[width=6.cm]{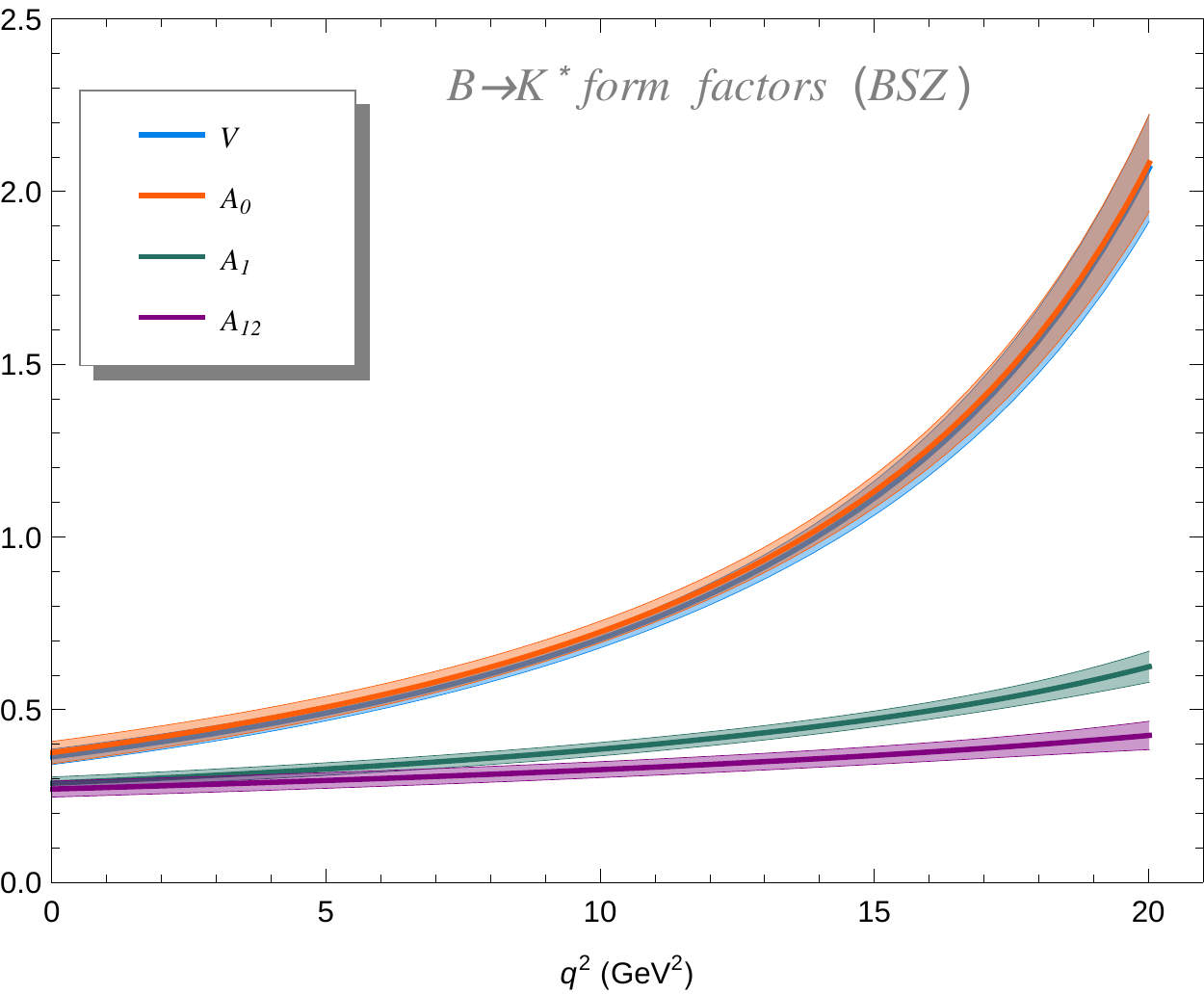}\qquad\includegraphics[width=6.cm]{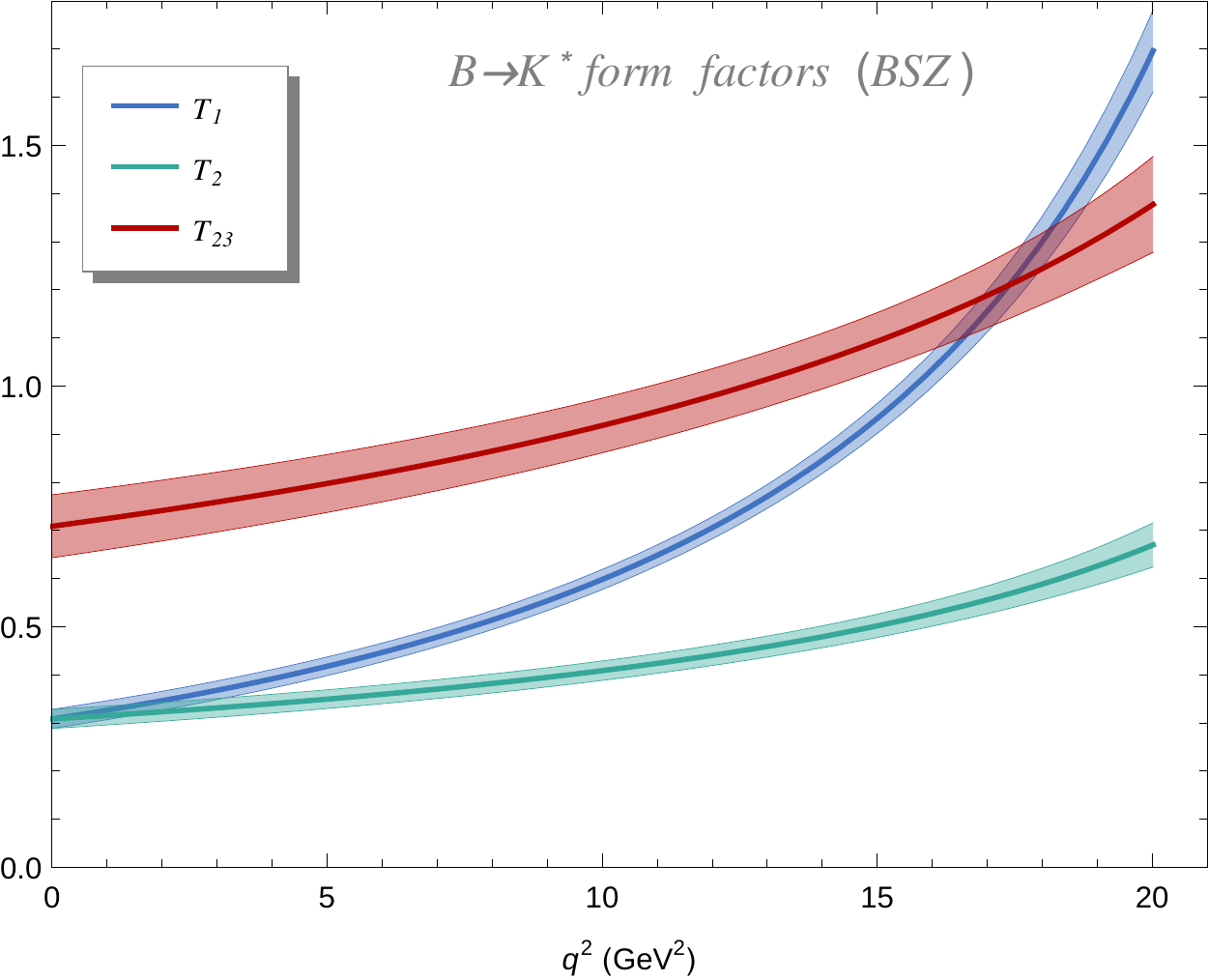}
\caption{LCSR results for the $B \to K^*$ form factors. In the upper row the KMPW results~\cite{Khodjamirian:2010vf} are shown
while in the lower row the BSZ results~\cite{Straub:2015ica} are presented. The relation among $T_{23}$ and $T_{2,3}$ as well as the
relation between $A_{12}$ and $A_{1,2}$ are defined in Ref.~\cite{Horgan:2013hoa}.}
\label{fig:FFcomparison}
\end{center}
\end{figure}

The LCSR results for the seven independent $B\to K^*$ form factors including 
their theoretical uncertainties are shown in Fig.~\ref{fig:FFcomparison}, 
where the KMPW form factors which are applicable only at low $q^2$ have been extrapolated to high-$q^2$ as well.
For the BSZ form factors we have presented the fit results of Ref.~\cite{Straub:2015ica} which are applicable for both the low- and high-$q^2$ regions.

\section{SM predictions of $B\to K^* \mu^+ \mu^-$ observables with different theoretical approaches
and form factors}\label{app:SoftFullFF}

\begin{figure}[!t]
\centering
\includegraphics[width=0.49\textwidth]{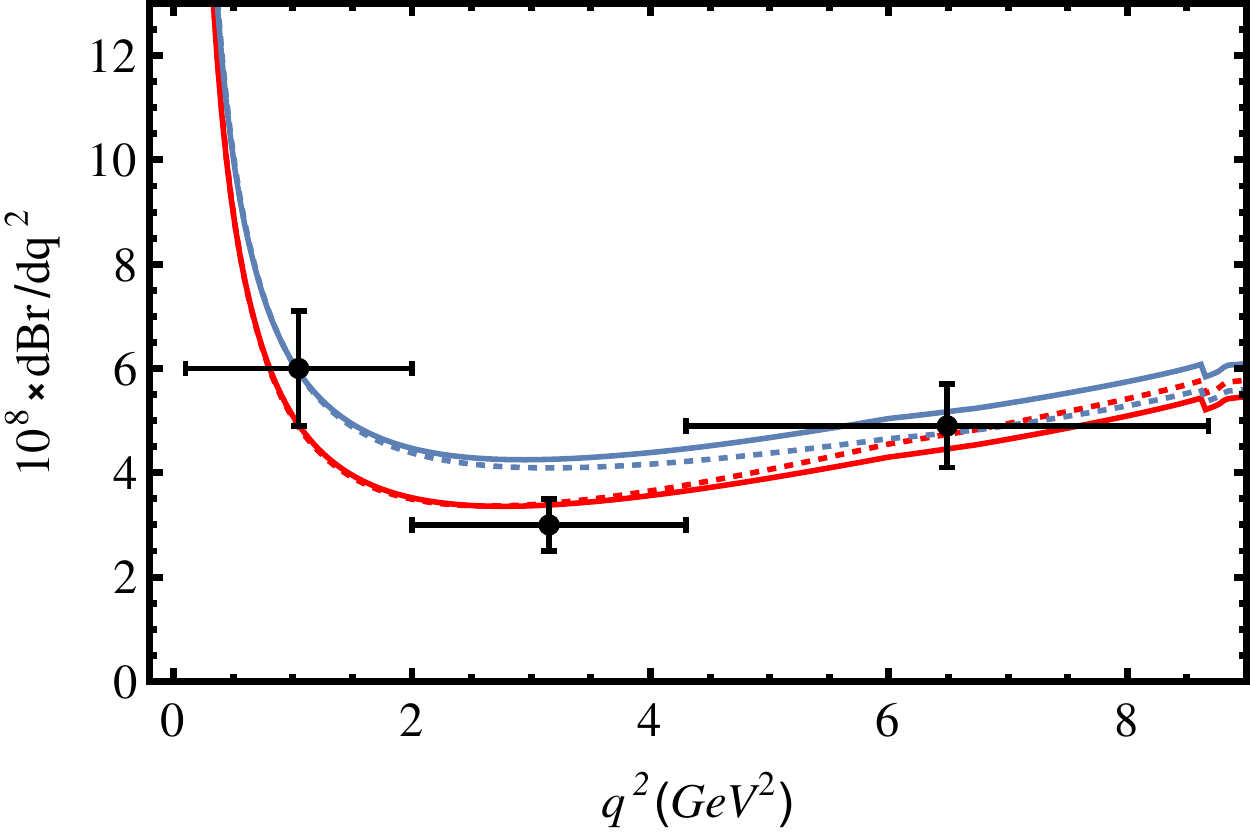}
\includegraphics[width=0.49\textwidth]{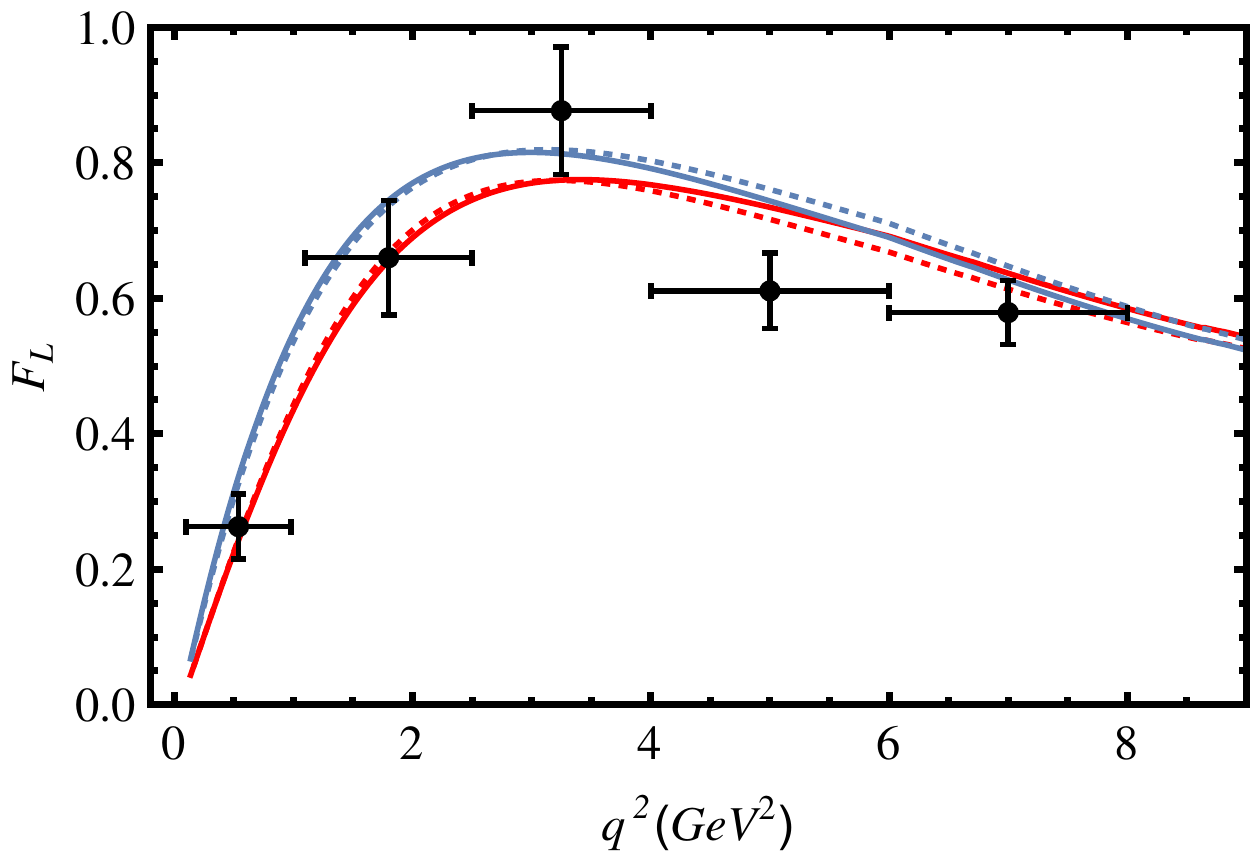}
\includegraphics[width=0.49\textwidth]{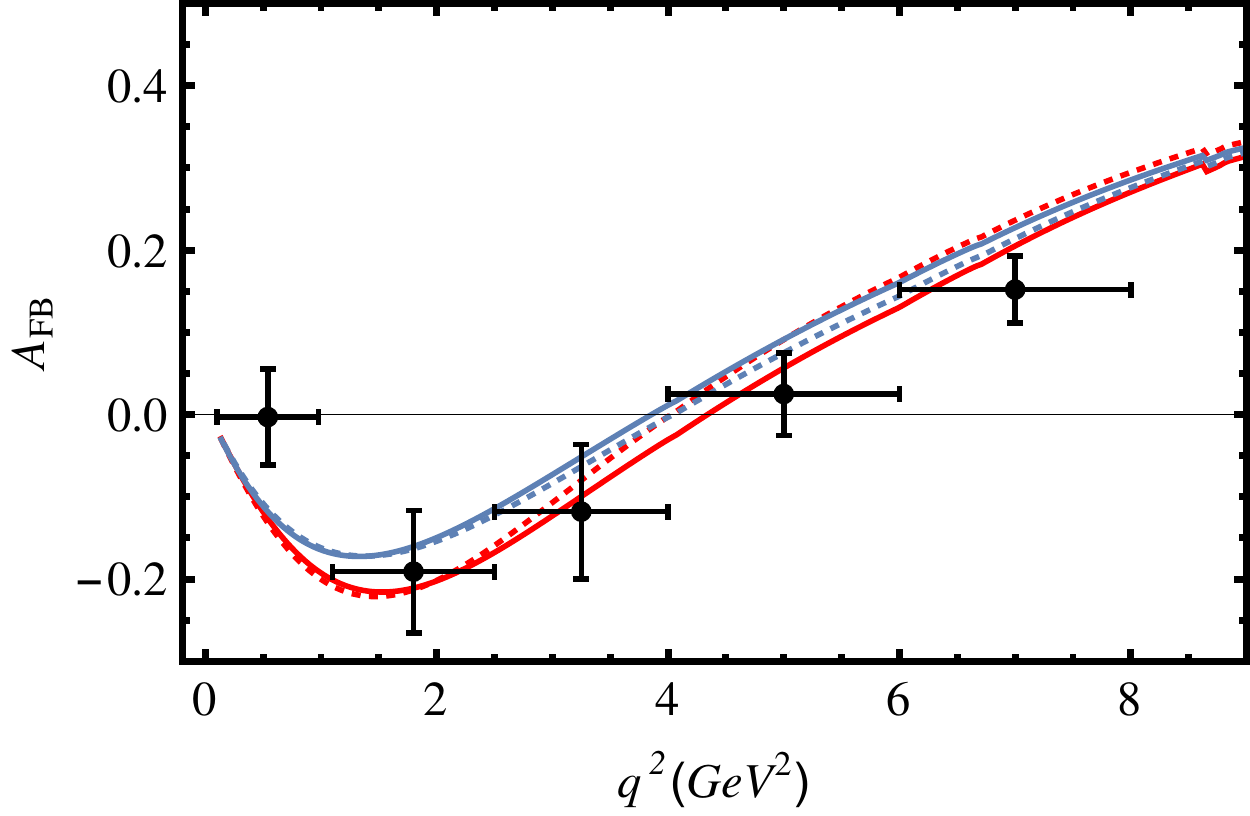}
\includegraphics[width=0.49\textwidth]{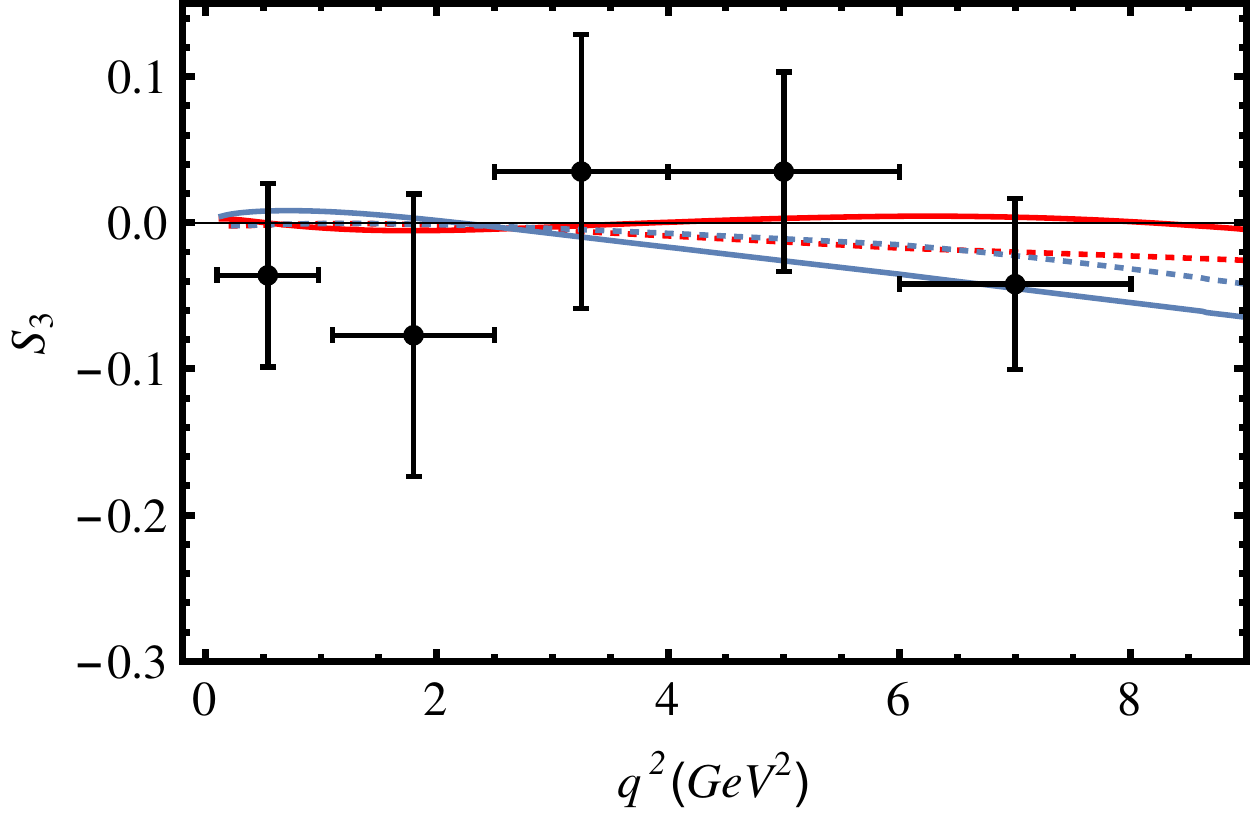}
\includegraphics[width=0.49\textwidth]{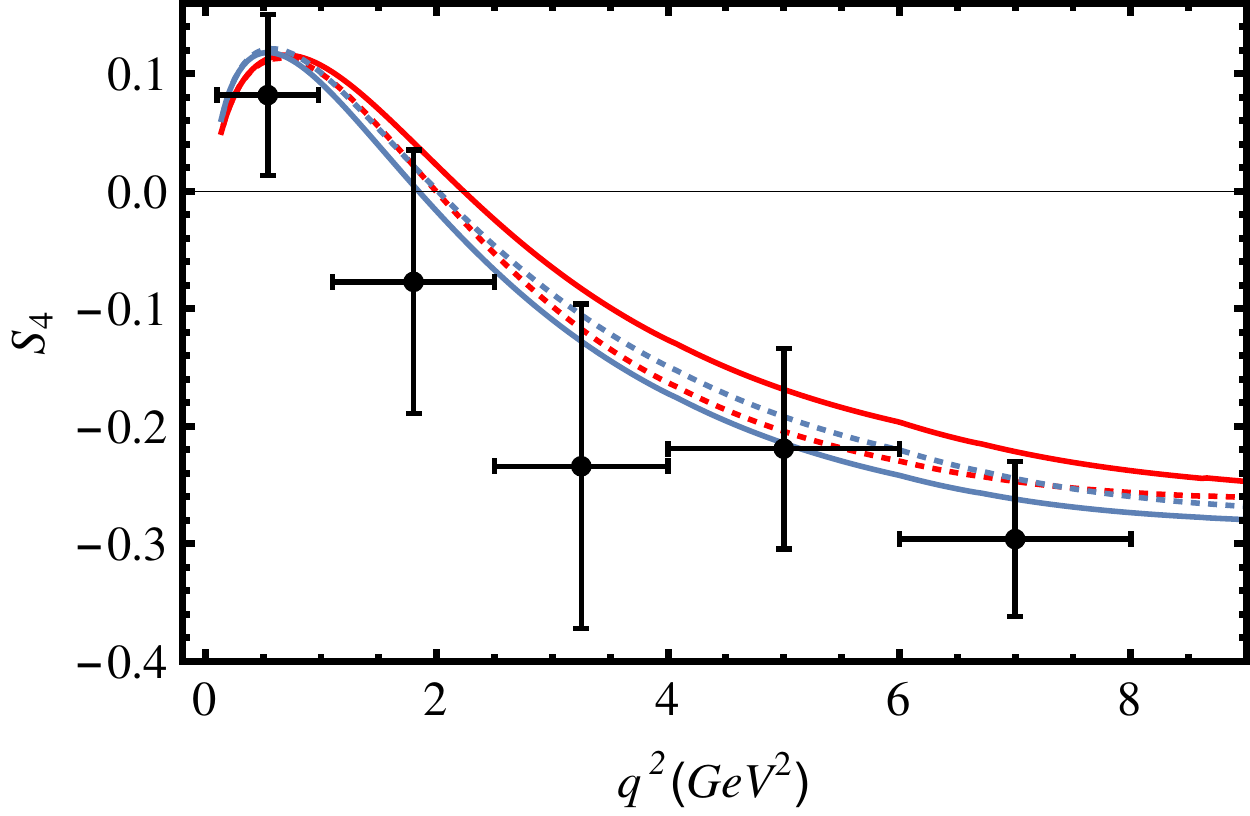}
\includegraphics[width=0.49\textwidth]{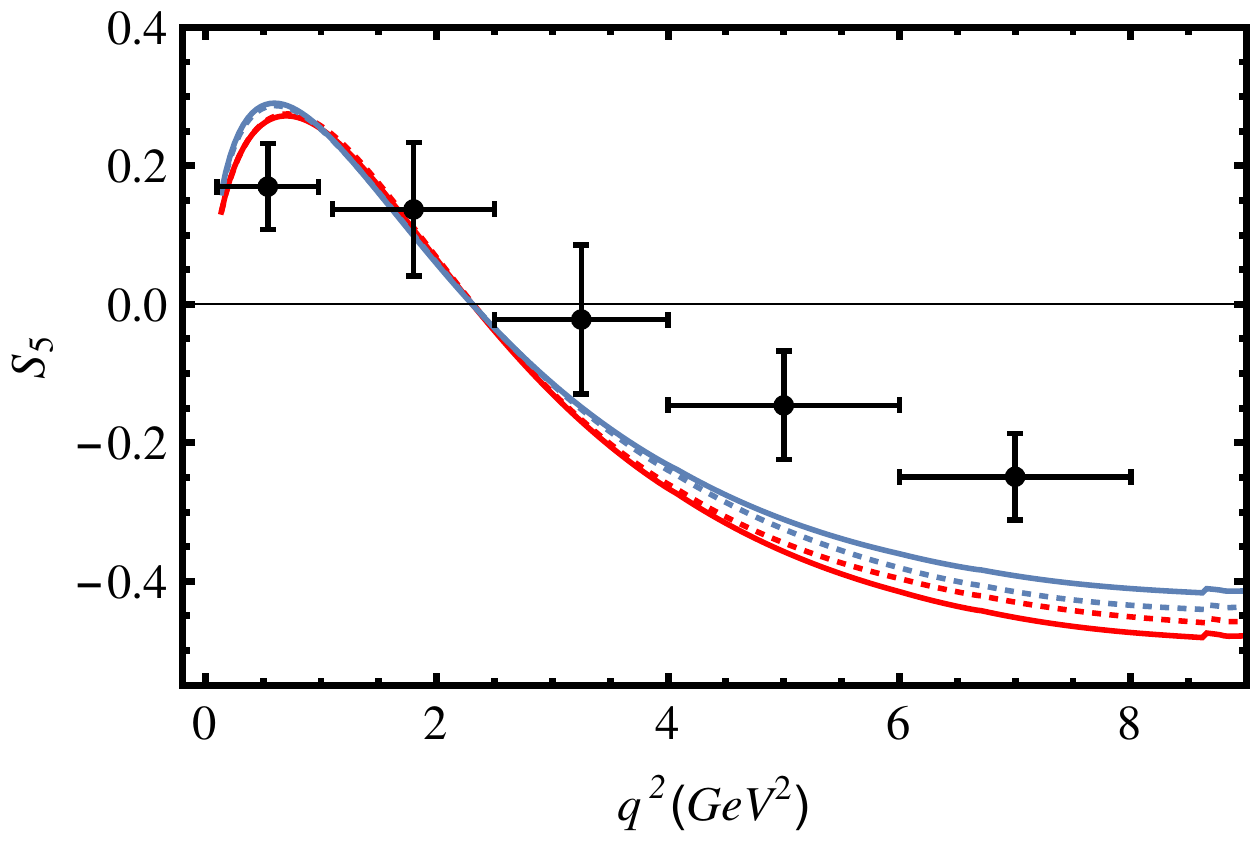}
\caption{Comparison of the SM predictions for the central values 
of $B\to K^* \mu^+ \mu^-$ observables: $10^8\times dBR/dq^2, F_L, A_{FB}$ and $S_{3,4,5}$
within different approaches and using different form factors. The blue crosses represent the LHCb measurements~\cite{Aaij:2015oid}.
The  solid and dotted blue lines correspond to SM predictions using BSZ form factors 
within the full form factor and soft form factor approaches, respectively. 
The solid and dotted red lines correspond to SM predictions using KMPW form factors 
within the full form factor and soft form factor approaches, respectively. 
\label{fig:Comparison1}}
\end{figure}

In order to compare with the results of Ref.~\cite{Descotes-Genon:2015uva}, we reproduced the SM predictions
using KMPW form factors within the soft FF approach and we also added the fit results of Ref.~\cite{Descotes-Genon:2014uoa}
which are meant to take into account the missing $1/m_b$ factorisable corrections.
These predictions coincided nicely with central values quoted in Ref.~\cite{Descotes-Genon:2015uva} up to deviations in 
observables which are very small or in the bins where there is a zero-crossing (the difference being mostly due to slight
disagreements in the choice of SM Wilson coefficient values).
A similar comparison was done with the results of Ref.~\cite{Straub:2015ica}, when using the BSZ form factors in 
the full FF approach. Again the results were in good agreement, and only in the bins with zero crossings and 
for observables having very small values, there were some slight  differences.
 
\begin{figure}[!t]
\centering
\includegraphics[width=0.49\textwidth]{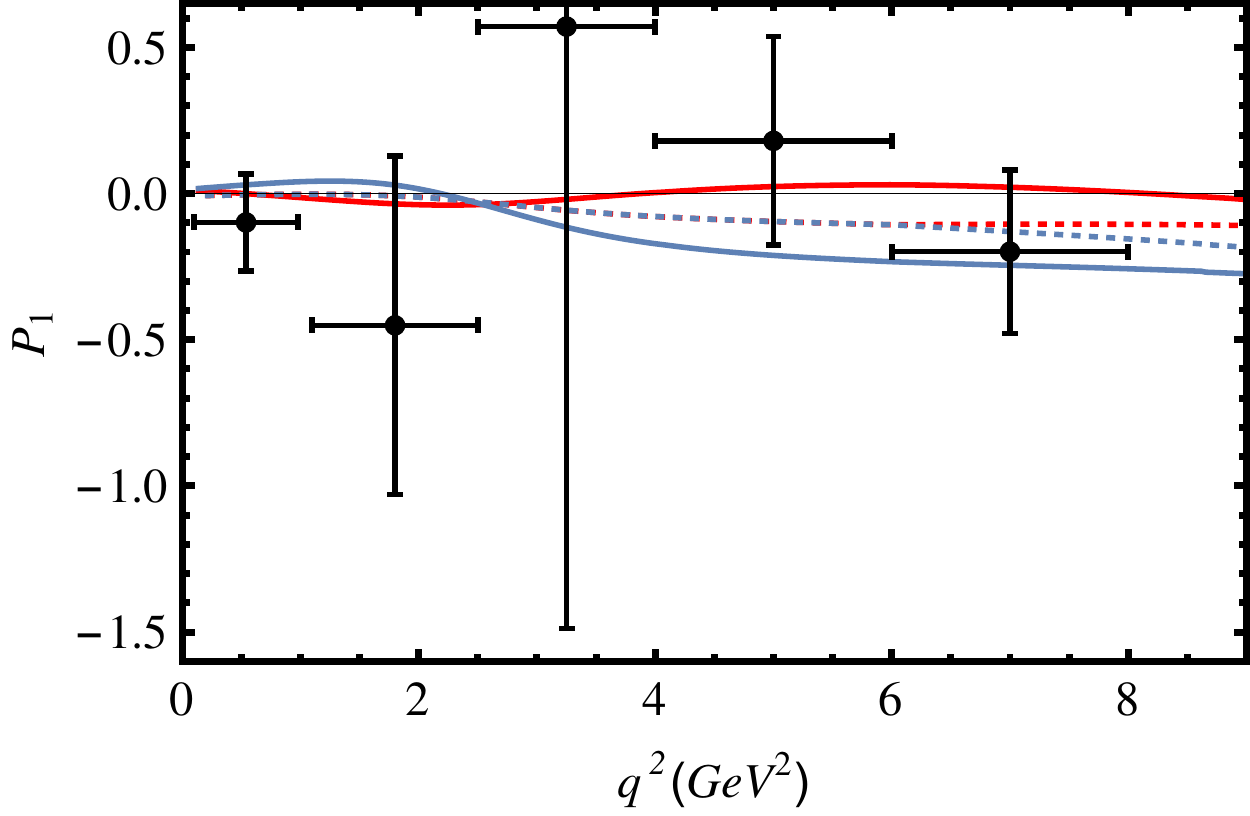}
\includegraphics[width=0.49\textwidth]{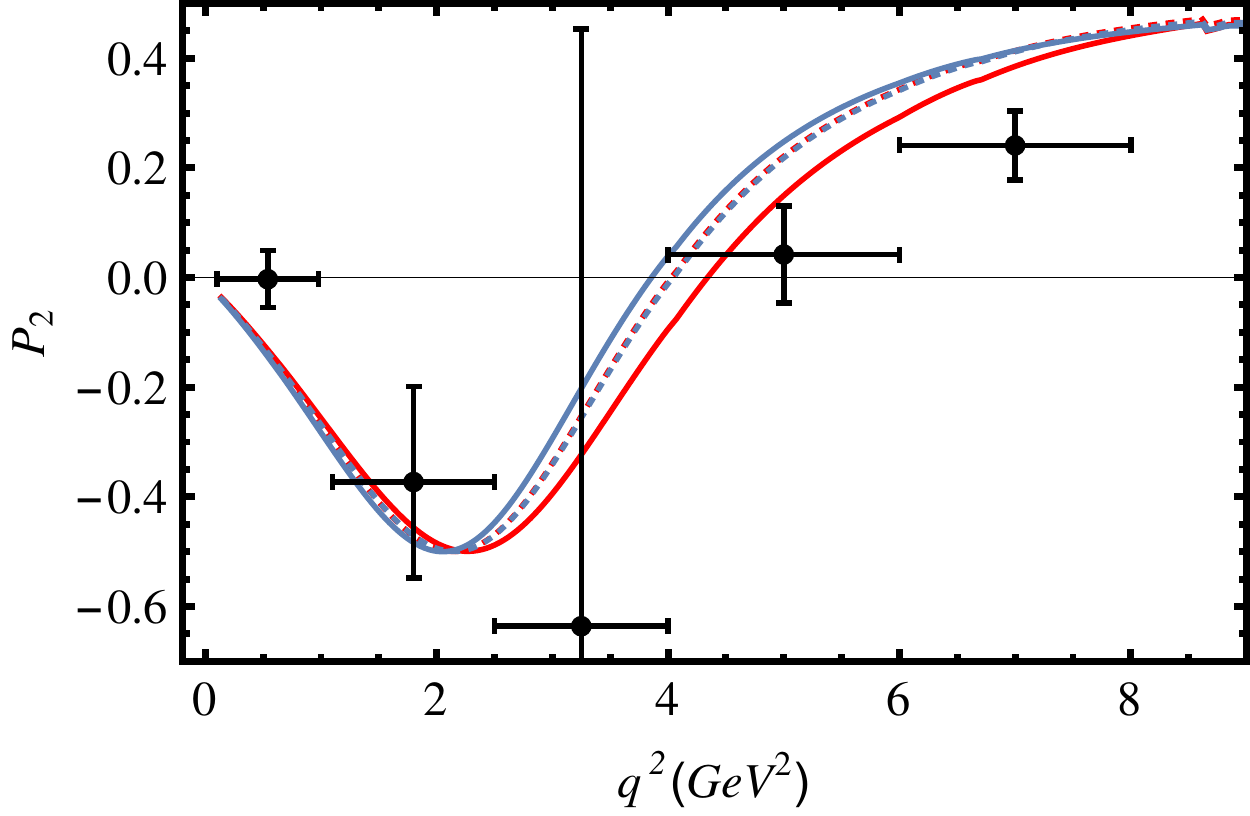}
\includegraphics[width=0.49\textwidth]{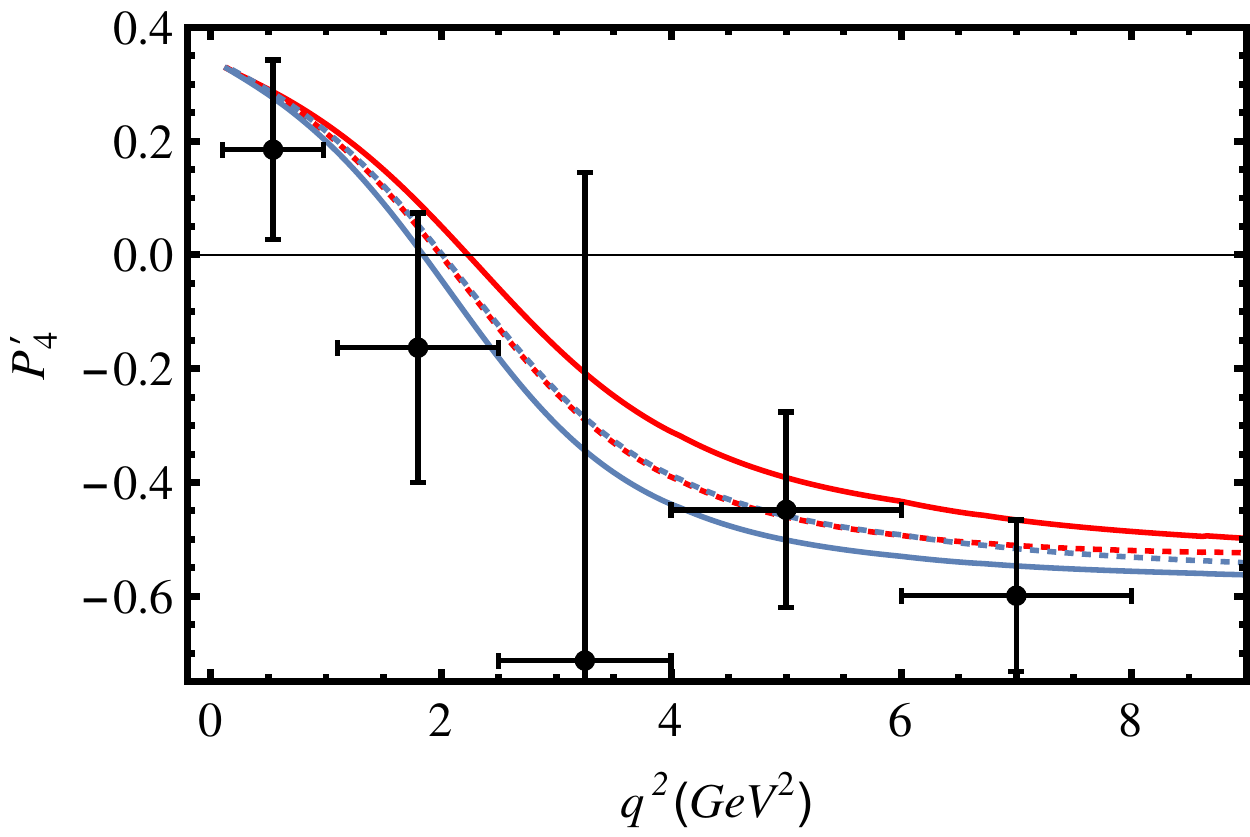}
\includegraphics[width=0.49\textwidth]{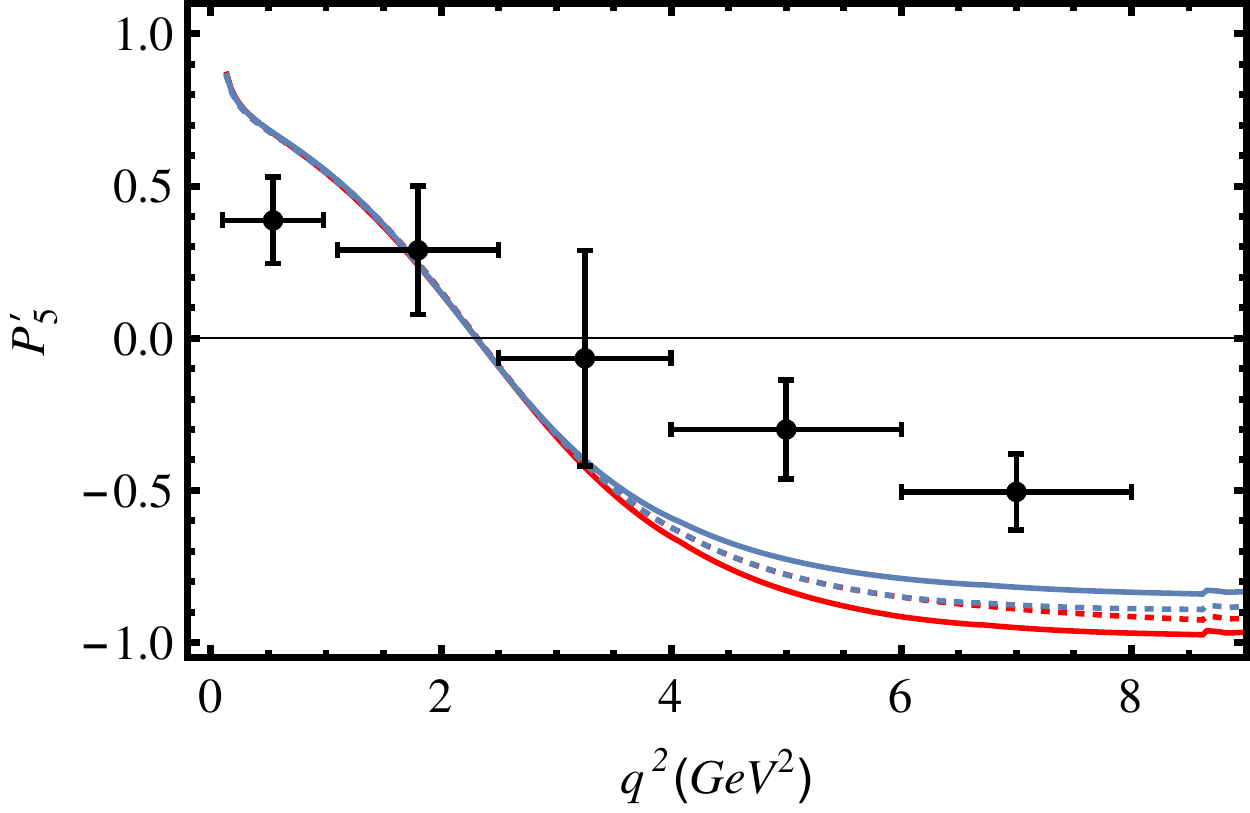}
\caption{Comparison of the SM predictions for the central values 
of $P_{1,2,4,5}^{(\prime)} (B\to K^* \mu^+ \mu^-)$
within different approaches and using different form factors as described in the caption of Fig.~\ref{fig:Comparison1}.
\label{fig:Comparison3}}
\end{figure}

In Figs.~\ref{fig:Comparison1},\ref{fig:Comparison3}, we have presented SM predictions of the central values for the most relevant $B\to K^* \mu^+ \mu^-$ observables,
employing the soft FF and full FF approaches both when using KMPW and BSZ form factors (Appendix~\ref{app:FF}).
In these figures, for the soft FF approach we have not included the $1/m_b$ power corrections which have been estimated through fitting with ad hoc functions in Ref.~\cite{Descotes-Genon:2014uoa} (for the KMPW form factors). 
In Figs.~\ref{fig:Comparison1},\ref{fig:Comparison3} it can be seen that while there 
are good agreement between the different approaches and the different form factor choices, 
the significance of the tension between central value of the SM predictions and the experimental data depends on the particular choice of the theoretical approach as well as which set of form factors are used.
E.g., both for the SM prediction of $S_5$ in Fig~\ref{fig:Comparison1} 
and $P_5^{\prime}$ in Fig.\ref{fig:Comparison3} the tension with experimental data in the [4.0,6.0] and [6.0,8.0] GeV$^2$ bins are smaller when using the full FF approach with BSZ form factors, and larger when using the soft FF approach with KMPW form factors.
The situation is reversed for the $P_2$ observables in the [4.0,6.0] and [6.0,8.0] GeV$^2$ bins as shown in Fig.~\ref{fig:Comparison3}.

%%%

\section{Dependency of $B\to K^* \mu^+ \mu^-$ observables on modification of a single Wilson coefficient}\label{app:PiShapes}

\begin{figure}[h!]
\centering
\includegraphics[width=0.49\textwidth]{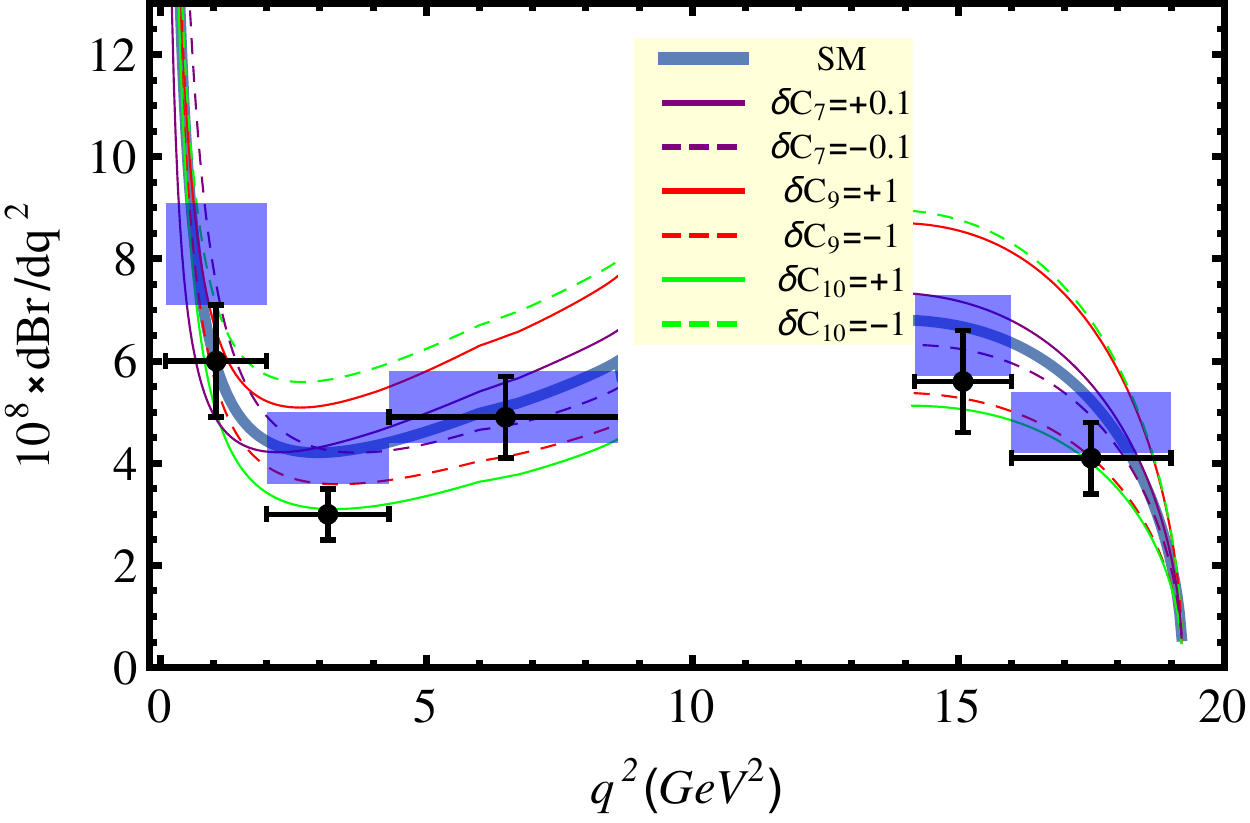}
\includegraphics[width=0.49\textwidth]{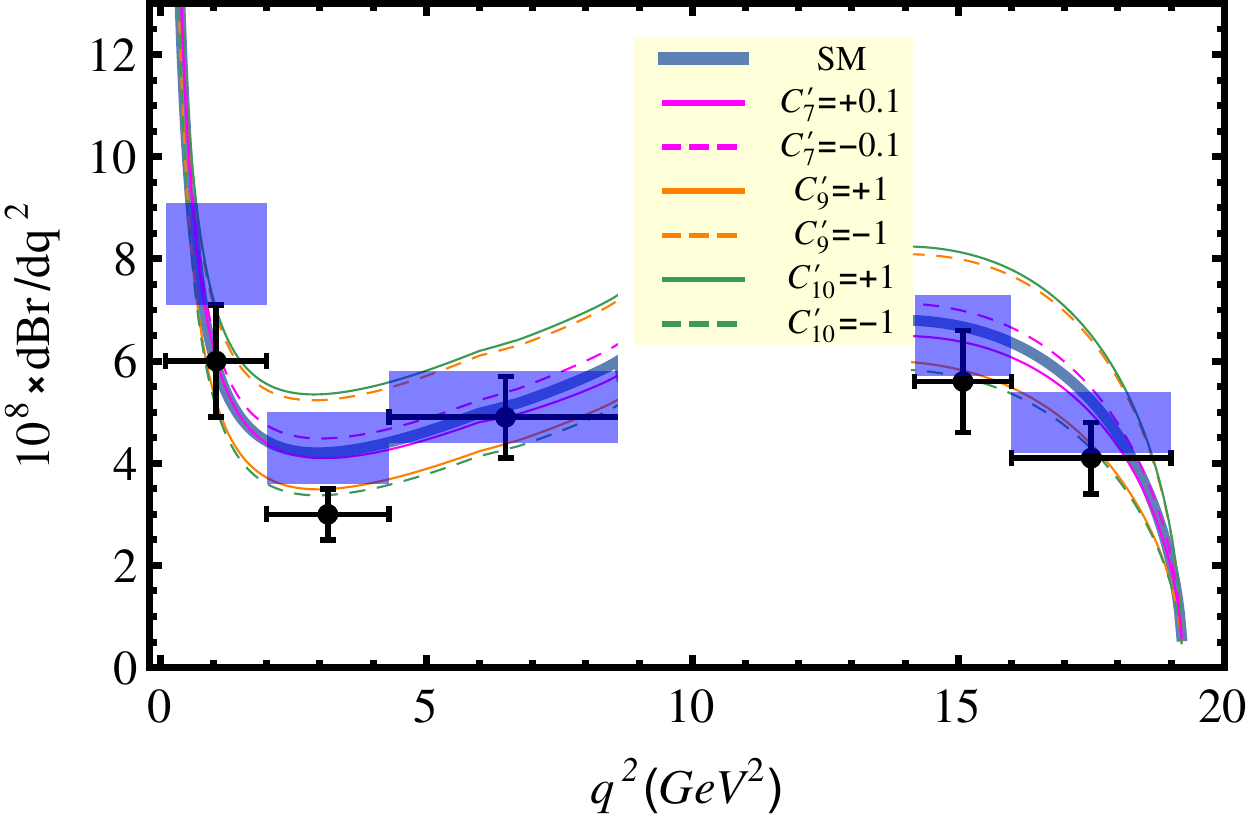}
\includegraphics[width=0.49\textwidth]{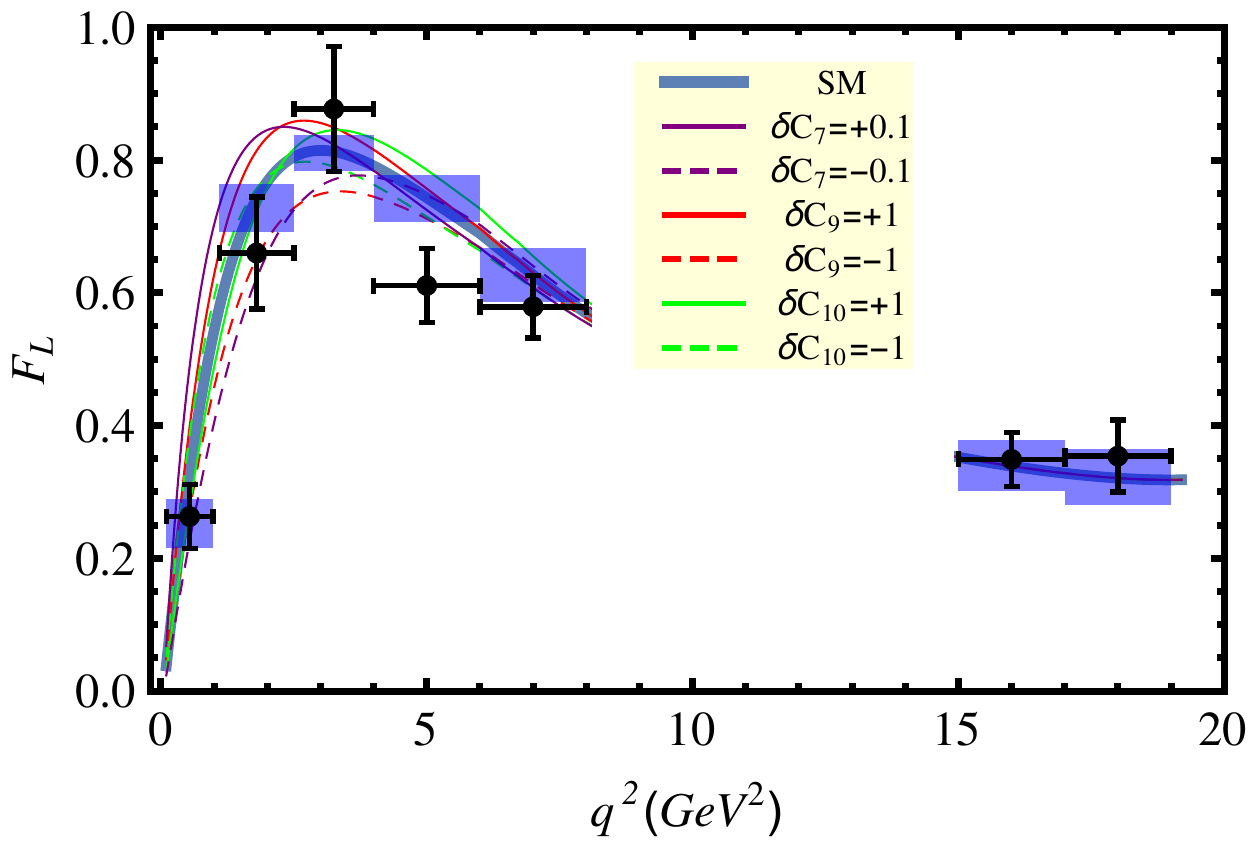}
\includegraphics[width=0.49\textwidth]{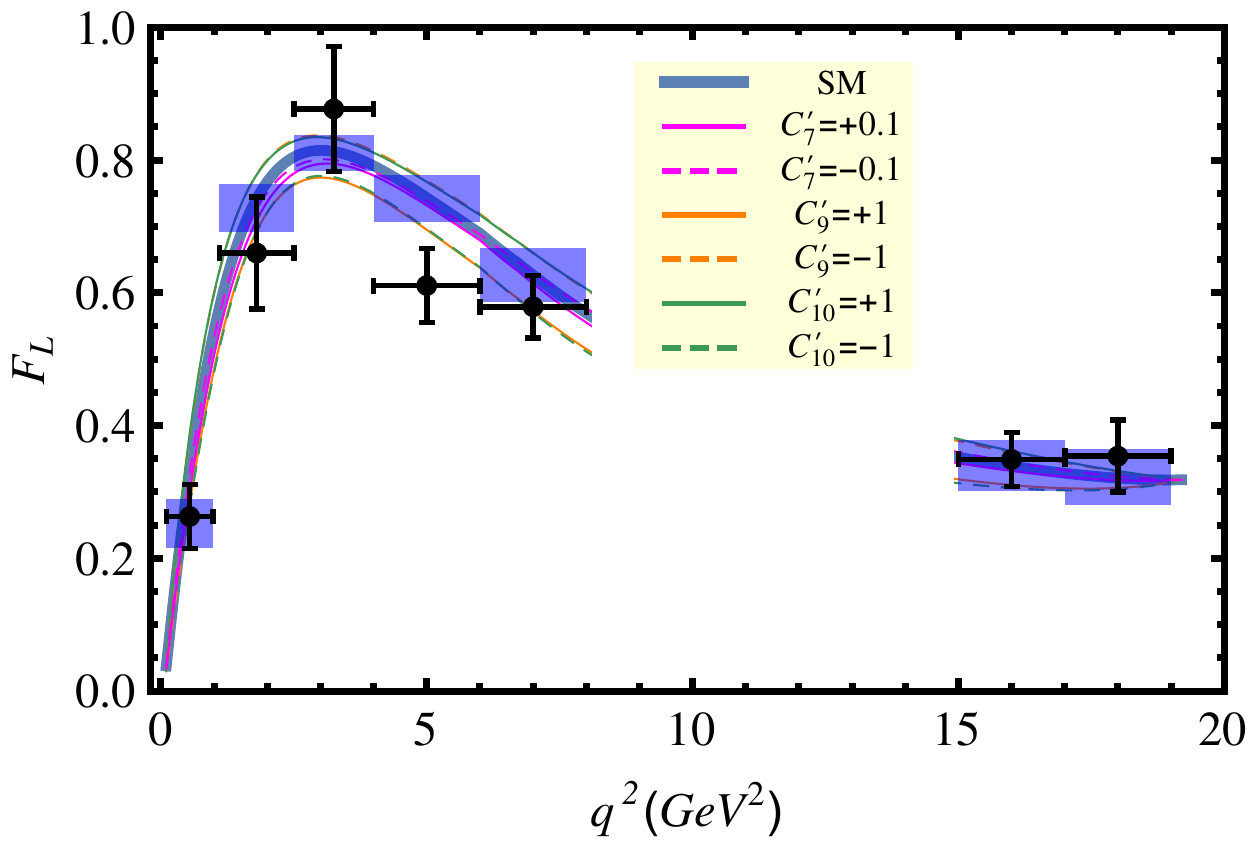}
\includegraphics[width=0.49\textwidth]{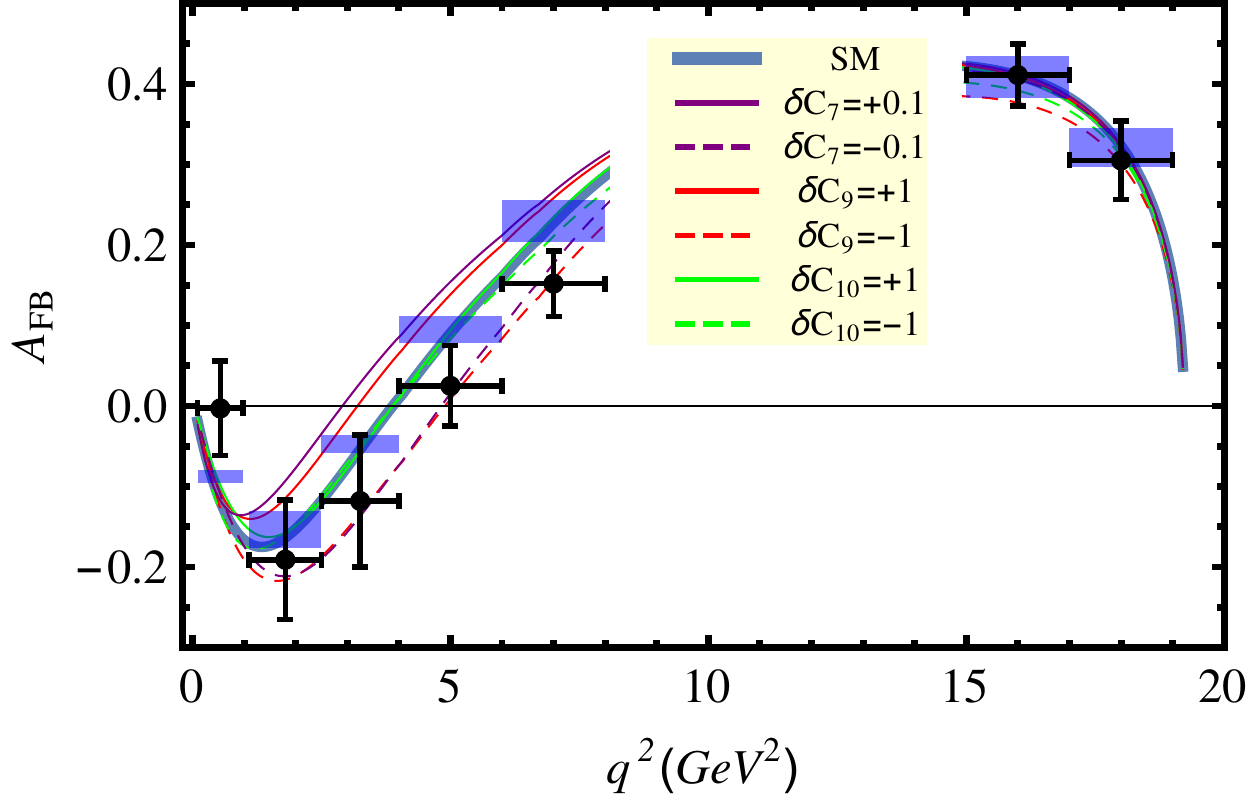}
\includegraphics[width=0.49\textwidth]{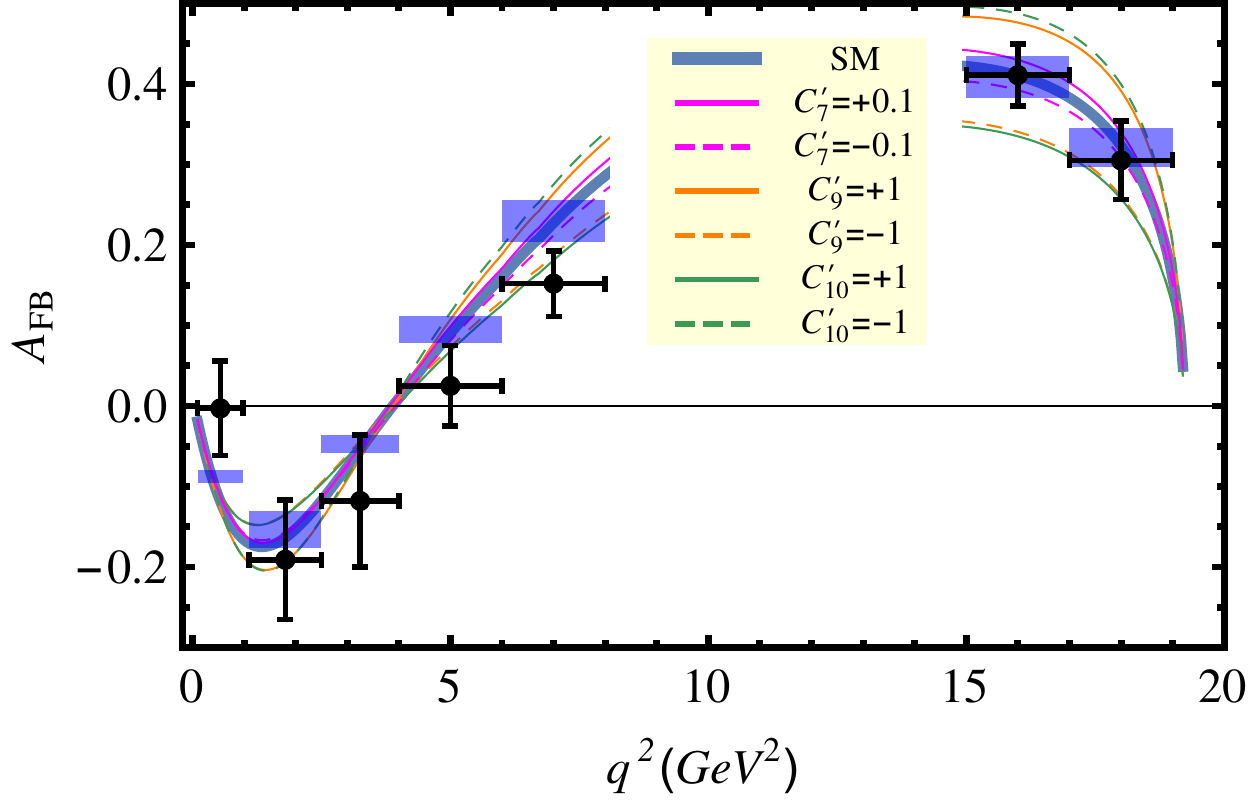}
\caption{$B\to K^* \mu^+ \mu^-$ observables: $10^8\times dBR/dq^2, F_L$ and $A_{FB}$. 
On the left side the behaviour of the modified Wilson coefficients 
($\delta C_7\!\! =\! \pm 0.1,\; \delta C_{9}\!\!=\! \pm 1.0,\; \delta C_{10}\!\! =\! \pm 1.0$) are shown and on the right side the
behaviour of the modified \emph{primed} Wilson coefficients ($C_7^\prime\!\! =\! \pm 0.1,\; C_{9}^\prime \!\! =\! \pm 1.0,\;  
C_{10}^\prime \!\!=\! \pm 1.0$).
The black crosses correspond to the LHCb measurements where $dBR/dq^2$ is from the 1 fb$^{-1}$ of data~\cite{Aaij:2013iag} 
and the angular observables are from the 3 fb$^{-1}$ of data~\cite{Aaij:2015oid}. 
The blue bands correspond to the binned SM predictions with their relevant uncertainties.
\label{fig:BKstarmumu:AFB}}
\end{figure}

\begin{figure}
\centering
\includegraphics[width=0.49\textwidth]{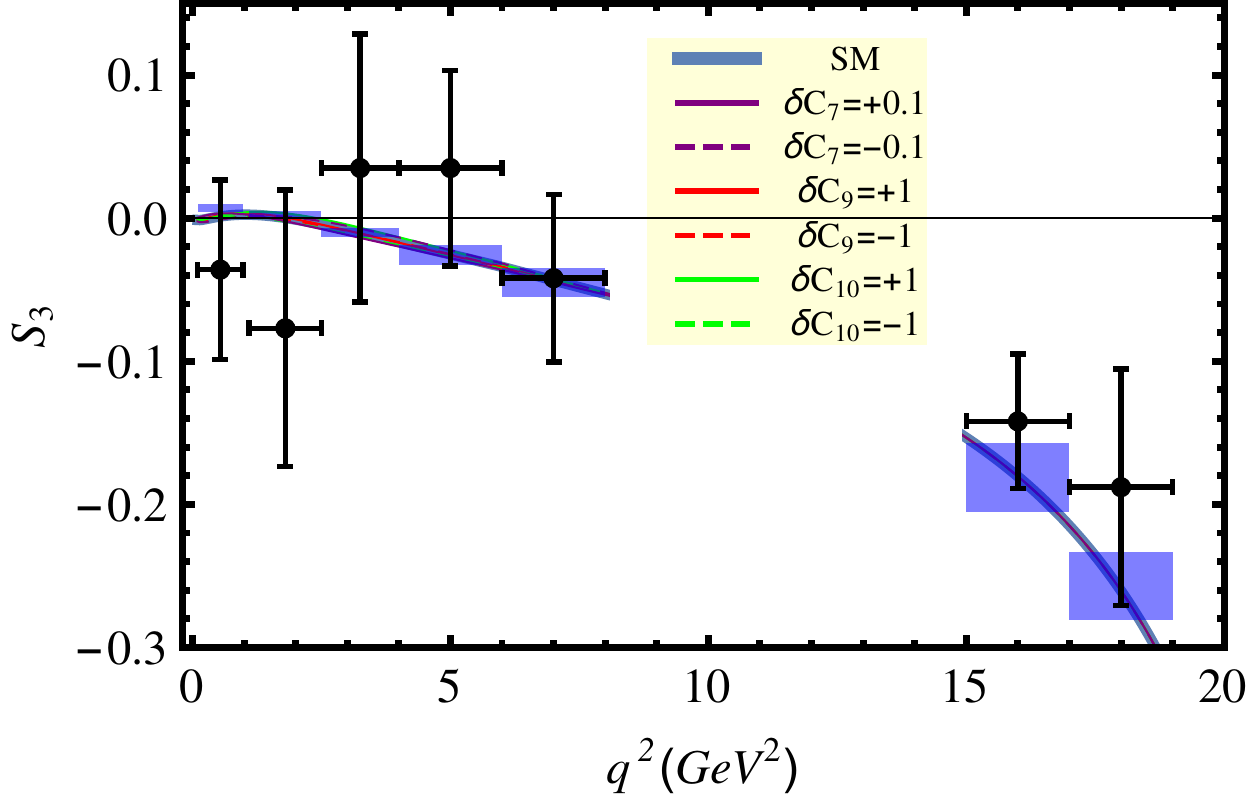}
\includegraphics[width=0.49\textwidth]{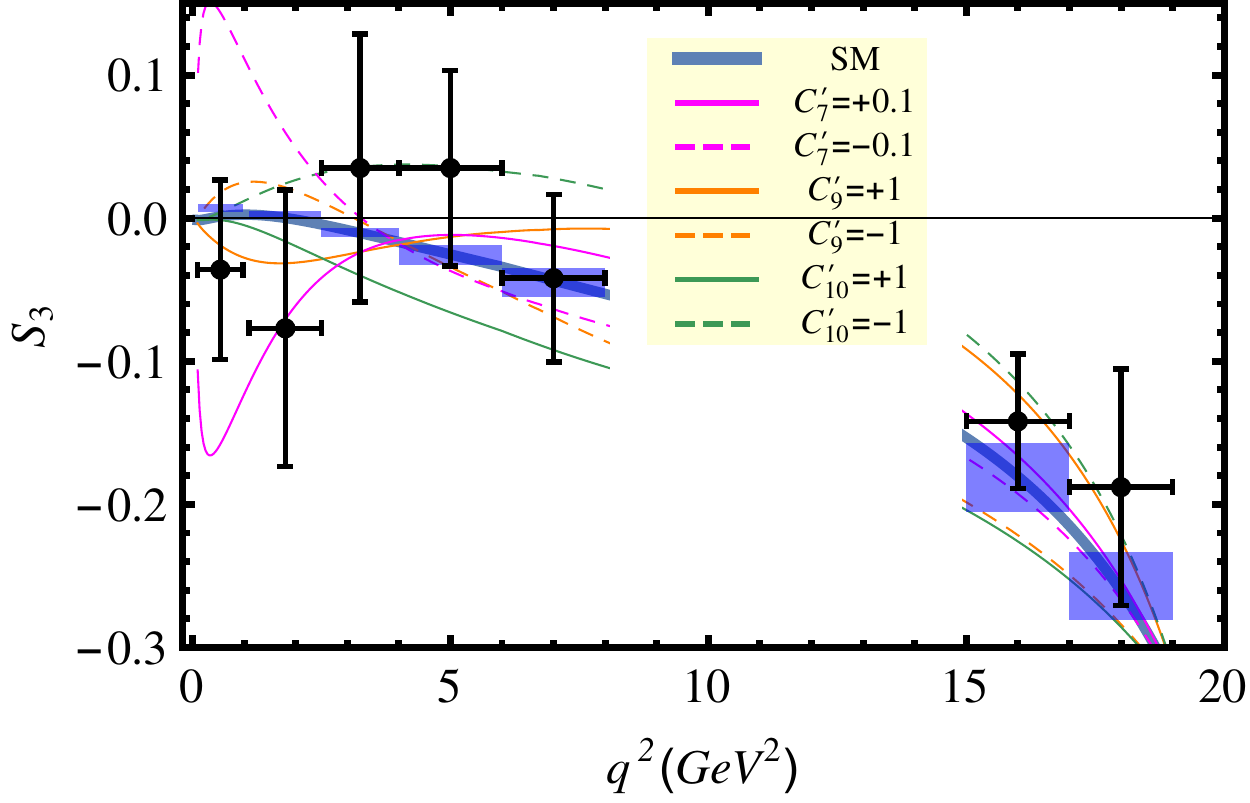}
\includegraphics[width=0.49\textwidth]{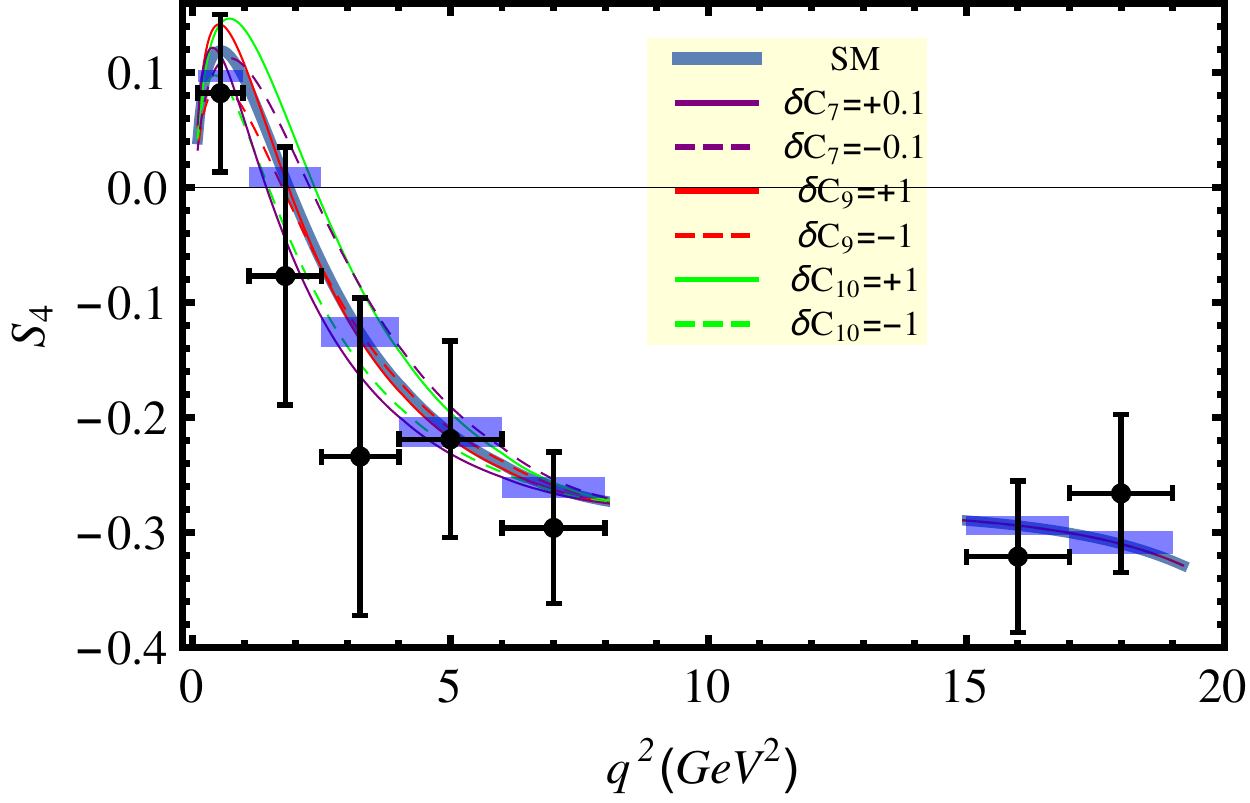}
\includegraphics[width=0.49\textwidth]{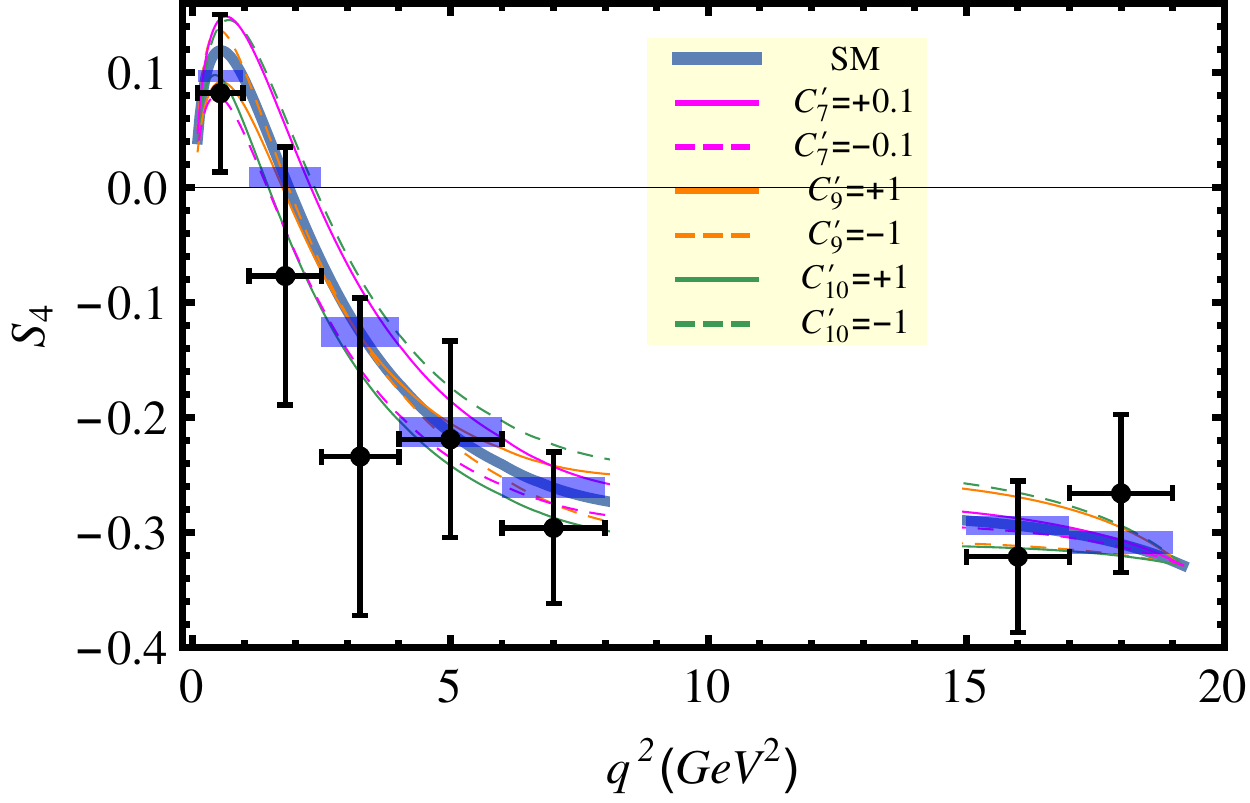}
\includegraphics[width=0.49\textwidth]{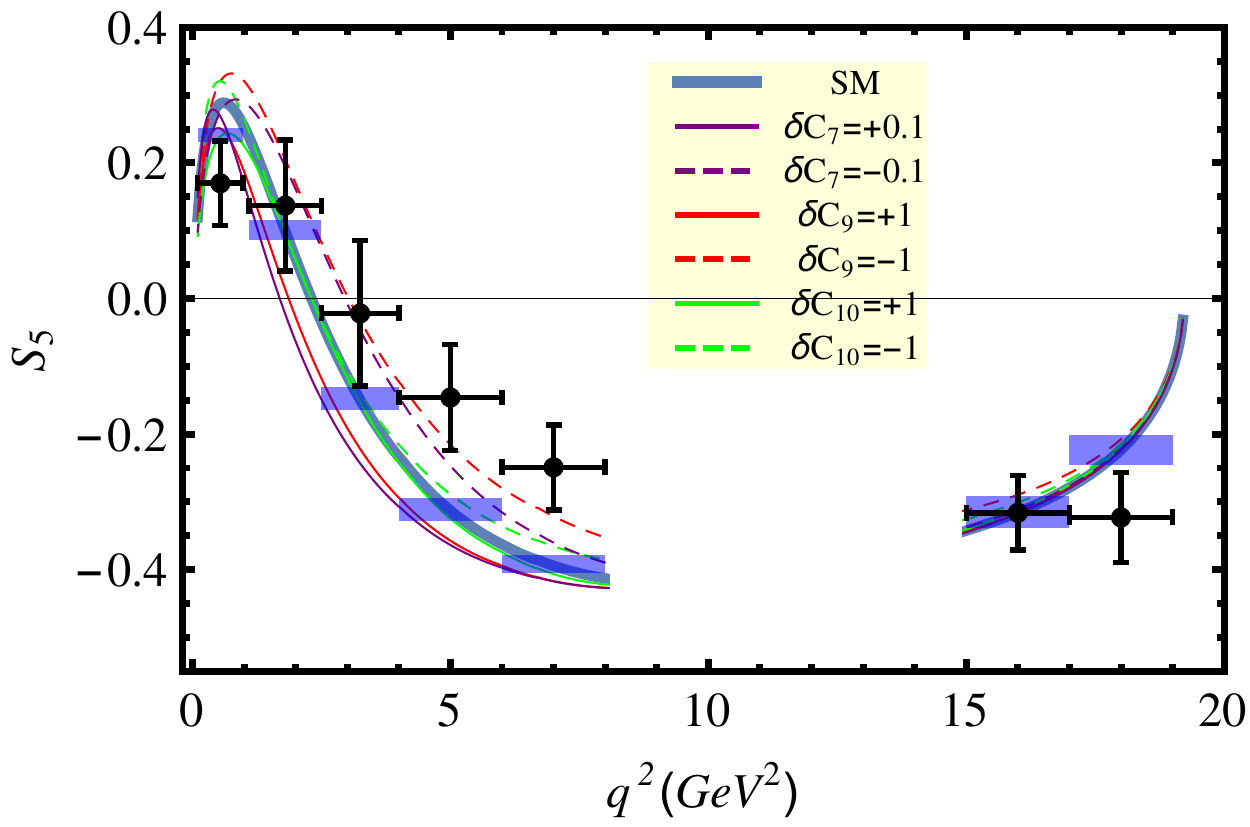}
\includegraphics[width=0.49\textwidth]{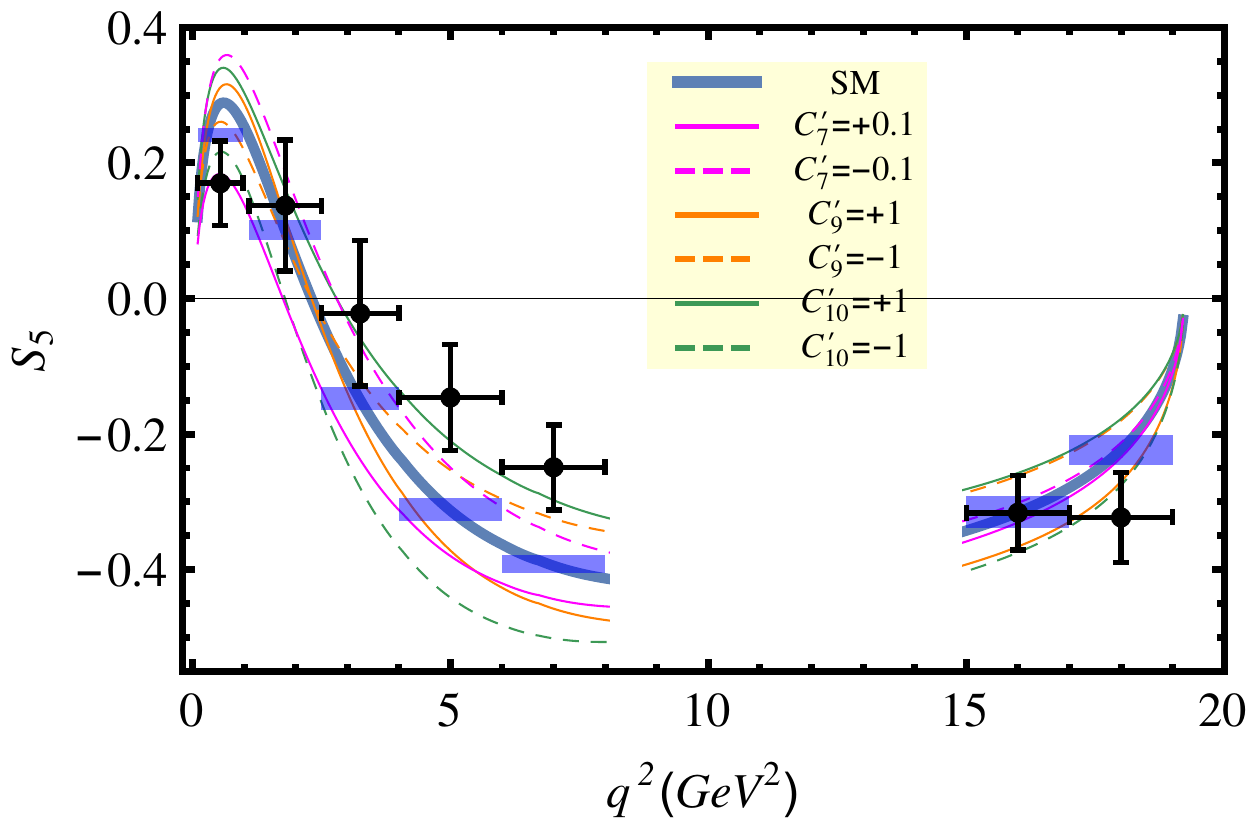}
\caption{$S_{3,4,5} (B\to K^* \mu^+ \mu^-)$, as described in Fig.~\ref{fig:BKstarmumu:AFB}. \label{fig:BKstarmumu:S3-S5}}
\end{figure}

\begin{figure}
\centering
\includegraphics[width=0.49\textwidth]{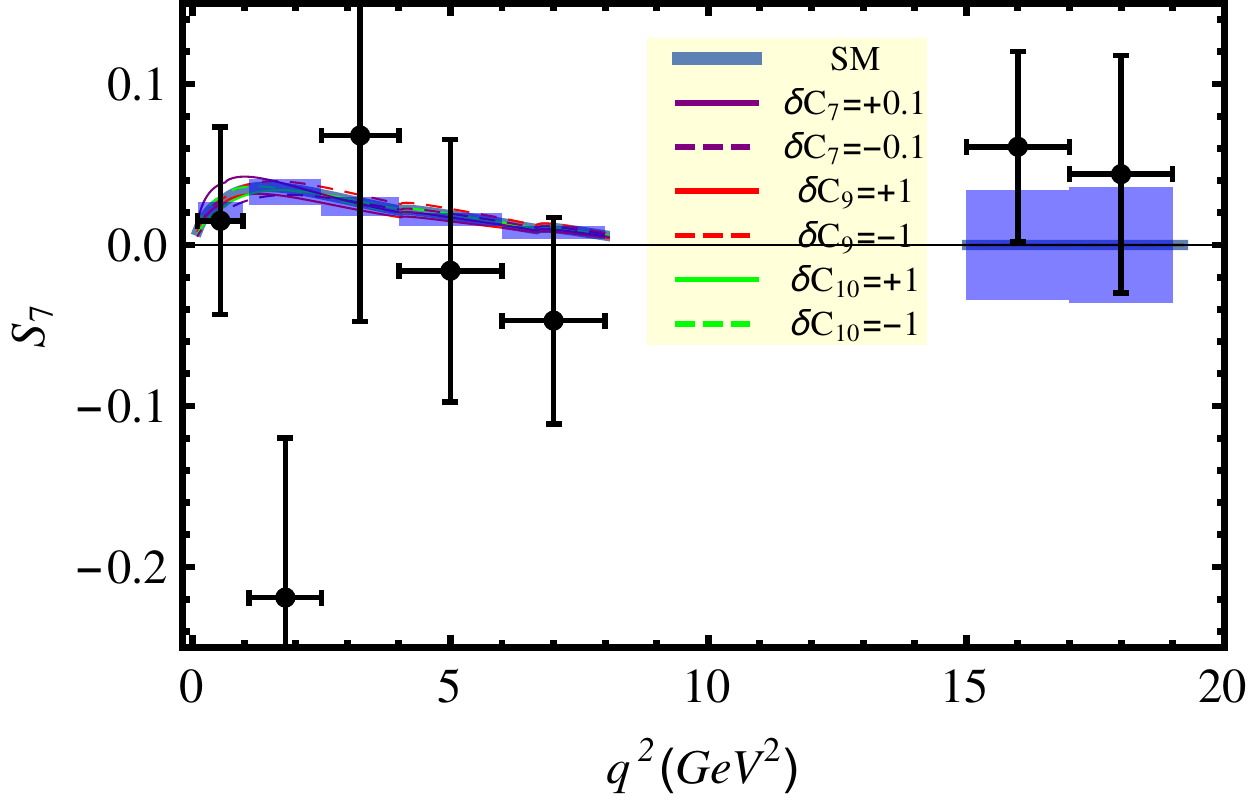}
\includegraphics[width=0.49\textwidth]{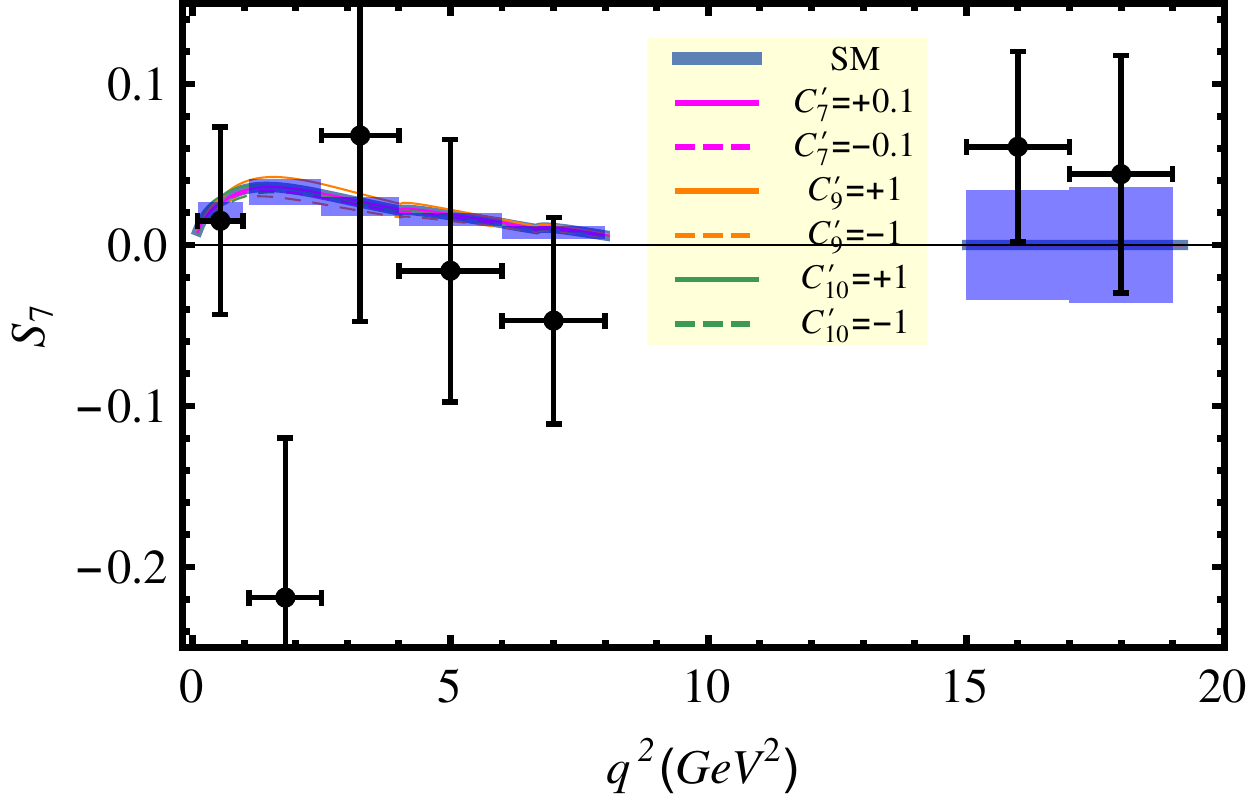}
\includegraphics[width=0.49\textwidth]{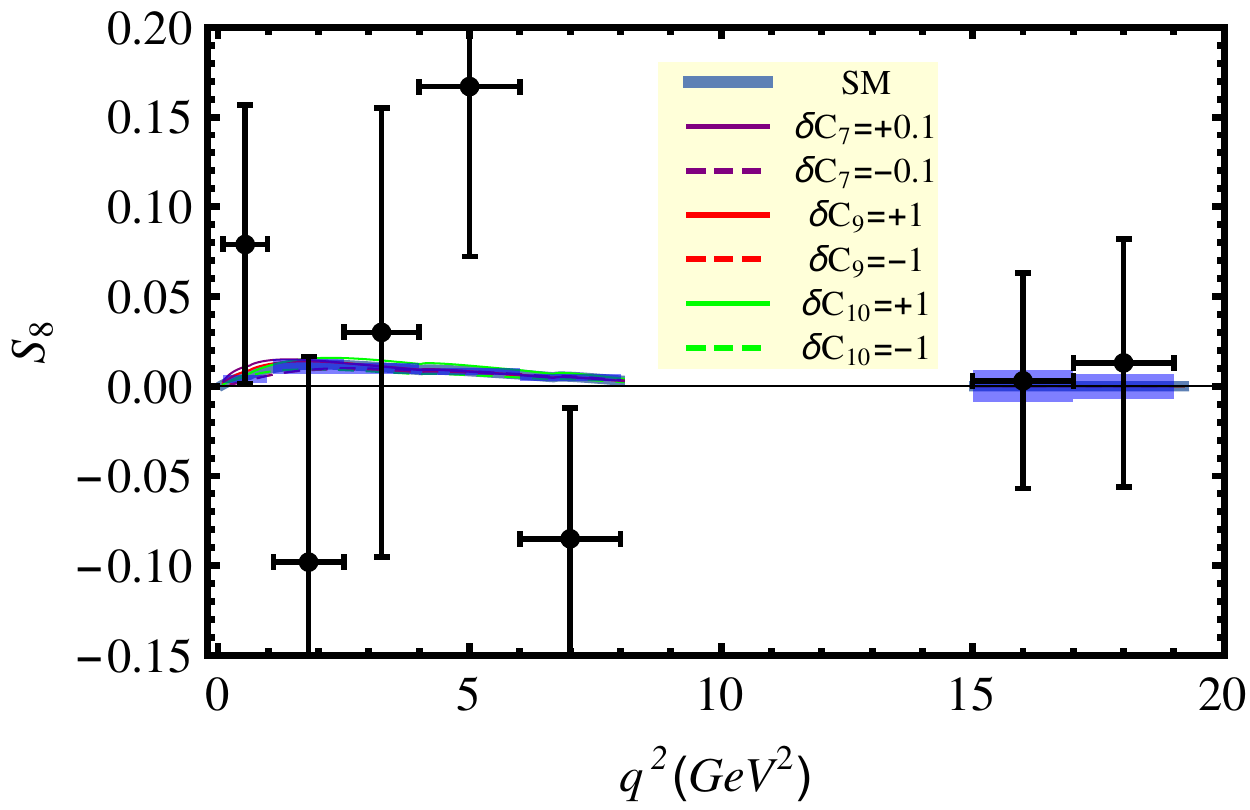}
\includegraphics[width=0.49\textwidth]{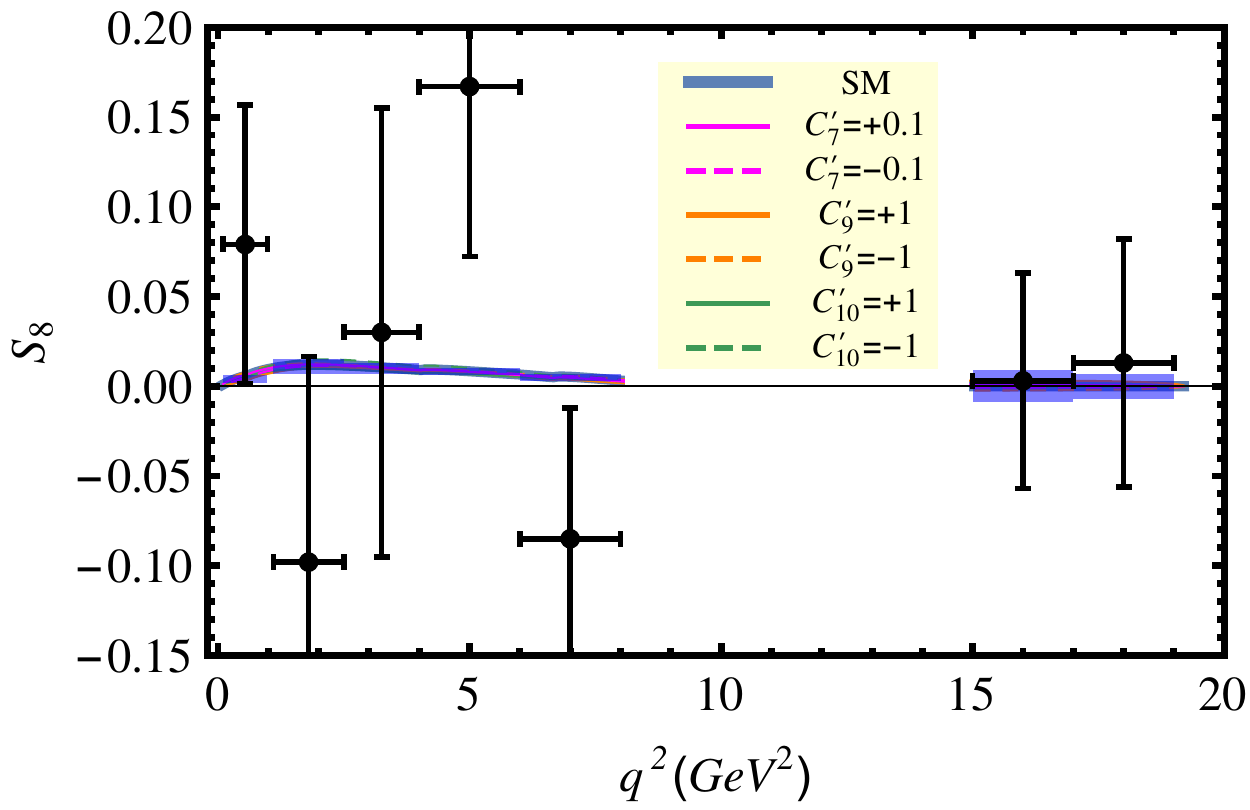}
\includegraphics[width=0.49\textwidth]{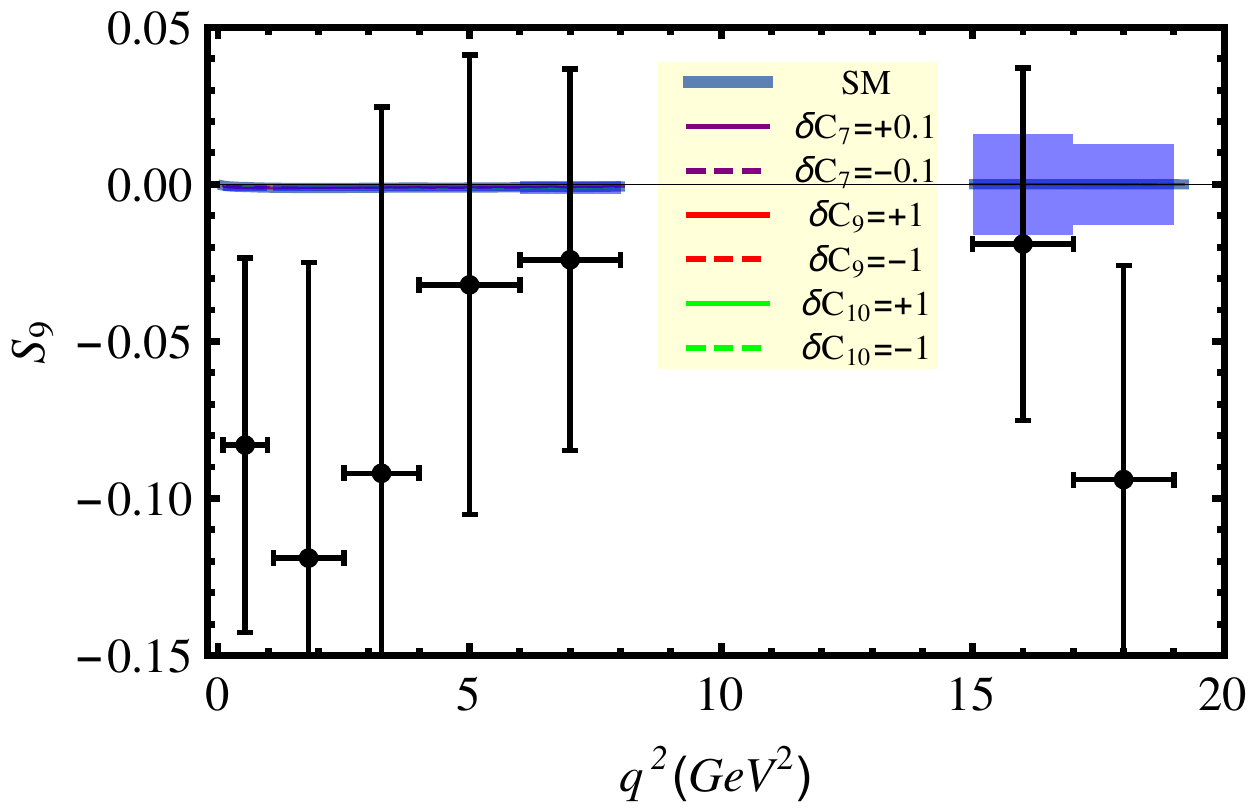}
\includegraphics[width=0.49\textwidth]{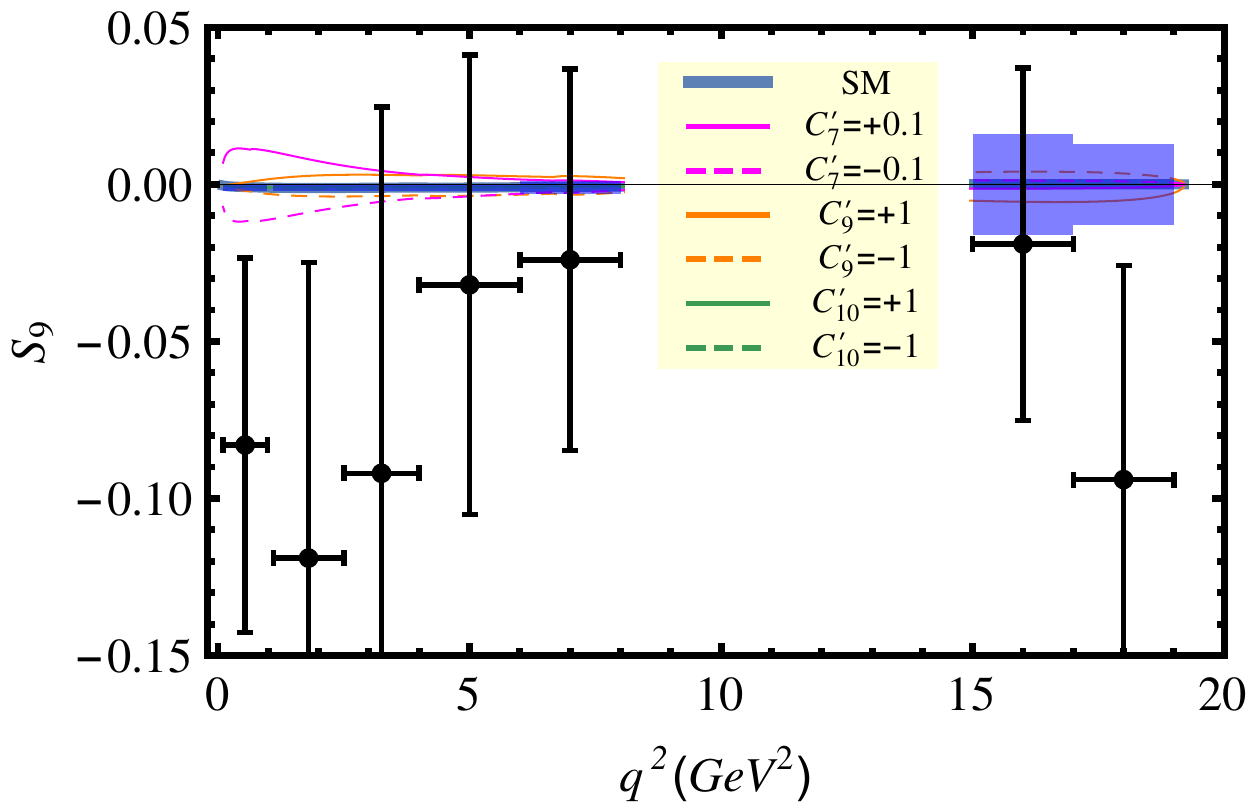}
\caption{$S_{7,8,9} (B\to K^* \mu^+ \mu^-)$, as described in Fig.~\ref{fig:BKstarmumu:AFB}. \label{fig:BKstarmumu:S7-S9}}
\end{figure}

\begin{figure}[!h]
\centering
\includegraphics[width=0.49\textwidth]{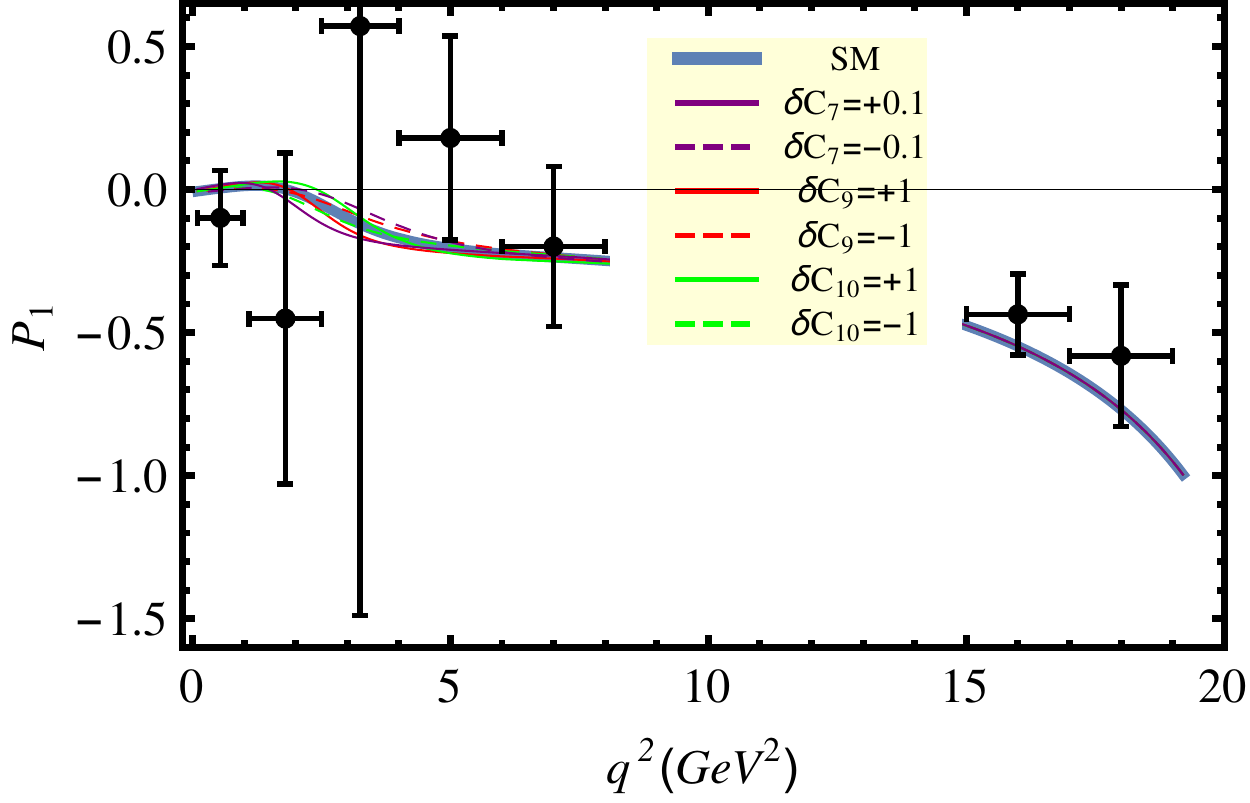}
\includegraphics[width=0.49\textwidth]{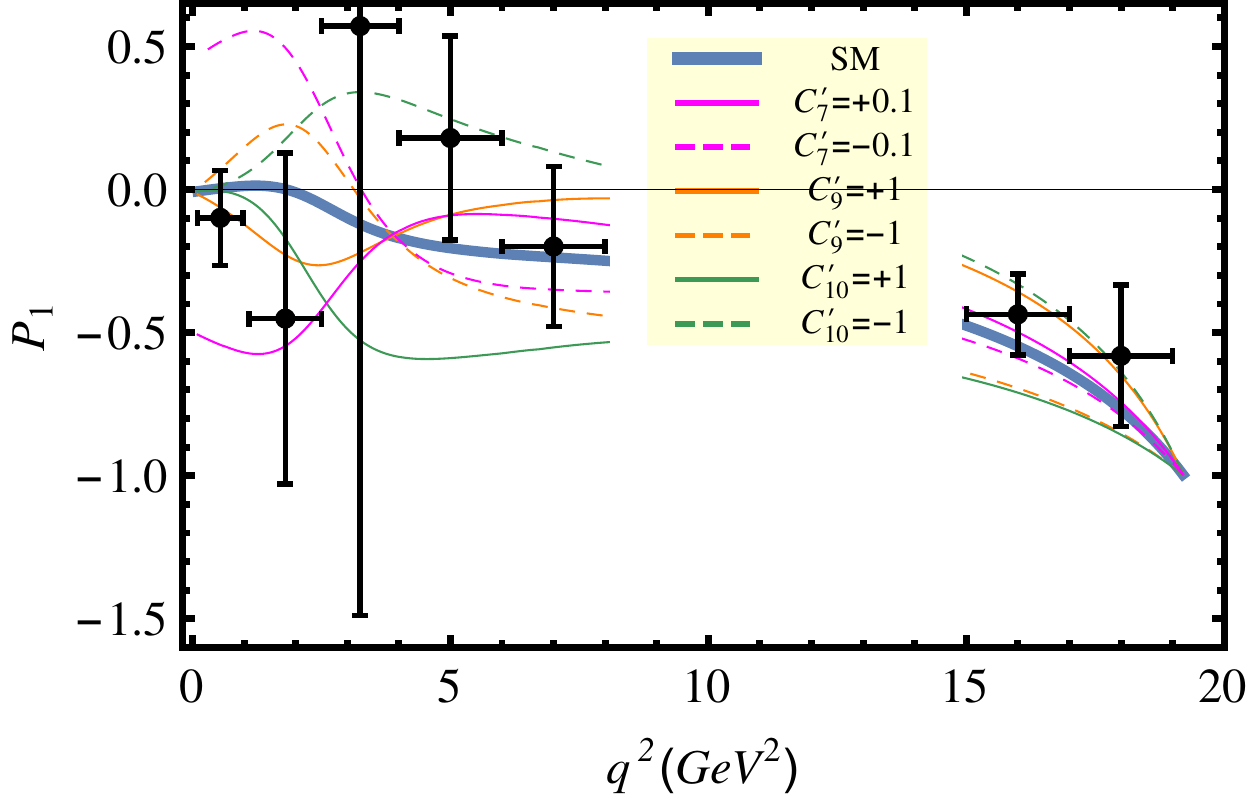}
\includegraphics[width=0.49\textwidth]{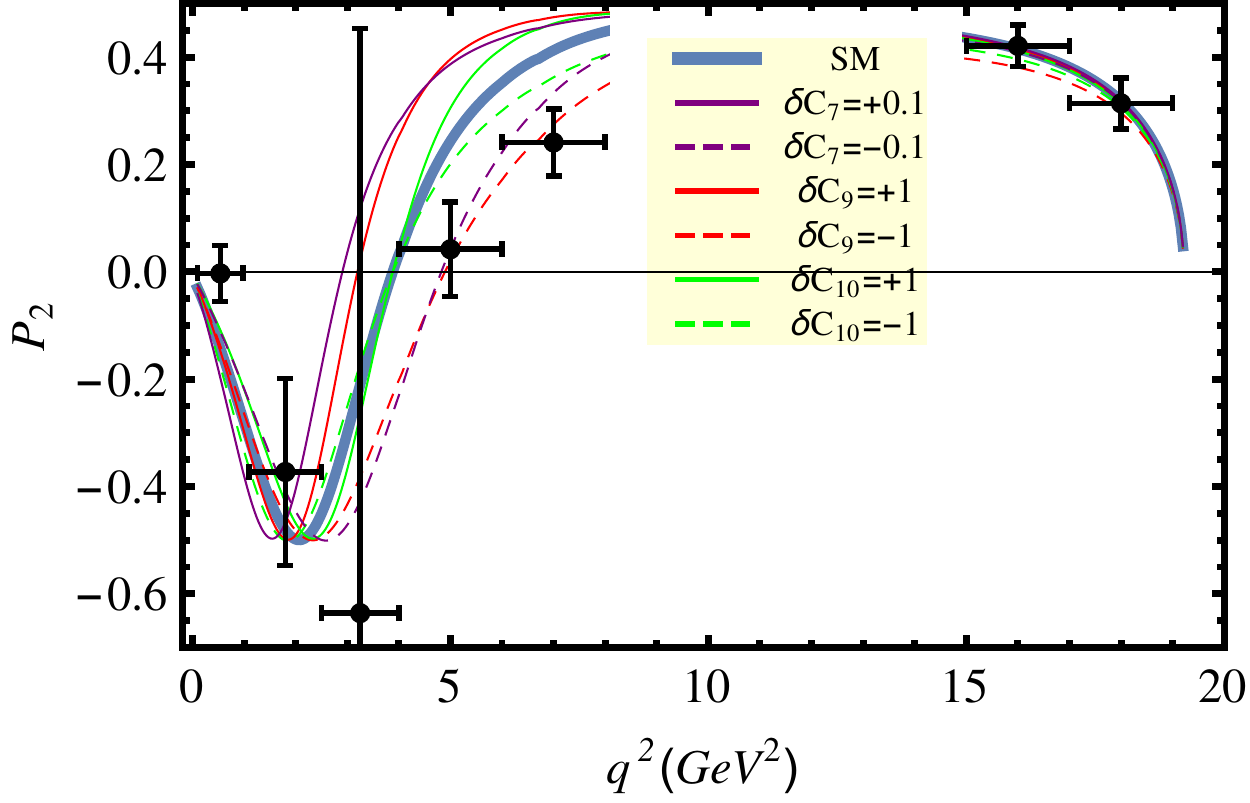}
\includegraphics[width=0.49\textwidth]{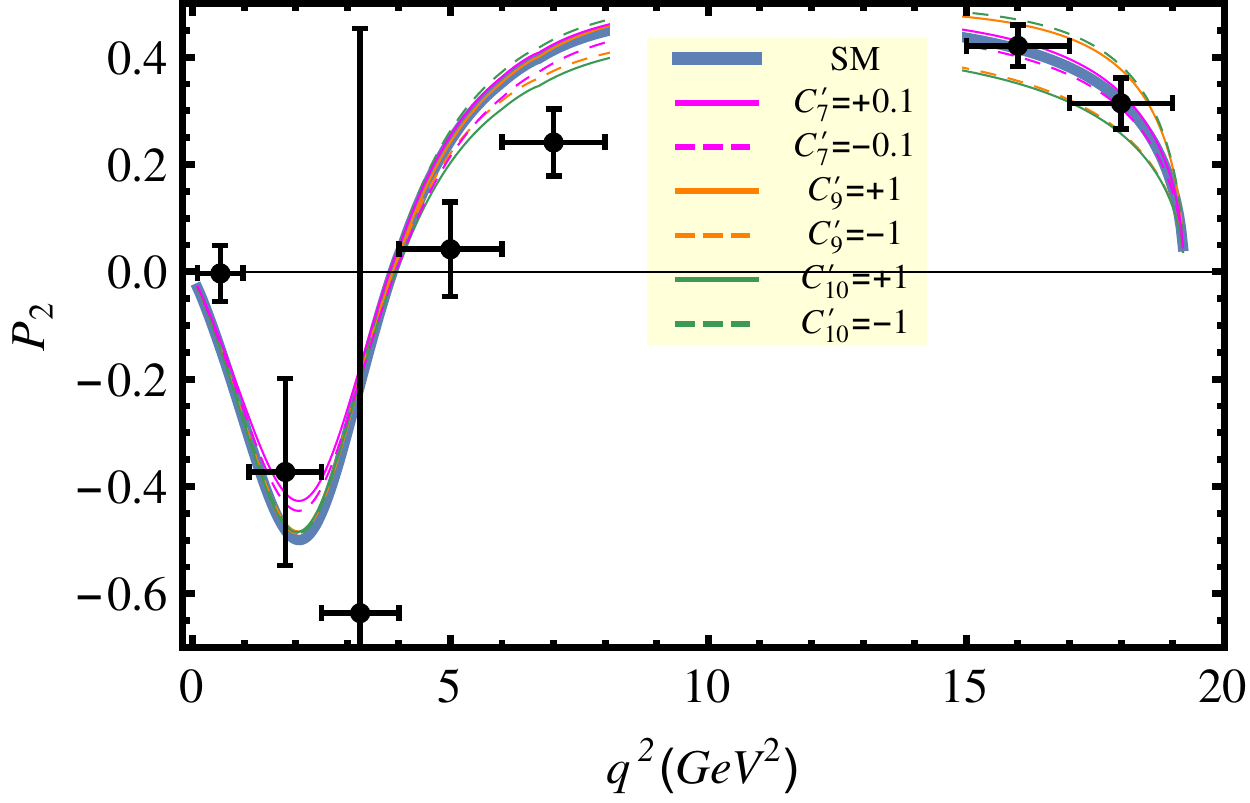}
\includegraphics[width=0.49\textwidth]{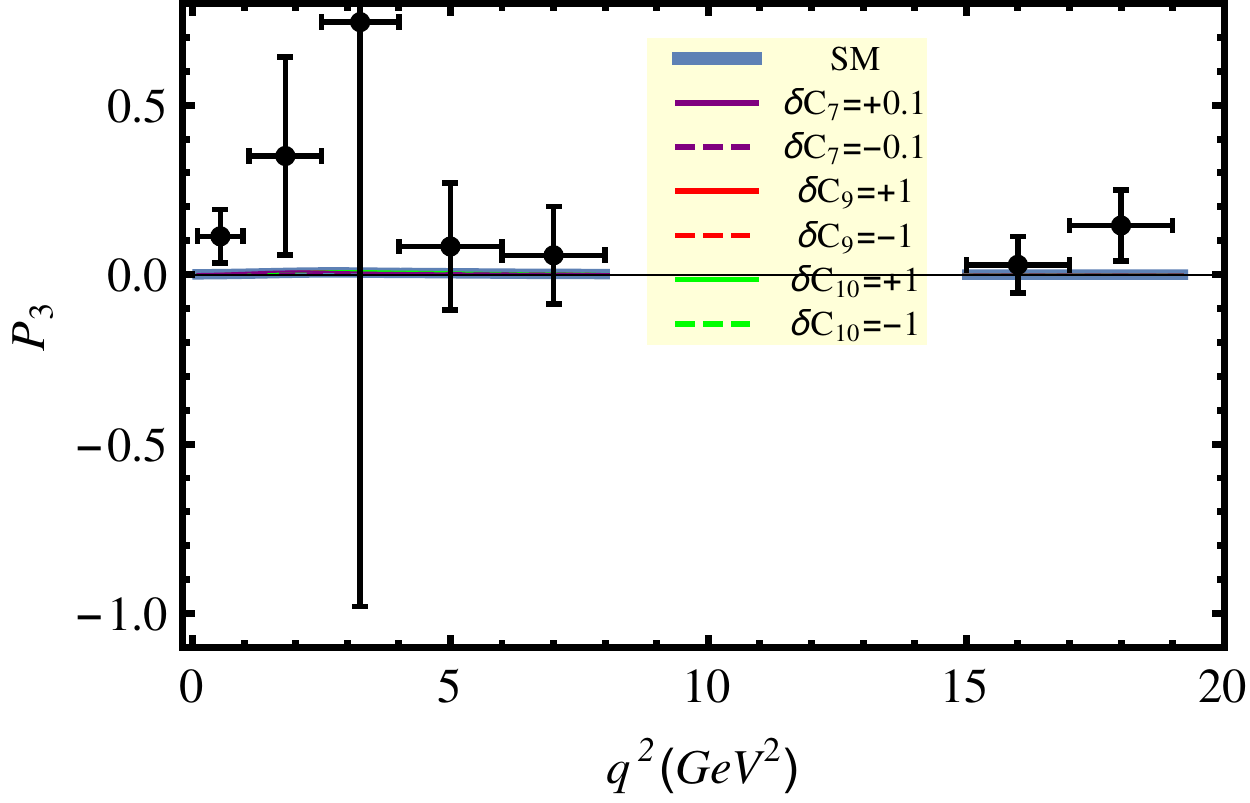}
\includegraphics[width=0.49\textwidth]{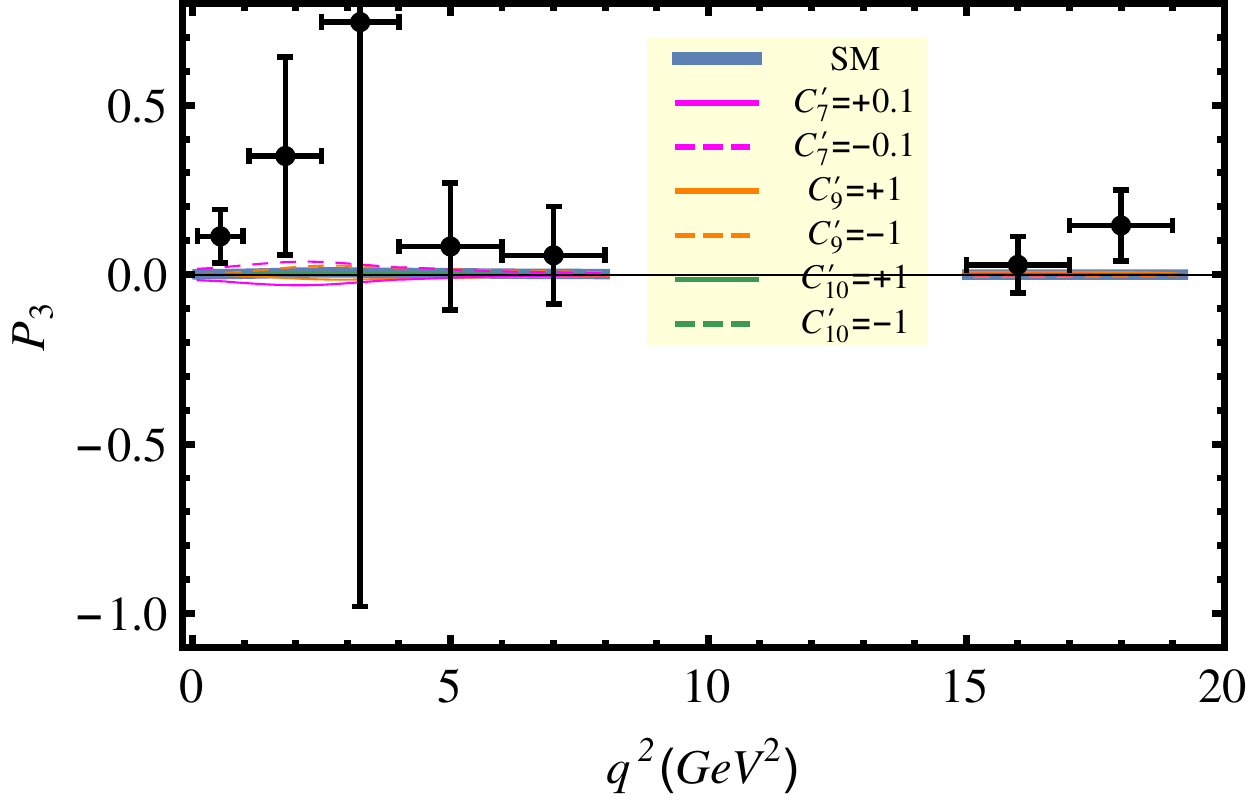}
\caption{$P_{1,2,3} (B\to K^* \mu^+ \mu^-)$, as described in Fig.~\ref{fig:BKstarmumu:AFB}. \label{fig:BKstarmumu:P1-P3}}
\end{figure}

\begin{figure}
\centering
\includegraphics[width=0.49\textwidth]{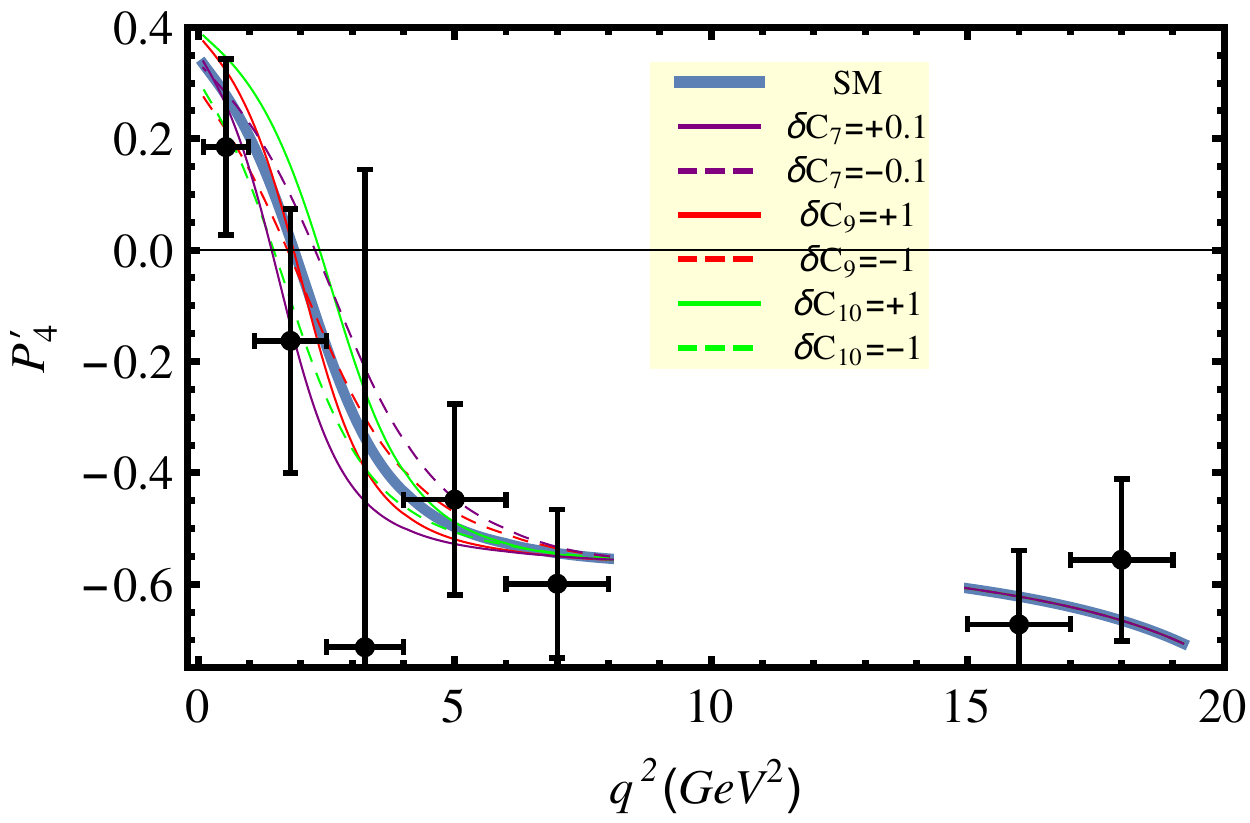}
\includegraphics[width=0.49\textwidth]{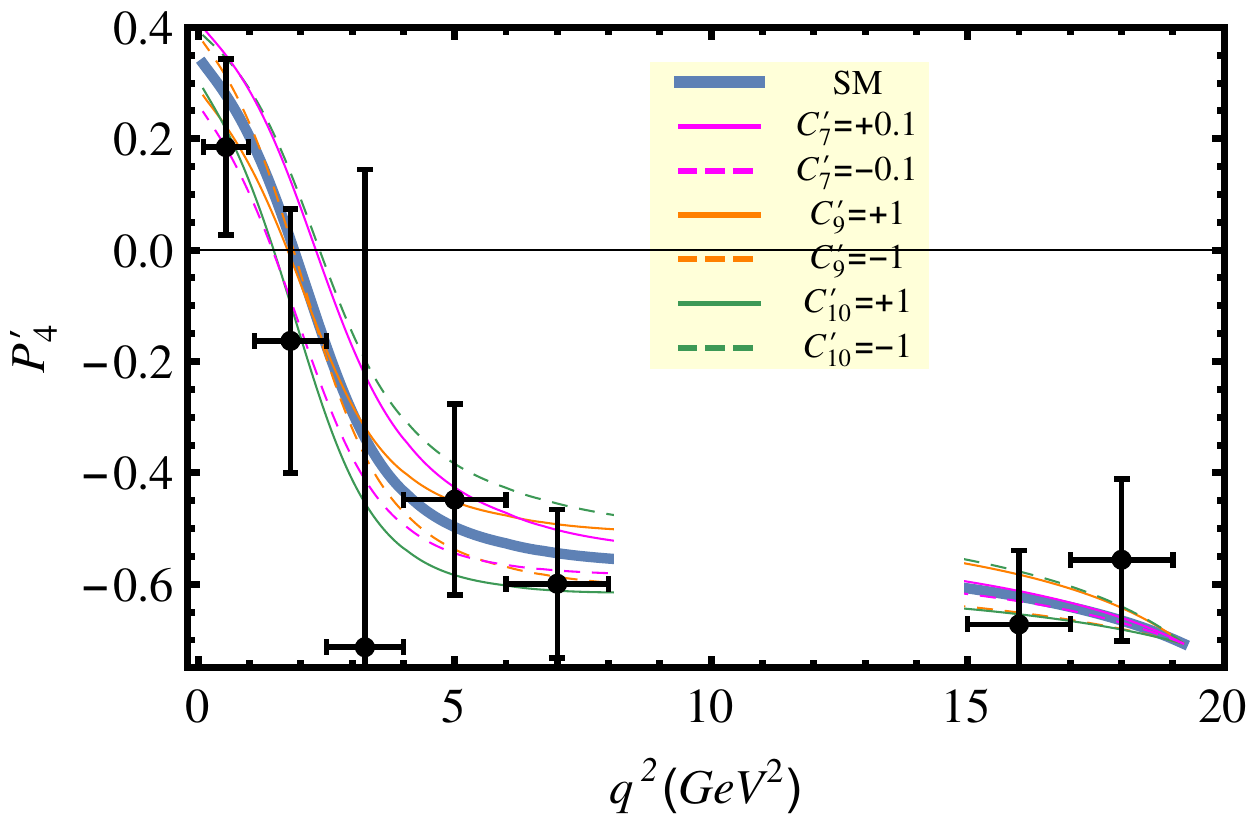}
\includegraphics[width=0.49\textwidth]{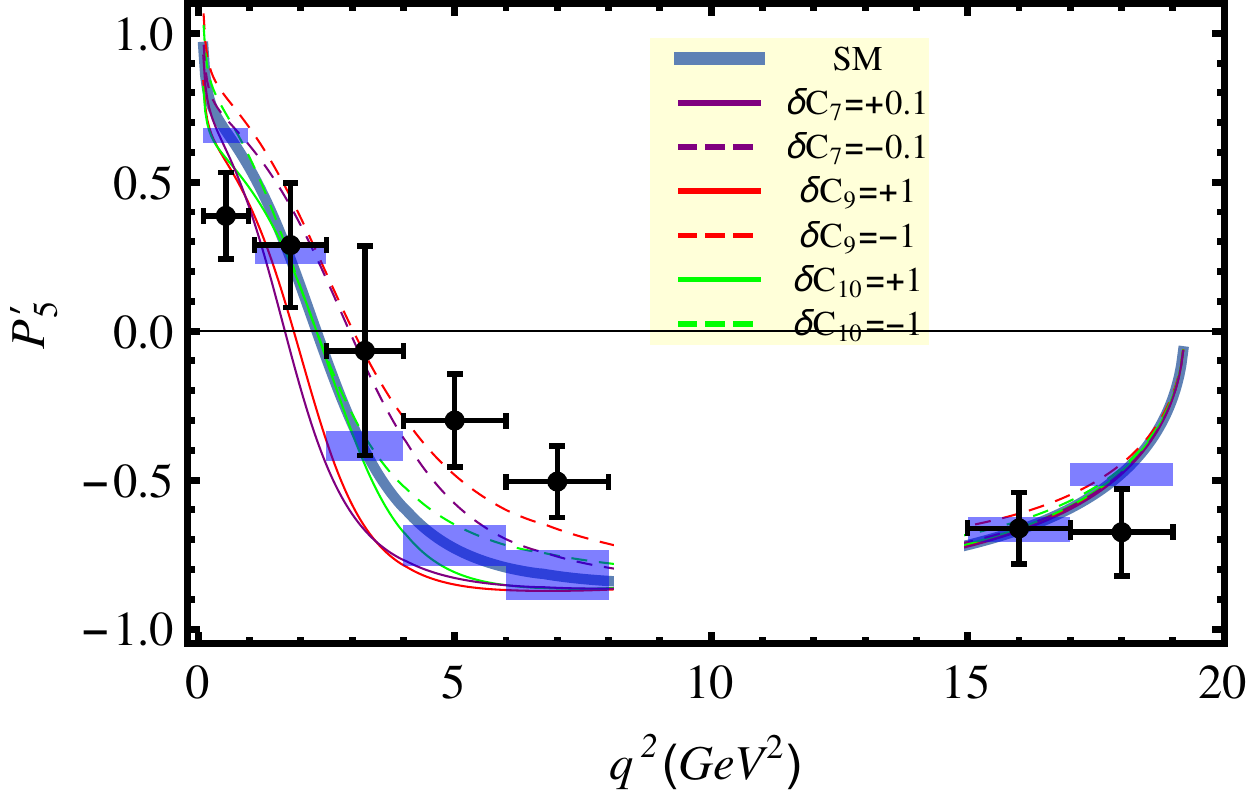}
\includegraphics[width=0.49\textwidth]{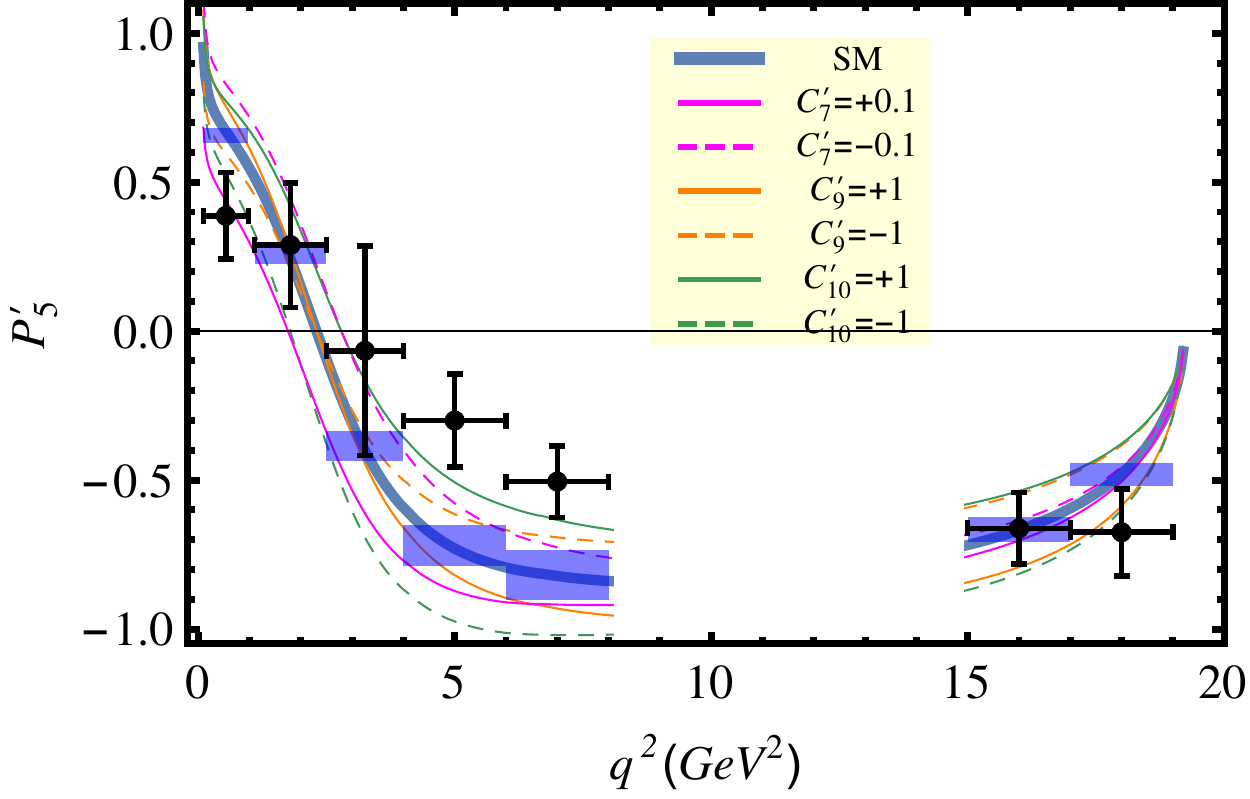}
\includegraphics[width=0.49\textwidth]{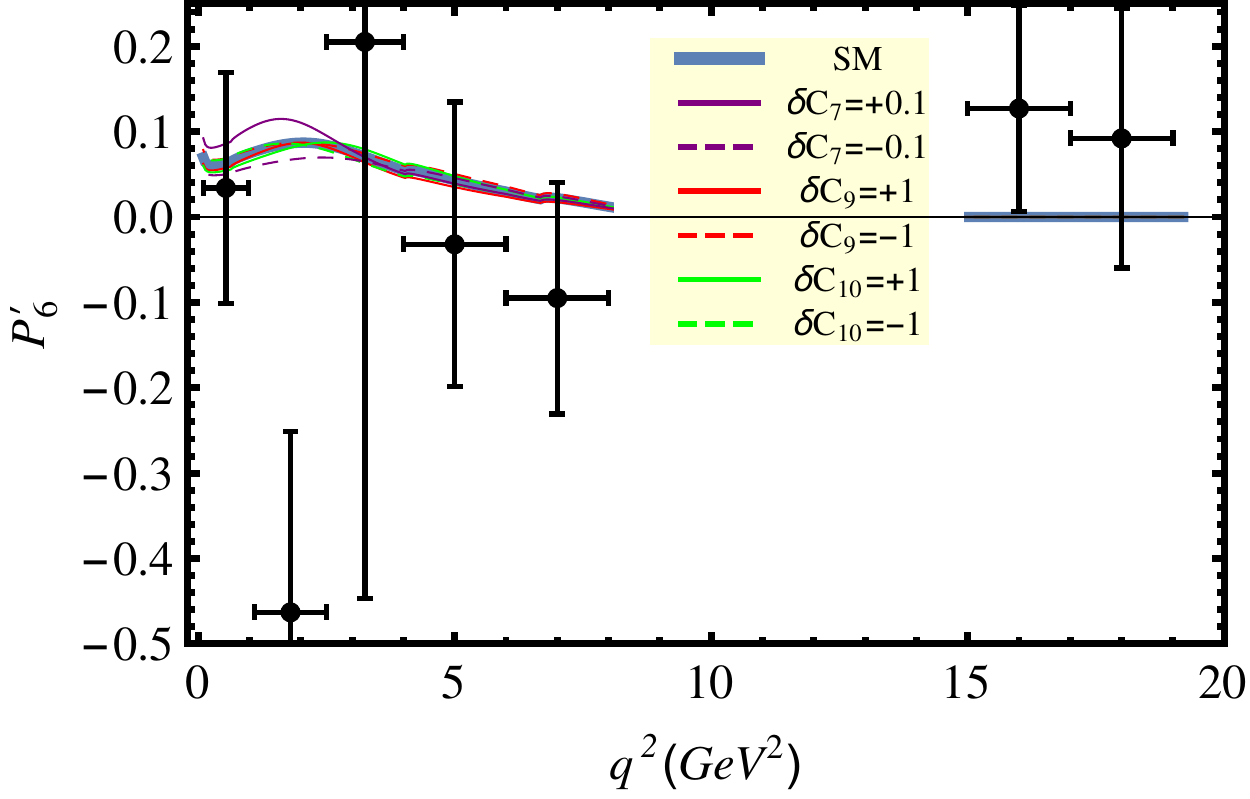}
\includegraphics[width=0.49\textwidth]{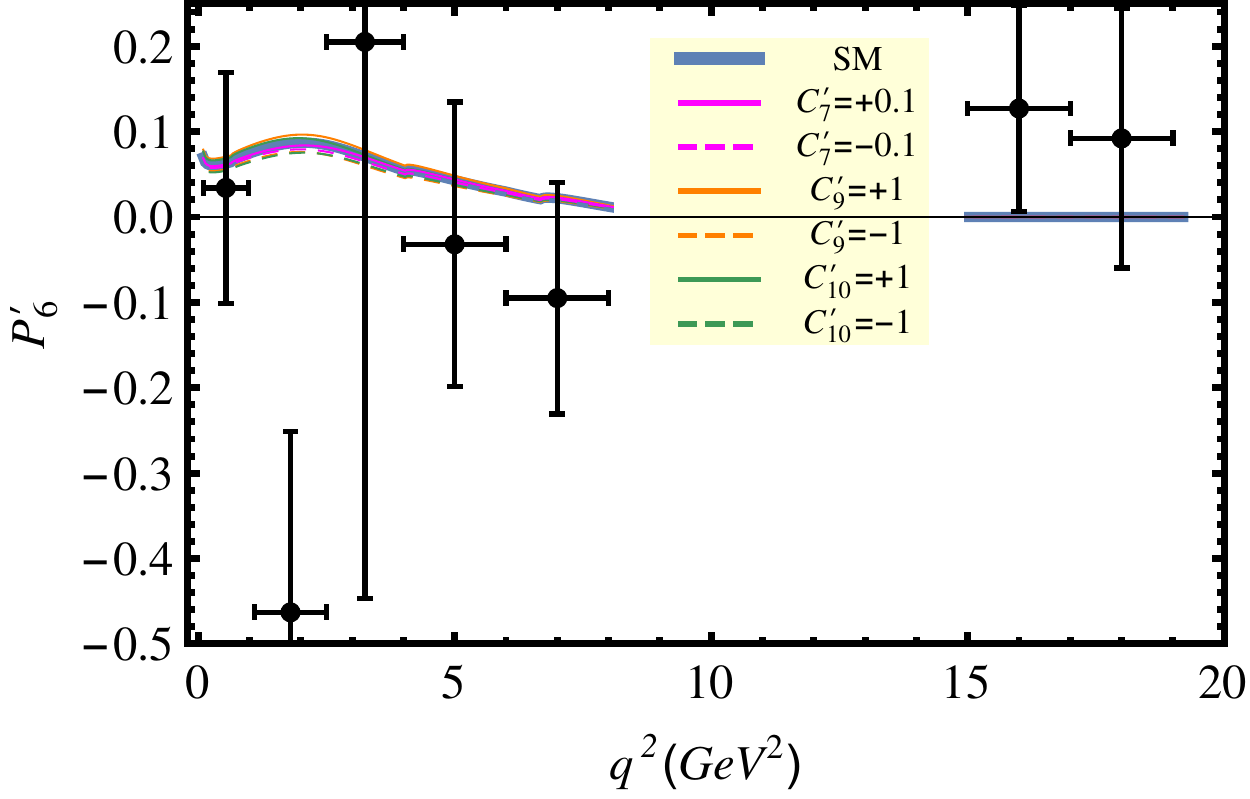}
\includegraphics[width=0.49\textwidth]{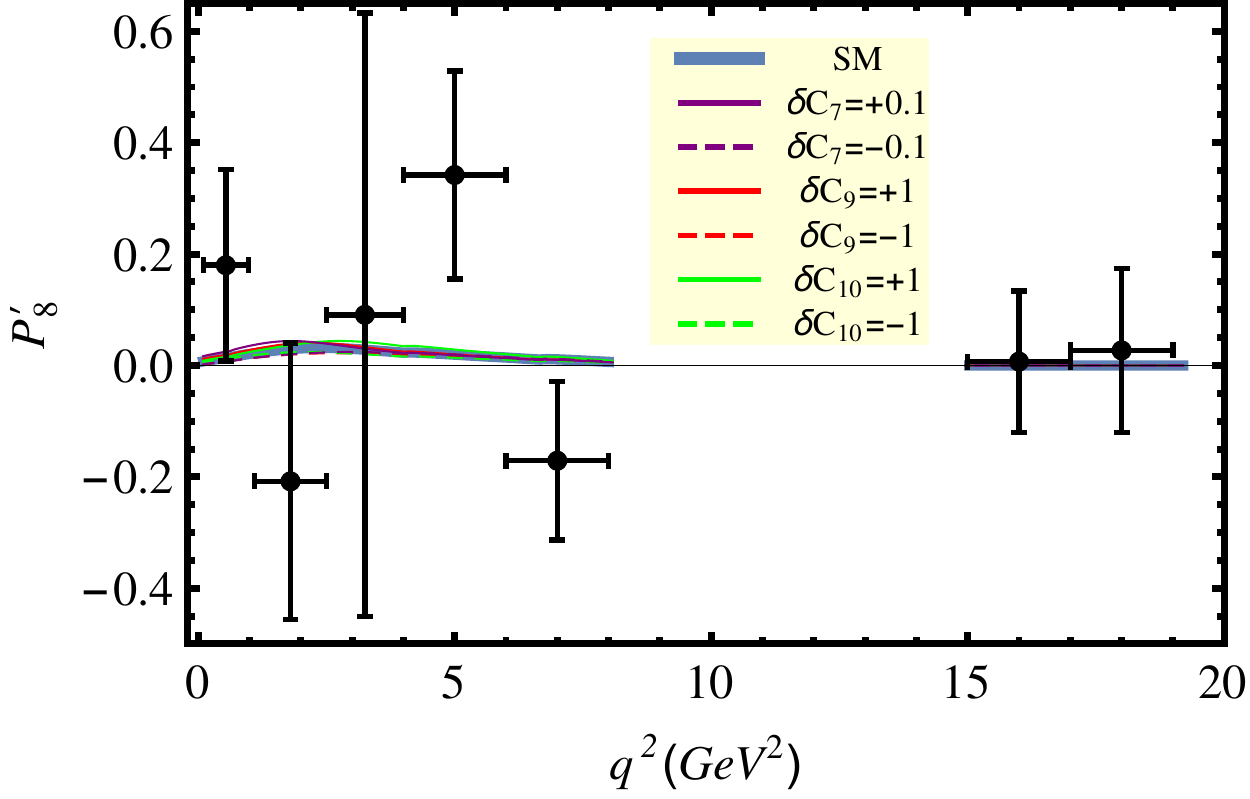}
\includegraphics[width=0.49\textwidth]{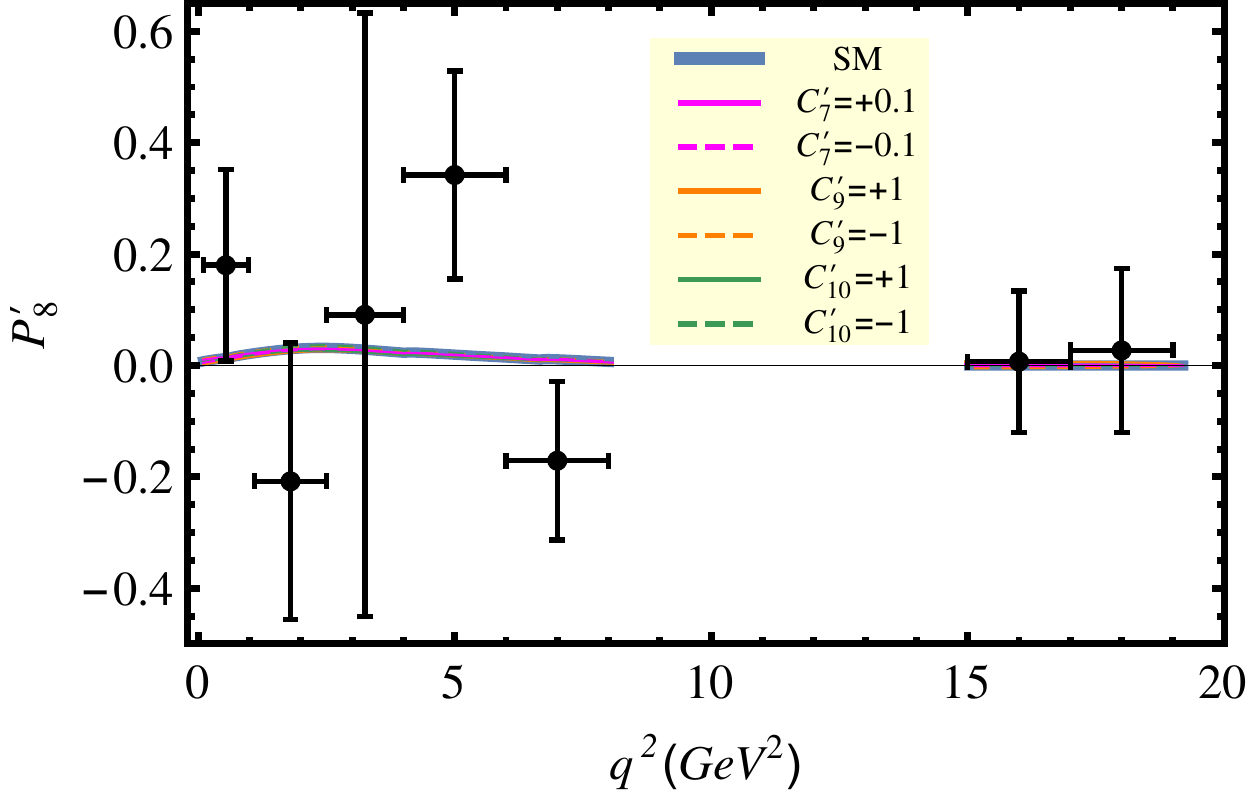}
\caption{$P_{4,5,6,8}^\prime (B\to K^* \mu^+ \mu^-)$, as described in Fig.~\ref{fig:BKstarmumu:AFB}. \label{fig:BKstarmumu:P4p-P8p}}
\end{figure}

The effects of single modified Wilson coefficients on the $S_i$ observables are shown in Figs.~\ref{fig:BKstarmumu:AFB}-\ref{fig:BKstarmumu:S7-S9}
while their effects on optimised observables $P_i$ are shown in Figs.~\ref{fig:BKstarmumu:P1-P3},\ref{fig:BKstarmumu:P4p-P8p}.

%%%%%%%%%%%%%%%%%%%%%%%%%%%%%%%%%%%%%%%%%%%%%%%%%%%%%%%%%%%%%%%%%%%%%%%%%%%%%%%%%%%%%%%%%%
%%%%%%%%%%%%%%%%%%%%%%%%%%%%%%%%%%%%%%%%%%%%%%%%%%%%%%%%%%%%%%%%%%%%%%%%%%%%%%%%%%%%%%%%%%
%%%%%%%%%%%%%%%%%%%%%%%%%%%%%%%%%%%%%%%%%%%%%%%%%%%%%%%%%%%%%%%%%%%%%%%%%%%%%%%%%%%%%%%%%%
%%%%%%%%%%%%%%%%%%%%%%%%%%%%%%%%%%%%%%%%%%%%%%%%%%%%%%%%%%%%%%%%%%%%%%%%%%%%%%%%%%%%%%%%%%
%%%%%%%%%%%%%%%%%%%%%%%%%%%%%%%%%%%%%%%%%%%%%%%%%%%%%%%%%%%%%%%%%%%%%%%%%%%%%%%%%%%%%%%%%%
%%%%%%%%%%%%%%%%%%%%%%%%%%%%%%%%%%%%%%%%%%%%%%%%%%%%%%%%%%%%%%%%%%%%%%%%%%%%%%%%%%%%%%%%%%

\clearpage

\providecommand{\href}[2]{#2}\begingroup\raggedright\endgroup

\end{document}